\documentclass[twocolumn]{aastex63}

\usepackage{gensymb,amsmath}

\usepackage{tabularx}

\newcommand\clearrow{\global\let\rowmac\relax}
\clearrow
\shorttitle{CH$_3$NCO and CH$_3$SH in G31.41+0.31}
\shortauthors{}

\begin{document}
\title{{Identification of Methyl Isocyanate and Other Complex Organic Molecules in a Hot Molecular Core, G31.41+0.31}}
\email{ankan.das@gmail.com,\\
prasanta.astro@gmail.com}
\author{Prasanta Gorai}
\affiliation{Department of Space, Earth \& Environment, Chalmers University of Technology, SE-412 96 Gothenburg, Sweden}
\affiliation{Indian Centre for Space Physics, 43 Chalantika, Garia Station Road, Kolkata 700084, India}
\author{Ankan Das}
\affiliation{Indian Centre for Space Physics, 43 Chalantika, Garia Station Road, Kolkata 700084, India}
\author{Takashi Shimonishi}
\affiliation{Center for Transdisciplinary Research, Niigata University, Ikarashi-ninocho 8050, Nishi-ku, Niigata, 950-2181, Japan}
\affiliation{Environmental Science Program, Department of Science, Faculty of Science, Niigata University, Ikarashi-ninocho 8050, Nishi-ku, 
Niigata, 950-2181, Japan}
\author{Dipen Sahu}
\affiliation{Academia Sinica Institute of Astronomy and Astrophysics, 11F of AS/NTU Astronomy-Mathematics Building, No.1, Section 4, Roosevelt Rd, Taipei 
10617, Taiwan, R.O.C.}
\author{Suman Kumar Mondal}
\affiliation{Indian Centre for Space Physics, 43 Chalantika, Garia Station Road, Kolkata 700084, India}
\author{Bratati Bhat}
\affiliation{Indian Centre for Space Physics, 43 Chalantika, Garia Station Road, Kolkata 700084, India}
\author{Sandip K. Chakrabarti}
\affiliation{Indian Centre for Space Physics, 43 Chalantika, Garia Station Road, Kolkata 700084, India}

\begin{abstract}

G31.41+0.31 is a well known chemically rich hot molecular core (HMC). Using Band 3 observations of Atacama Large Millimeter Array (ALMA), we 
have analyzed the chemical and physical properties of the source. We have identified methyl isocyanate (CH$_3$NCO), a precursor 
of prebiotic molecules, towards the source. In addition to this, we have reported complex organic molecules (COMs) like methanol (CH$_3$OH), 
methanethiol (CH$_3$SH), and methyl formate (CH$_3$OCHO). Additionally, we have used transitions from molecules like HCN, HCO$^+$, SiO to 
trace the presence of infall and outflow signatures around the star-forming region. For the COMs, we have estimated the column 
densities and kinetic temperatures, assuming molecular excitation under local thermodynamic equilibrium (LTE) conditions. From the estimated 
kinetic temperatures of certain COMs, we found that multiple temperature components may be present in the HMC environment. Comparing the 
obtained molecular column densities between the existing observational results toward other HMCs, it seems that the COMs are favourably 
produced in the hot-core environment ($\sim 100$~K or higher). Though the spectral emissions towards G31.41+0.31 are not fully resolved, 
we find that CH$_3$NCO and other COMs are possibly formed on grain/ice phase and populate the gas environment similar to other hot cores 
like Sgr B2, Orion KL, and G10.47+0.03, etc.

%We report a detailed observational analysis of a hot molecular core (HMC), G31.41+0.31, by analyzing ALMA archival data in band
%3. For the first time, we detected the prebiotic molecule, methyl isocyanate in G31.41+0.31.  In addition two well abundant molecules, 
%methanol and methyl formate, we have also detected complex organic molecule (COM) like . Assuming LTE conditions, using the
%emission features of the molecules, we estimated the presence of two temperature components; one is associated with $\sim$33 K and 
%another is associated with excitation temperature of $\sim$180 K. Due to the coarser beam size of the observation, the molecular 
%emissions were not resolved. Due to this limitation, we can not estimate the origin of the two temperature components. To understand 
%the chemical origin of the molecules, in addition to continuum dust emission, we studied outflows, infall using molecular tracers 
%like SiO, HCN and H$^{13}$CO$^{+}$. We estimated column density/abundance of molecules and compared the results with other hot-cores
%and found that the chemical origin of COMs are similar to other hot cores.

\end{abstract}


\section{Introduction}
Chemistry in high-mass star-forming regions (HMSFRs) is very rich, and it has a significant impact on the evolution of the Interstellar Medium (ISM) 
\citep[e.g.,][]{vand98,tan14}. Over the years, numerous complex organic molecules (COMs) were identified in hot molecular cores (HMCs) 
\citep[e.g.,][]{herb09}. Hot molecular cores are compact, hot, dense, and associated with luminous infra-red (IR) or with ultra-compact (UC) HII 
regions \citep{cesa10,bonf17}. To date, a number of HMCs were discovered \citep{Kurt00}, while a limited number 
of circumstellar disks were found around B type stars such as the disks in Cepheus A HW2 \citep{pate05}, HH80-81 \citep{gira18} and the disk in the 
late-O-type star, IRAS 13481-6124 \citep{krau10}. Their dust emission has identified the majority of these disk structures. However, they are 
also associated with molecular emission. Complex organic molecules, CH$_3$CN, was observed in Cepheus A HW2 and P-V diagram of CH$_3$CN transitions 
used to understand the velocity gradient that explains gravitationally bound rotational motion. 

Many complex physical processes, such as accretion, infall, and outflows are present in the earliest phase of HMSFRs during their various evolutionary 
stages. Therefore the chemistry of these regions is used as a diagnostic tool to examine the nature of them. Complex organic molecules are omnipresent 
in the HMCs and show rich chemistry. Chemistry of these regions plays an essential role, which helps trace the physical parameters associated with
particular regions, such as density, temperature, and ionization rate. Since the abundances of chemical species are time dependent and collapsing 
time of a HMSFR is short, a comparison of observed abundance with modeling can be used to trace the chemical diversity of those region. For HMCs, both 
grain-surface and gas-phase processes are efficient. Various desorption mechanisms, such as thermal (active in hot core temperature) and non-thermal 
desorption (dominant at low temperature) processes help release the grain surface species to the gas phase.

G31.41+0.31 (hereafter G31) is an HMC that is thought to be located at a 7.9 kpc distance from the Earth \citep{chur90}.  This source has a luminosity of about 
$\rm{\sim 3\times10^{5}}$\(L_\odot\). It is supposed to be heated by O and B type stars \citep{cesa10,belt09}. Previous studies \citep{cesa11,cesa17} found
some hints of the existence of disk in G31.41+0.31. \cite{osor09} performed detailed modeling of the SED of dust to obtain the physical parameters of 
the star and the core. Using these parameters, they estimated the ammonia emission and compared it with the VLA observation having a sub-arcsec 
resolution of G31. They obtained the mass of the central star $\sim$ 20-25 M$_{\odot}$ and significantly higher mass (1400-1800 M$_{\odot}$) of 
the envelope by considering envelope size, R$_{env}$ = 30000 AU. Recently, \cite{belt18} observed G31 with ALMA at $1.4$ mm resolved G31 into two 
cores, namely main and NE. They observed the physical properties such as accelerating infall, rotational spin up, red-shifted absorption, and 
possible outflow directions associated with this source. The UC HII region is situated at the North-East side of G31 and separated by $\sim 5^{''}$ 
from the main core \citep{cesa10}. The main core's estimated mass is 120 M$_{\odot}$, which is based on old distance 7.9 kpc \citep{belt18}. However, 
new parallax measurement suggests that G31 is located at a distance of 3.7 kpc. Therefore the luminosity of the source would be
$\rm{\sim 4\times10^{4}}$\(L_\odot\) and estimated mass of the main core ($M_c$) of G31 is $\sim$ 26 M$_{\odot}$ \citep{ried19,esta19,belt19}. However, 
this estimation should be considered the lower limit because high angular resolution might have filtered spatially extended emission \citep{belt19}.

A wide variety of species (around $70$) were already identified in G31. The simplest form 
of sugar, glycolaldehyde (HOCH$_2$CHO), was observed in G31 with IRAM PdBI at $1.4$, $2.1$, and $2.9$ mm observations \citep{belt09}. This is the 
earliest evidence of the presence of sugar outside our Galactic center. A large  number of COMs such as methanol (CH$_3$OH), 
methyl cyanide (CH$_3$CN), ethanol (C$_2$H$_5$OH), ethyl cyanide (C$_2$H$_5$CN), methyl formate (CH$_3$OCHO), dimethyl ether (CH$_3$OCH$_3$), ethylene 
glycol (CH$_2$OH)$_2$, and acetone (CH$_3$COCH$_3$) had already been identified in G31 by using both single dish (IRAM 30m, GBT) and Submillimeter Array 
(SMA) observations \citep[and references therein]{rivi17,isok13}.

The presence of a large number of COMs, including branched chain molecule might be an indication of the existence of prebiotic molecules and 
essential building blocks of life in space \citep{bell14,chak15,maju15,sil18,sahu20}.  \cite{goes15} detected various COMs (e.g., methyl isocyanate, formamide, glycolaldehyde, ethylamine etc.) in a Jupiter 
family comet, 67P/Churyumov-Gerasimenko (67P/C-G). Among them, methyl isocyanate (CH$_3$NCO) was found to be relatively abundant as compared to other 
observed COMs. However, \cite{altw17} reported the resulting CH$_3$NCO abundance as an upper limit. The presence of CH$_3$NCO in space
is supposed to be important because of its peptide-like bond. Detection of CH$_3$NCO was first claimed by \cite{half15} in a massive hot molecular core, Sgr B2. It was also detected in Orion KL \citep{cern16}. 
It has recently been observed in G10.47+0.03 \citep{gora20}. Here, in G31, we have identified some transitions of this species.

Methanethiol ($\rm{CH_3SH}$) is a sulfur analog of methanol, where the S atom is replaced by an O atom. To date, this is the only complex sulfur-bearing species 
(having 6 atoms) which has been firmly detected in the ISM. Tentative identification of higher-order thiols, ethanethiol 
($\rm{C_2H_5SH}$) has been reported in Sgr B2N \citep{mull16}. Identification of $\rm{CH_3SH}$ was also reported toward the Galactic center \citep{Link79,mull16}, 
Orion KL \citep{kole14} and chemically rich massive hot core, G327.3-6 \citep{gibb00} and 67P/C-G comet \citep{calm16}. 

In this work, we present the interferometric (ALMA archival data) observation towards G31. We report first detection of CH$_3$NCO and CH$_3$SH in G31 and 
discuss the kinematics associated with this HMC. We also discuss the chemistry of other complex species in this source. The remainder of this paper is summarized as follows.
In Section 2, we describe observational details and data reduction processes. All results are presented and discussed in Section 3 and Section 4 respectively. Finally, 
we draw our conclusion in Section 5.

\section{ALMA archival data}
In this paper, we used Atacama Large Millimeter/submillimeter Array (ALMA) cycle 3 archival data 
(\#2015.1.01193.S) to study COMs in G31. The observation was performed during June, 2016 using forty 12 m antennas. 
The shortest and the longest baseline  was $15$ m, and $783.0$ m, respectively, and synthesized beam size was
$\sim$ 1.1 $''$. Four spectral windows (86.559-87.028 GHz, 88.479-88.947 GHz, 
98.977-99.211 GHz, and 101.076-101.545 GHz) of width $~0.47$~GHz was set up to observe the target source, G31; phase center was 
$\alpha$(J2000) = 18$^h$47$^m$34.309$^s$, $\delta$ =-1$^{o}$12$'$46.00$''$. The flux calibrator was Titan, the phase calibrator 
was J1824+0119, and the bandpass calibrator was J1751+0939. 
The data cube has a spectral resolution of 244 kHz ($\sim$0.84 km s$^{-1}$) with a synthesized 
beam of $\sim$ $\rm{1.19^{''}\times 0.98^{''}}$ having 
position angle (PA) $\sim$ 76$^{o}$. The systematic velocity of the source is known to be
$\sim$ $97$ km s$^{-1}$ \citep{cesa10,rivi17}. 
Image analysis was done using CASA 4.7.2 software \citep{mcmu07}. Implementing the first-order baseline 
fit and by using `uvcontsub' command available in the CASA program, 
we separated each spectral window into two data: continuum and line emission. 
The identification of lines of all the observed species presented in this paper
is carried out using CASSIS (This software was developed by 
IRAP-UPS/CNRS, http://cassis.irap.omp.eu) together with the spectroscopic database 
`Cologne Database for Molecular Spectroscopy' \citep[CDMS,][]{mull01,mull05} 
\footnote{\url{(https://www.astro.uni-koeln.de/cdms)}} and Jet Propulsion Laboratory 
\citep[JPL,][]{pick98}\footnote{\url{(http://spec.jpl.nasa.gov/)}} database.

\section{Results}
\subsection{Continuum emission and $\rm{H_2}$ column density estimation}

\begin{figure}
\centering
\includegraphics[height=9cm,width=10cm]{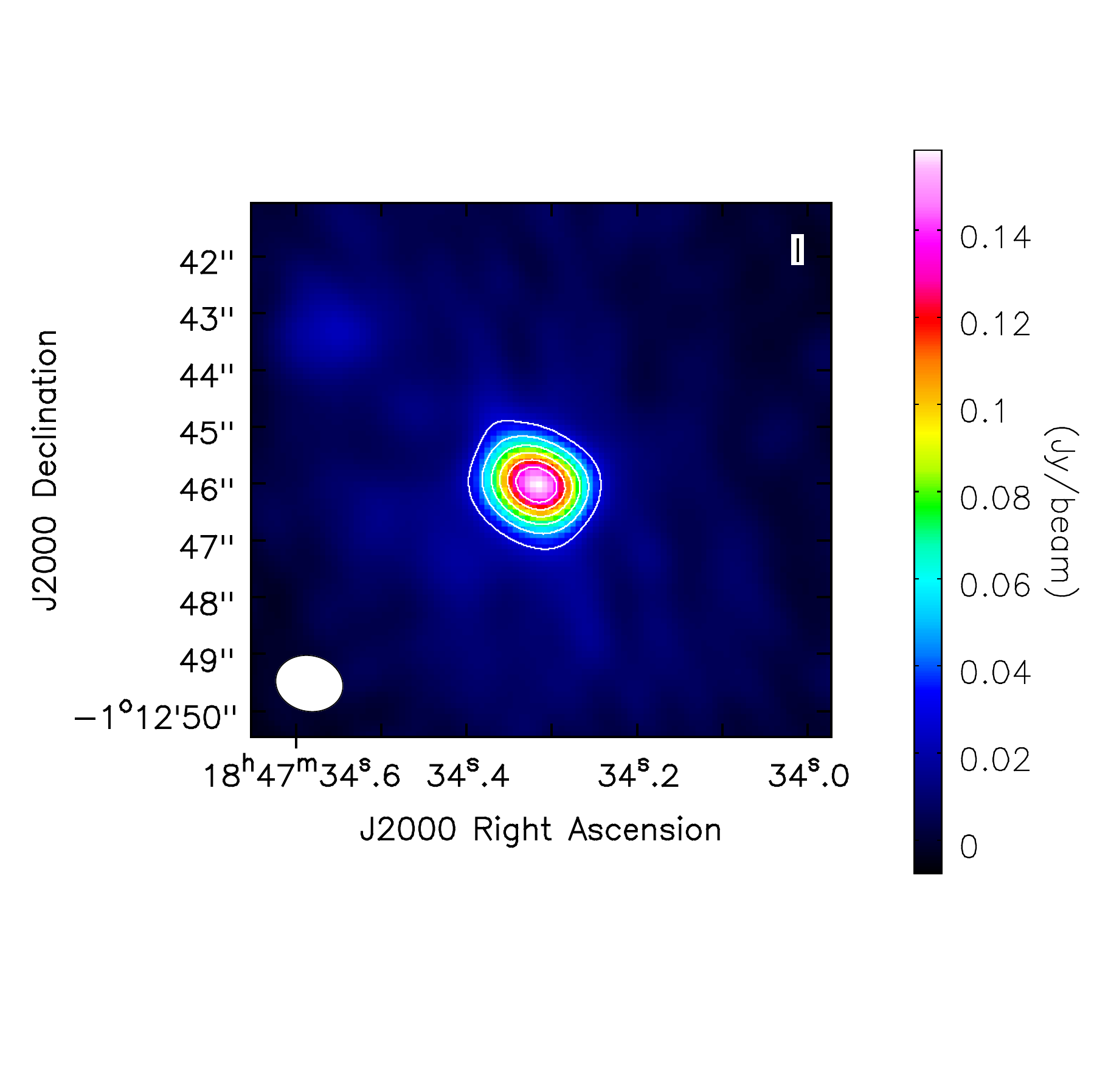}
\caption {Continuum map of G31 obtained with ALMA at $94$ GHz. Contour levels start at 4$\sigma$ (1$\sigma$=7 mJy/beam) and steps are in  
3$\sigma$. Ellipse at the below left corner of the image shows the synthesized beam ($\rm{1.19'' \times 0.98''}$) with PA=76.83$^{o}$.} 
\label{fig:cont}
\end{figure}

The continuum image of G31 at $94.3$ GHz is shown in Figure \ref{fig:cont}. 
The synthesized beam size of the continuum is $\rm{1.19^{''}\times 0.98^{''}}$ with 
PA $\sim$76$\degree$. Using two-dimensional Gaussian fitting over the dust continuum emission, we obtained the deconvolved size of G31 $\sim 0.90 '' \times 0.68''$ .

Assuming the distance to the source $7.9$ kpc, we obtained a source size $\sim 7110 \times5372 $ AU. This source size changes to $3330 \times 2516$ AU,
while $\sim 3.7$ kpc distance is used.

\cite{belt18} observed G31 with an angular resolution of 0.22$''$ and found the continuum brightness temperature  $\sim$132 K, which is comparable to 
the dust temperature ($\sim 150$ K). This suggests that the dust continuum emission observed at $217$ GHz is optically thick. 
\cite{rivi17} also observed that dust opacity is high ($\tau$ = 2.6) at 220 GHz.  

We have observed peak intensity of the continuum is $\sim$158 mJy/beam which corresponds to a brightness temperature $18$ K (using Rayleigh-Jeans 
approximation, 1 Jy/beam $\equiv$ 118 K). This estimation is much smaller than the assumed dust temperature of 150 K, 
this indicates a small optical depth ($\rm{\tau_\nu}$) $\sim$ 0.12 (using $T_{mb}=T_d(1-exp(-\tau_\nu))$.

It should be noted that in our observation, the source was not resolved or at best marginally resolved; therefore, the estimated brightness temperature
could be underestimated.. After adding the beam dilution correction for the unresolved source, the brightness temperature is estimated to be $\sim$36 K. 
Considering this value of brightness temperature; dust emission remains optically thin ($\tau_\nu \sim 0.24 $). Due to the collapsing envelope, 
we can expect higher density and temperature toward the inner region than the outer envelope of this source. Since dust density is higher 
towards the inner region, opacity would be higher, as 
observed by \cite{belt18}. It may reduce the true line intensity of the observed transitions. However, due to the lower angular resolution, our present 
data would not provide further details about the dust opacity of the whole region of G31. 

Assuming the dust continuum to be optically thin, flux density can be written as,
%Flux density of the dust continuum for optically thin condition can be written as,

\begin{equation}
S_\nu^{beam} =\Omega_{beam} \tau_\nu B_\nu (T_d),
\end{equation}
{where $\Omega_{beam}$=$\frac{\pi}{4ln2}$ $\times \theta_{major} \times \theta_{minor}$} is solid angle of the synthesized beam, 
S$_\nu^{beam}$ is the peak flux density measured in unit of mJy/beam, $\tau_\nu$ is optical depth, and $\rm{T_d}$ is dust temperature 
and $\rm{B_\nu(T_d)}$ is Planck function 
\citep{whit92}. Optical depth can be expressed as,
\begin{equation}
 \tau_\nu =\rho_d\kappa_\nu L,
\end{equation}
where $\rho_d$ is the mass density of dust, $\kappa_\nu$ is the mass absorption coefficient, and L is the path length.
Using the dust to gas mass ratio ($Z$), the mass density of the dust can be written as,
\begin{equation}
\rho_d = Z\mu_H\rho_{H_2}=2Z{  \mu_H}N_{H_2}m_H/L,
\end{equation}
where $\rho_{H_2}$ is the mass density of hydrogen molecule, $\rm{N_{H_2}}$ is the column density of hydrogen, $\rm{m_H}$ is the hydrogen mass,
and $\rm{\mu_H}$ is the mean atomic mass per hydrogen atom. Here, we used $Z=0.01$, $\mu_H=1.41$ \citep{cox00}, and dust temperature 150 K.
Two dimensional Gaussian fitting of the continuum yields a peak intensity (S$_\nu^{beam}$) of dust continuum of 158.4 mJy/beam and integrated 
flux (F$_\nu$) of 242.4 mJy. RMS noise of the continuum map was found to be $5$ mJy/beam. From the above equations column density of molecular 
hydrogen can be written as,

\begin{equation}
N_{H_2} = \frac{S_\nu^{beam} /\Omega_{beam}}{2\kappa_\nu B_\nu(T_d)Z\mu_H m_H}.
\end{equation}

According to the interpolation of the data presented in \cite{osse94}, we estimated the mass absorption coefficient. By considering a thin ice condition, 
here, we obtained the mass absorption coefficient at 94 GHz ($3187.90$ micron) $\sim 0.176$ cm$^2$/g. By using equation 4, total hydrogen column density 
of G31 source is $\sim{\rm1.53\times10^{25}}$ cm$^{-2}$.

\subsection{Line analysis, molecular emission and spectral fitting}
 
Observed spectra are obtained within the circular region having a diameter of 1.1$''$ centered at RA (J2000) = 18$^h$47$^m$34.309$^s$, 
Dec (J2000) =-1$^{o}$12$'$46.00$''$. Line parameters of all the observed transitions are obtained 
using a single Gaussian fit. From fitting, we have estimated the line width (FWHM), LSR velocity, and the integrated intensity.
All the line parameters of observed molecules 
such as molecular transitions (quantum numbers) along with their  rest frequency ($\nu$), line width (FWHM, $\Delta V$), integrated intensity 
($\rm{\int T_{mb}dV}$), upper state energy (E$_u$), $\rm{V_{LSR}}$, and transition line intensity (S$\mu^{2}$, where S is the transition line 
strength and $\mu$ is the dipole moment) are presented in Table \ref{table:line-parameters}. Overall, we have observed a broad line width 
($\ge$ 5 km s$^{-1}$) and all lines are at the center around the systematic velocity of the source ($\rm{\sim97\ km\ s^{-1}}$) 
(see Table \ref{table:line-parameters}). In the Appendix, we have presented observed and Gaussian fitted spectra (see Figures 
\ref{Gfit-CH3OH}-\ref{Gfit-other}).

\begin{figure*}[t]
\centering
\includegraphics[height=5cm, width=14cm]{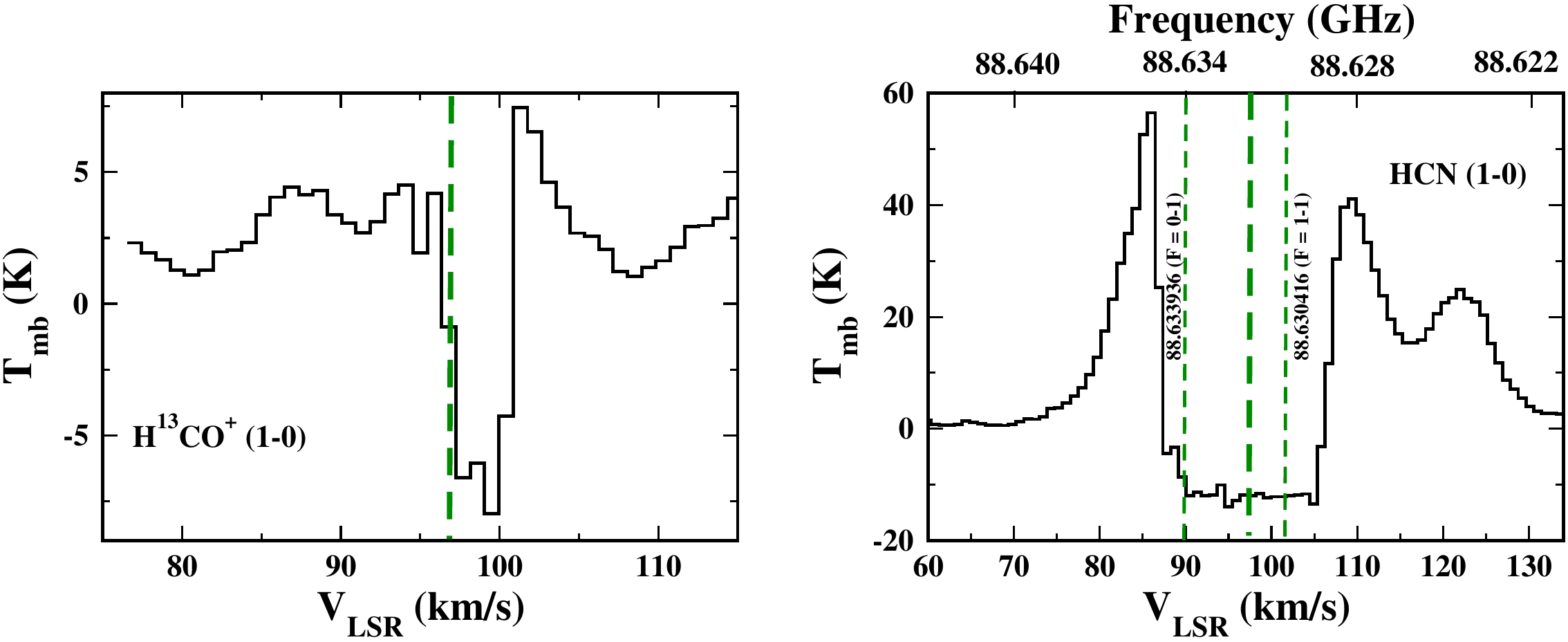}
\caption{Observed spectra of (a) $\rm{H^{13}CO^{+}}$ (inverse P-Cygni) and (b) HCN towards the dust continuum peak. 
86.631847 GHz transition of HCN is used to set the velocity scale. Other two lines (green dashed) are plotted in the same figure by taking 
the difference in frequency between the transitions which corresponds to have velocity shift by $\sim$ 7 $\rm{km\ s^{-1}}$. for transition 88.63393 and 
$\sim$ 5 $\rm{km\ s^{-1}}$ for 88.63041 GHz. Bold dashed green line shows the systematic velocity ($\rm{V_{LSR}}$) of the source 
$\rm{\sim97\ km\ s^{-1}}$.}
\label{abs-sp}
\end{figure*}

\subsubsection{Simple Molecules}
We have observed SO$_2$ ($86.63908$ GHz) and H$_2$CO ($101.33299$ GHz) in G31.
We have also identified some simple molecules such as HCN, H$^{13}$CO$^{+}$, and SiO, which are widely used to trace various physical 
processes of star-forming regions. The observed spectral profile of H$^{13}$CO$^{+}$ shows red-shifted absorption, which is shown in 
Figure \ref{abs-sp}a. This is a signature of inverse P-Cygni profile. {This suggests that the source has an infall nature}. Observed emission of 
SiO and HCN traces the outflows in this source.

Figure \ref{abs-sp}b shows the observed spectral profile of HCN. Here, we have identified three hyperfine transitions of HCN (F = 1$\rightarrow$1, 
2$\rightarrow$1, and 1$\rightarrow$0 hyperfine components of the J=1$\rightarrow$0 transition, see Table \ref{table:absorption}). Three hyperfine 
transitions of HCN are appeared in absorption and blended together. These three absorption lines together resulting in a broad feature. HCN traces the 
outflow associated with the source. The observed transitions of both HCN and H$^{13}$CO$^{+}$ are found to be optically thick. 
 
Here, we have also detected one transition of SiO at 86.84696 GHz (J=2$\rightarrow$1). SiO shows the absorption feature towards the 
center of the dust continuum and emission profile at all other positions. \citep{belt18} also observed similar spectral feature of SiO
(5-4, 217.10498 GHz). SiO traces outflows associated with this source.

\subsubsection{Complex Organic Molecules}
We have observed numerous COMs that were previously reported in G31 \citep[e.g.,][and references therein]{rivi17}. Additionally, for the 
first time in this source, we are reporting the detection of two new molecules, methyl isocyanate (CH$_3$NCO) and methanethiol (CH$_3$SH).
Altogether, we have identified six transitions of CH$_3$NCO with the ground vibrational state ($v_b=0$) and one transition with $v_b=1$. Among 
the six transitions of CH$_3$NCO, one transition (86.86674 GHz) is blended with ethylene glycol (confirmed from LTE modeling). 

We have identified three different peaks of CH$_3$SH, where each peak contains multiple transitions. 
These transitions are reported in Table \ref{table:line-parameters}. We are unable to separate these observed 
transitions. Because of closely spaced transitions, it is difficult to distinguish two or more different transitions from a single observed peak of 
CH$_3$SH. Thus, to obtain the observed line parameters of CH$_3$SH transitions, we have fitted multiple Gaussian  components to the observed spectra.
For multiple Gaussian fitting, we have used fixed values of velocity separation and expected line intensity ratio. Only FWHM is kept as a free
parameter. This method works well in the observed spectral profile 1 and 3 around 101.139 GHz and 101.167 GHz of CH$_3$SH, where there are two 
transitions around each profile (see Table \ref{table:line-parameters}). Two-component Gaussian fitting corresponding to the two transitions are used to fit those spectral profiles (see Table 
\ref{table:line-parameters} and Figure \ref{Gfit-ch3sh}). However, this method does not work well for spectral profile 2, around 101.159 GHz, where 
four transitions are present. The four transitions can 
be classified into two classes based on their expected peak intensity (equivalent to S$\mu^{2}$) and E$_{up}$. The expected intensity ratio between these four 
transitions is 1:0.58:0.58:0.58. We have used two-component Gaussian fitting corresponding to the two classes of transitions, one for 101.15933 GHz 
(S$\mu^{2}$ =4.96, E$_{up}$=31.26 K) transition and another component for the other three transitions, 101.15999 GHz, 101.16066 
GHz, 101.16069 GHz (S$\mu^{2}$ = 2.90, E$_{up}$ = 52.55 K) around 101.159 GHz line profile. For the fitting, we have used 1:1.74 (1.74 is the sum of intensity of 
three transitions) intensity ratio for these two components. Therefore, resultant Gaussian corresponding to the three transitions can be obtained 
via this method; as these transitions are equally intense and have similar excitation temperature, the area under Gaussian can be divided by three 
to get the contribution from each transition multiplets. We have used only one transition from these multiplets for the rotation diagram of CH$_3$SH 
since the integrated area of all these three transitions are the same and have similar upper state energy (E$_{up}$). Therefore, these fitting results 
are used meaningfully to estimate rotational temperature using the rotation diagram method. Obtained line parameters are further used in the rotation 
diagram of CH$_3$SH. For the rotation diagram of CH$_3$SH, we use 101.13915 GHz, 101.13965 GHz, 101.15933 GHz, 101.1599 GHz, 101.16715 GHz, and 
101.16830 GHz transitions. Gaussian fits of CH$_3$SH observed spectra are provided in Figure \ref{Gfit-ch3sh}.

\begin{figure*}[t]
\begin{center}
\includegraphics[width=15cm,height=10cm]{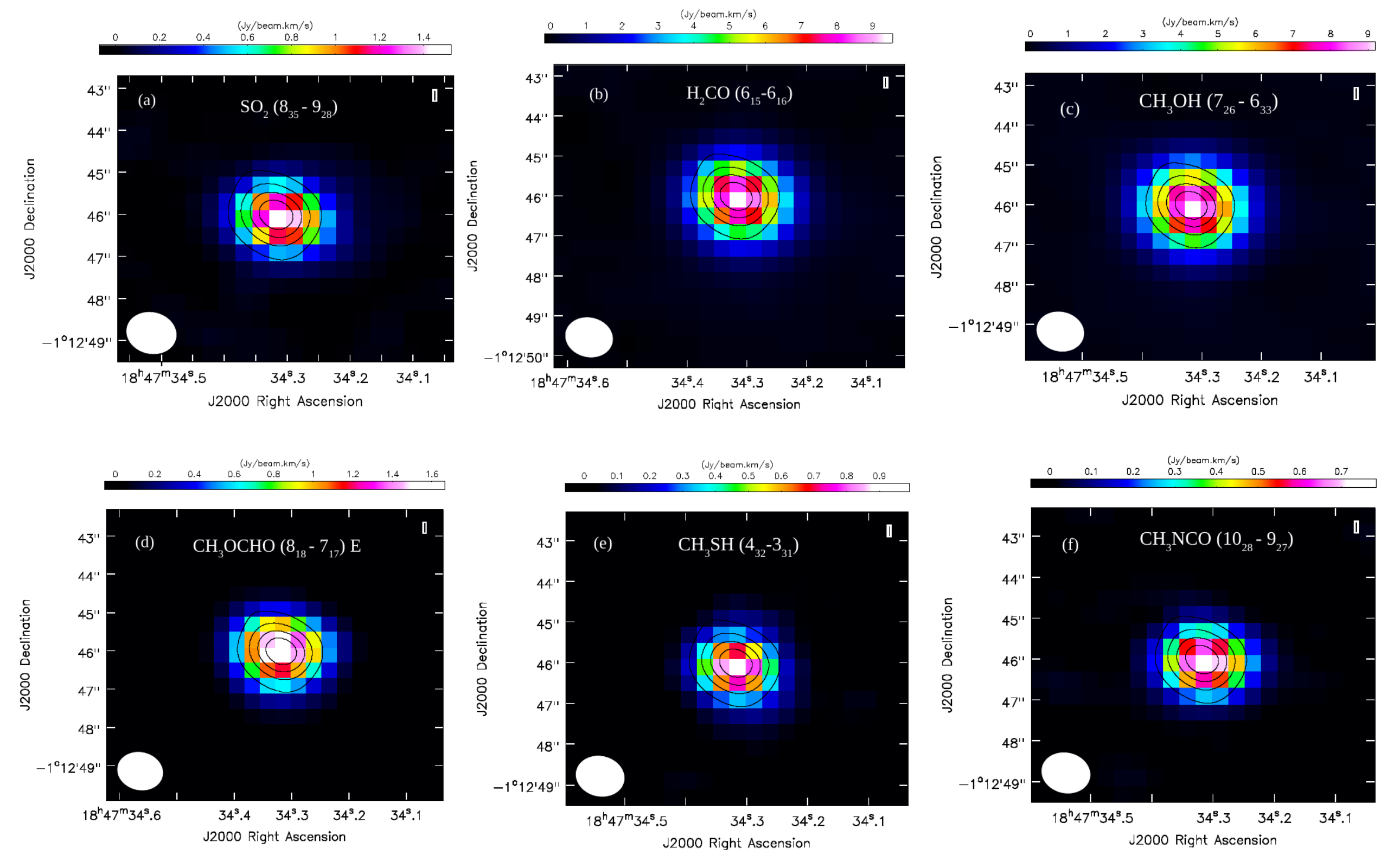}
\caption{The integrated intensity distribution (color) of (a) SO$_2$, (b) H$_2$CO, (c) CH$_3$OH, (d) CH$_3$OCHO, (e) CH$_3$SH, (f) and CH$_3$NCO 
lines overlaid of
the 3.1 mm continuum emission (black contours) is shown. Contour levels are at 20\%, 40\%, 60 \%, and 80\% of the peak flux. The synthesized beam
is shown in the lower left-hand corner of each figure. \label{int-int}}
\label{fig:integrated}
\end{center}
\end{figure*}
We also have identified the presence of two abundant COMs methanol and methyl formate in this source. Methanol is one of the most abundant COMs in 
star-forming regions.  Here, we have identified seven transitions of CH$_3$OH with different upper state energy in the ALMA band 3 regions. We have 
also identified CH$_3$OCHO transitions with a wide range of upper-level energies ($17-207$ K), which are listed in Table \ref{table:line-parameters}. 
The integrated intensity maps of some unblended transitions of SO$_2$, H$_2$CO, CH$_3$SH, CH$_3$OH, CH$_3$OHCO, and CH$_3$NCO are shown in 
Figures \ref{int-int}(a-f) where continuum is overlaid on the integrated intensity map, and all other maps correspond to different 
frequencies are given in the Appendix (see Figures \ref{int-int-ch3oh}-\ref{int-int-ch3sh}).
Using local thermodynamic equilibrium modeling and comparing with the integrated intensities of observed molecular transitions, we found that 
various transitions of CH$_3$OH, CH$_3$NCO, and CH$_3$OCHO are optically thin.

For example, in LTE and optically thin condition, the expected line ratio between the 87.01608 GHz and 86.68019 GHz transitions of CH$_3$NCO 
is 1.6 while the observed intensity (brightness temperature) ratio is 1.44. This suggests that these two lines are optically thin. Similarly for 
86.78078 GHz and 86.68019 GHz transitions of CH$_3$NCO, the observed intensity ratio between these two lines is 1.35, and the expected line ratio 
is 1.30. For the 101.12685 GHz and 101.46980 GHz transitions of CH$_3$OH, the observed intensity ratio is 2.3, which is consistent with the 
expected LTE line ratio of 2.2. For of CH$_3$OCHO, the observed line ratio between 88.68688 GHz and 88.85160 GHz transition is 4, which 
is similar to the observed integrated intensity ratio 3.7. The similarity between the expected and observed line ratios of these transitions 
suggest that they are optically thin.

\subsection{Rotation diagram analysis: estimation of gas temperature}
We have detected multiple transitions of CH$_3$OH, CH$_3$OCHO, CH$_3$SH, and CH$_3$NCO, with different $\rm{E_{u}}$. Thus, we employed 
the rotation diagram analysis to obtain the rotational temperatures. We performed the rotation diagram analysis by assuming the observed 
transitions to be optically thin and are in Local Thermodynamic Equilibrium (LTE). 
For optically thin lines, column density can be expressed as \citep{gold99},
\begin{equation}
\frac{N_u^{thin}}{g_u}=\frac{3k_B\int{T_{mb}dV}}{8\pi^{3}\nu S\mu^{2}},
\end{equation}
where g$_u$ is the degeneracy of the upper state, {  $\rm{k_B}$} is the Boltzmann constant, $\rm{\int T_{mb}dV}$ is the integrated intensity, $\nu$ 
is the rest frequency, $\mu$ is the electric dipole moment, and S is the transition line strength. 
Under LTE conditions, the total column density can be written as,

\begin{equation}
\frac{N_u^{thin}}{g_u}=\frac{N_{total}}{Q(T_{rot})}\exp^{-E_u/k_{B}T_{rot}},
\end{equation}
where $T_{rot}$ is the rotational temperature, E$_u$ is the upper state energy, $\rm{Q(T_{rot})}$ is the partition function at rotational
temperature. The equation 6 can be rearranged as,
\begin{equation}
log\Bigg(\frac{N_u^{thin}}{g_u}\Bigg)=-\Bigg(\frac{log\ e}{T_{rot}}\Bigg)\Bigg(\frac{E_u}{k_B}\Bigg)+log\Bigg(\frac{N_{total}}{Q(T_{rot})}\Bigg).
\end{equation}
Equation 7 shows that there is a linear relationship between E$_u$ and $\rm{log(N_u/g_u)}$. Two parameters, column density and rotational temperature 
can be obtained by fitting a straight line to the values of $\rm{log(N_u/g_u)}$ plotted as a function of E$_u$, where $\rm{N_u/g_u}$ is obtained from 
the observations through equation 5.

\begin{figure*}[t]
\begin{minipage}{0.45\textwidth}
\includegraphics[width=\textwidth]{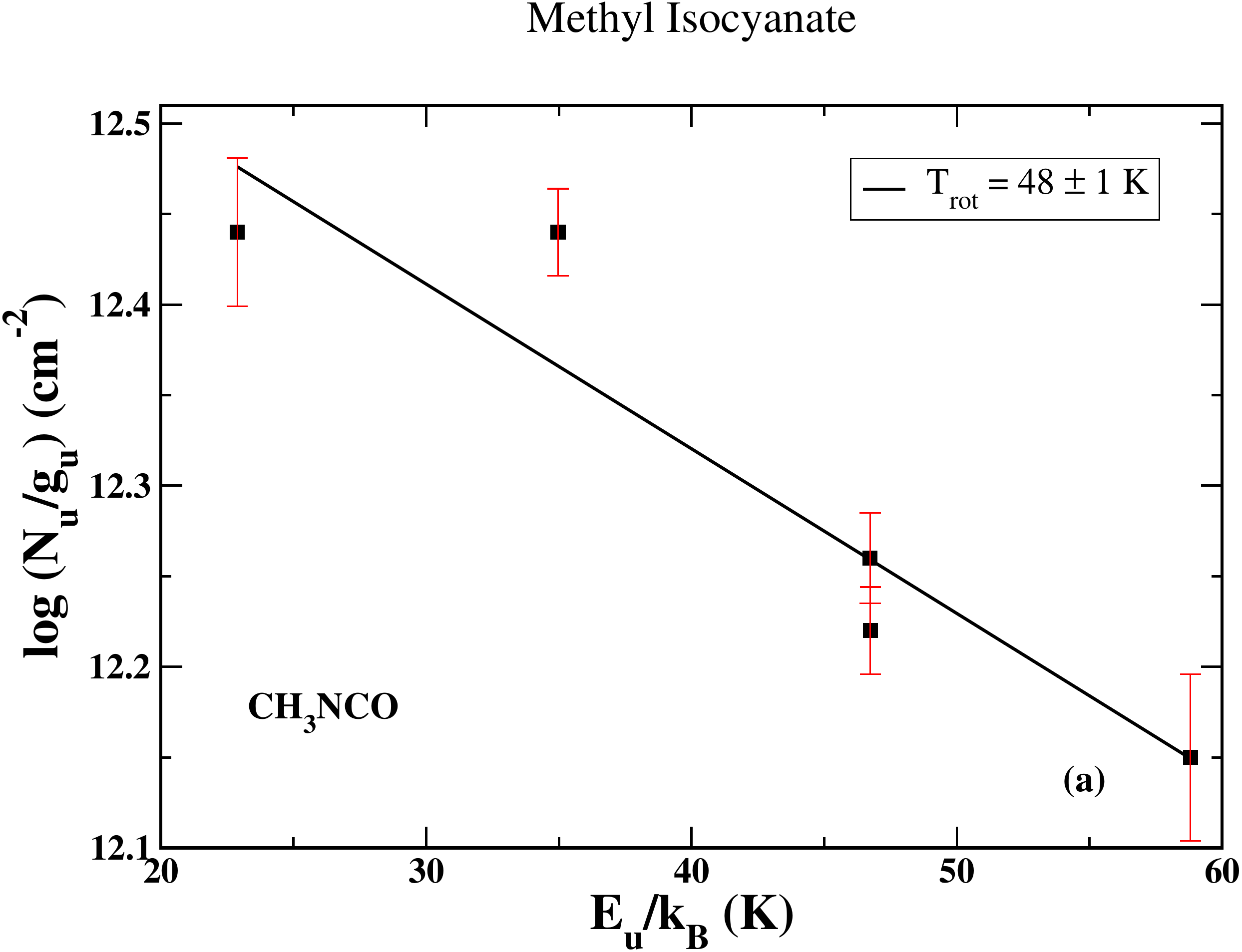}
\end{minipage}
% \hskip -0.8cm
\begin{minipage}{0.45\textwidth}
\includegraphics[width=\textwidth]{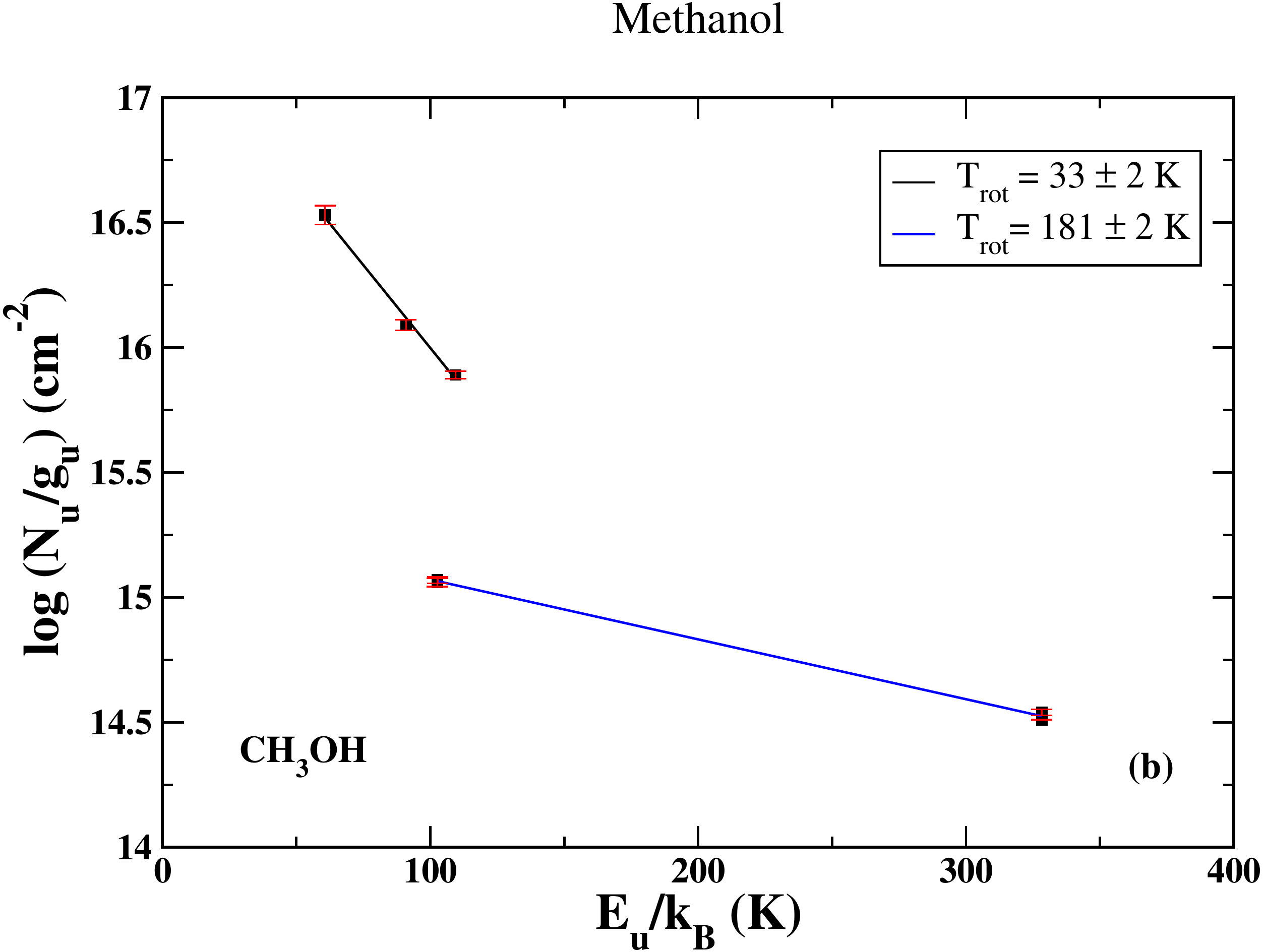}
\end{minipage}
\begin{minipage}{0.45\textwidth}
\includegraphics[width=\textwidth]{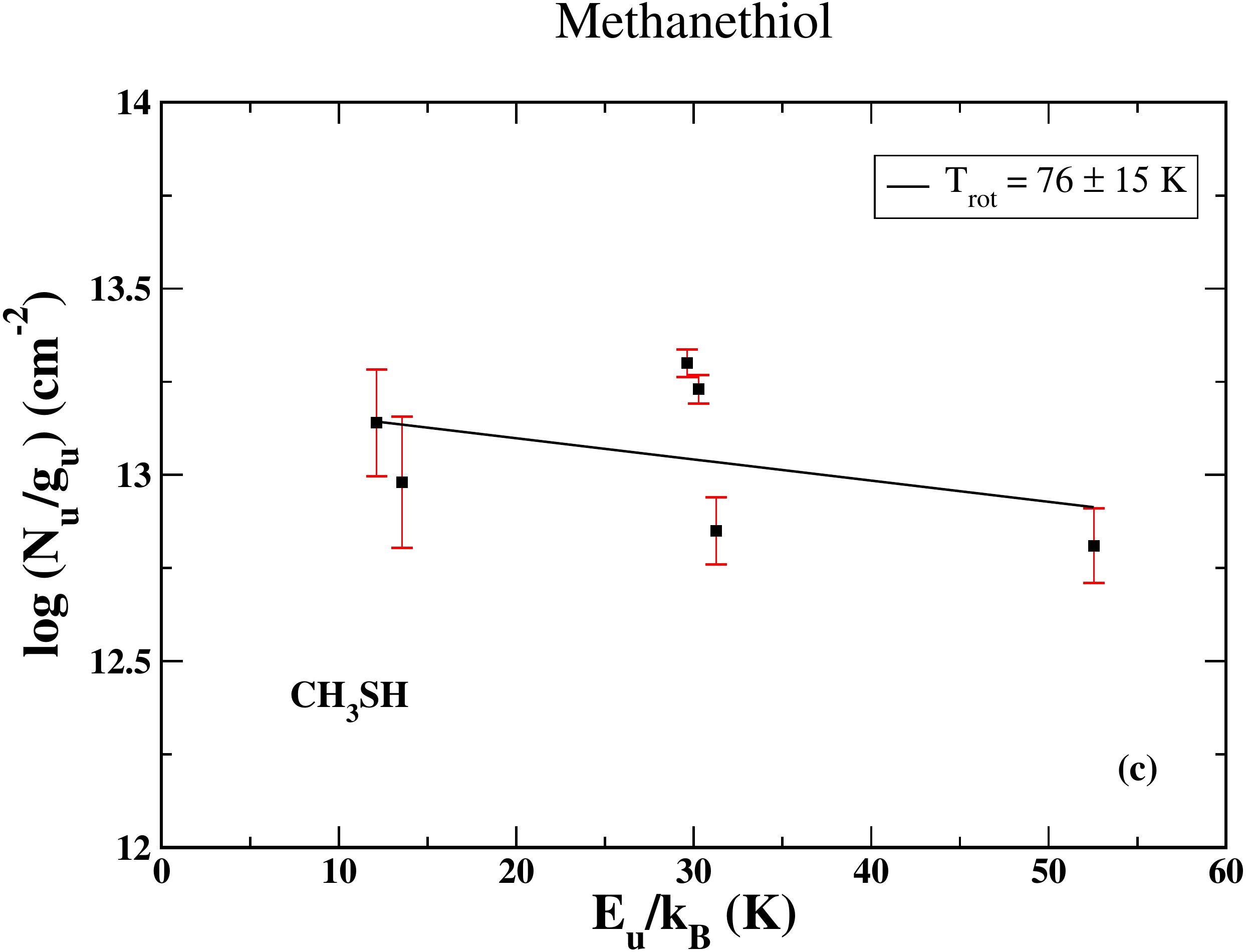}
\end{minipage}
\hskip 1.7cm
\begin{minipage}{0.45\textwidth}
\includegraphics[width=\textwidth]{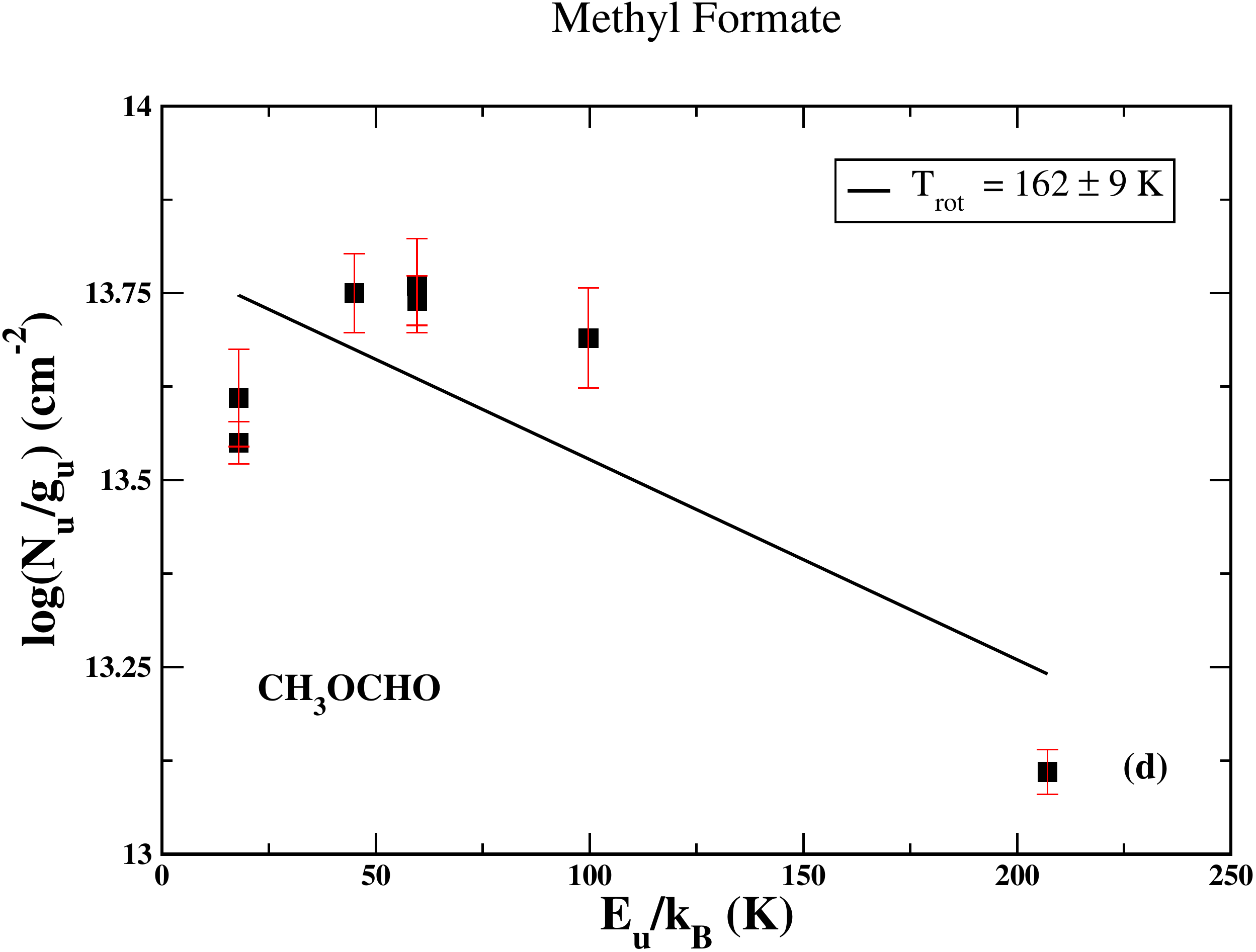}
\end{minipage}
\caption{Rotational diagram of (a) CH$_3$NCO, (b) CH$_3$OH, (c) CH$_3$SH, and (d) CH$_3$OCHO. Black filled square shows the data points, and 
the solid black and blue lines represent the fitted curve.}
\label{rot-diagram}
\end{figure*}

Rotational diagrams of methyl isocyanate, methanol, methanethiol, and methyl formate are presented in Figure \ref{rot-diagram}.
For CH$_3$OCHO, we have several transitions, but we do not see two components. Hence we fit all 
the data points with a line, which are not aligned. Though we did not find a very good fitting, the 
obtained rotational temperature of methyl formate $\sim 162$ K is consistent with the previously measured 
value in this source for the same molecule \citep{belt05,rivi17}. We have obtained a rotational temperature of 76 K for CH$_3$SH
and 48 K for CH$_3$NCO. The rotational diagram analysis suggests two temperature components of methanol, one is a cold component, 
and another is a hot component.  Cold component (solid black line, see Figure \ref{rot-diagram}b) 
temperature is obtained 33 K. In contrast, hot component (solid blue line, see Figure \ref{rot-diagram}b) temperature is 181 K. Rotational diagram of
CH$_3$OH suggests that the transitions associated with the higher $\rm{E_{u}}$ transitions (86.61557 GHz, 86.90291 GHz, 88.59478 GHz, 88.93997
GHz; hot component) are mainly arising from the hot gas surrounding the G31. In contrast, the emissions associated with the lower $E_{u}$
(101.12685 GHz, 101.29341 GHz,101.46980 GHz; cold component) transitions are from the cold region of the HMC. However, due to our observation's 
lower angular and spatial resolution, we marginally resolved these transitions. Therefore, it is not possible to provide a clear insight into the 
spatial distribution of methanol in this source.

The error bars (vertical red bars) in Figure \ref{rot-diagram} are the absolute uncertainty in a log of (N$_u$/g$_u$), and this arises from the error 
of the observed integrated intensity that we measured using a single Gaussian fitting to the observed profile of each transition. The integrated intensities with their
uncertainties of all observed transitions are provided in Table \ref{table:line-parameters}. Our estimated rotational temperatures for methyl formate and the
hot component of methanol are similar to the typical hot core temperature. On the other hand, methyl isocyanate and the cold component 
of methanol have lower rotational temperatures (33-48 K). However, due to poor angular resolution, we could not provide further details about these two 
temperature components' spatial distribution.

To obtain the molecular column density of the observed chemical species, we used LTE modeling with CASSIS. For this modeling, we used the estimated 
rotational temperature discussed above. We calculated beam average column density (N$_b$), where the effect of beam dilution was not taken into 
consideration. For the model, we used five parameters; (i) column density, (ii) excitation temperature (T$_{ex}$), (iii) line width ($\Delta$V), 
(iv) velocity (V$_{LSR}$), and (v) source size ($\theta_s$). To obtain the best fit with the observation, only column density was varied by
keeping all other parameters (Tex, source size, FWHM, and hydrogen column density) fixed and we choose the best fitted-results when LTE model
spectra are the same as the observed spectra. For LTE fitting, we used the source sizes for molecular emission  similar to the deconvolved size
as we obtained from 2D Gaussian fitting. Line width (FWHM) for molecular transitions are estimated from the Gaussian fitting of spectral profiles, and 
excitation temperature (Tex) is assumed to be $\sim$~150 K, a typical hot core temperature. Fixed parameters are noted in 
Table \ref{table:lte-para}. We did not have a rotation diagram for all the species (e.g., SO$_2$, H$_2$CO) 
presented here. Thus, we followed the LTE model to derive the column density of all the species. We have also compared the column densities of CH$_3$OH, 
CH$_3$SH, CH$_3$NCO, and CH$_3$OCHO between rotation diagram analysis and LTE model (see Table \ref{table:frac-abun} and Section 3.6).
Observed column densities and fractional abundances of all the species are presented in Table \ref{table:frac-abun}. Abundances of all 
species are noted w.r.to H$_2$.

\begin{table*}[t]
\centering{
\scriptsize{
\caption{Summary of the line parameters of observed molecules towards G31.41+0.31. \label{table:line-parameters}
}
\begin{tabular}{|>{\rowmac}c|>{\rowmac}c|>{\rowmac}c|>{\rowmac}c|>{\rowmac}p{1.2 cm}|>{\rowmac}p{1.2cm}|>{\rowmac}p{1.2 cm}|>{\rowmac}c|>{\rowmac}c<{\clearrow}|}
%\begin{tabular}{|c|c|c|c|p{1.2 cm}|p{1.2 cm}|c|c|c|}
\hline
\hline
Species&(${\rm J^{'}_{K_a^{'}K_c^{'}}}$-${\rm J^{''}_{K_a^{''}K_c^{''}}}$)&Frequency (GHz)&E$_u$/k$_B$ (K)&FWHM (km s$^{-1}$)&Peak Intensity, 
I$_\nu$ (K)&V$_{LSR}$ (km s$^{-1}$)&S$\mu^{2}$(Debye$^{2}$)&${\int{T_{mb}}dv\ (K.km/s)}$\\
\hline\hline
SO$_2$&$\rm{8_{3,5}-9_{2,8}}$&86.63908&55.20&8.46&14.63&97.27&3.01&{131.9$\pm$3.6}\\
H$_2$CO&$\rm{6_{1,5}-6_{1,6}}$&101.33299&87.57&10.20&75.90&97.26&5.05&{824.4$\pm$22.4}\\
&&&&&&&&\\
CH$_3$SH&$\rm{4_{0,4}-3_{0,3}}$ (A$+$)&101.13915$^{1}$&12.14&8.70&6.00&97.00&6.61&55.2$\pm$17.7\\
&$\rm{4_{0,4}-3_{0,3}}$ (E$+$)&101.13965$^{1}$&13.56&6.10&5.90&95.45$^{i}$&6.61&38.5$\pm$14.8\\
&&&&&&&&\\
&$\rm{4_{2,3}-3_{2,2}}$ (A$-$)&101.15933$^{2}$&31.26&6.75&3.50&97.00&4.96&26.1$\pm$4.7\\
&$\rm{4_{3,2}-3_{3,1}}$ (E$-$)&101.15999$^{2}$&52.39&7.09&1.56&95.03$^{ii}$&2.90&13.1$\pm$2.9\\
&$\rm{4_{3,1}-3_{3,0}}$ (A$+$)&101.16066$^{2}$&52.55&7.09&1.56&93.05$^{ii}$&2.90&13.1$\pm$2.9\\
&$\rm{4_{3,2}-3_{3,1}}$ (A$-$)&101.16069$^{2}$&52.55&7.09&1.56&92.94$^{ii}$&2.90&13.1$\pm$2.9\\
&&&&&&&&\\
&$\rm{4_{2,3}-3_{2,2}}$ (E$-$)&101.16715$^{3}$&29.62&6.94&8.20&97.00&4.96&60.6$\pm$5.2\\
&$\rm{4_{2,2}-3_{2,1}}$ (E$+$)&101.16830$^{3}$&30.27&6.20&7.70&93.58$^{iii}$&4.96&50.8$\pm$4.5\\
&&&&&&&&\\
CH$_3$OH&$\rm{7_{2,5^{-}}-6_{3,3^{-}}, v_t=0}$&86.61557&102.70&9.06&84.55&97.02&1.35&{815.3$\pm$26.8}\\
&$\rm{7_{2,5^{+}}-6_{3,4^{+}}, v_t=0}$&86.90291&102.72&9.13&85.16&97.00&1.35&{827.8$\pm$24.9}\\
&$\rm{15_{3,13^{+}}-14_{4,10^{+}}, v_t=0}$&88.59478&328.26&8.98&82.26&97.10&4.20&{786.4$\pm$14.6}\\
&$\rm{15_{3,12^{+}}-14_{4,11^{-}}, v_t=0}$&88.93997&328.28&8.55&79.78&97.16&4.20&{726.5$\pm$9.9}\\
&$\rm{5_{-2,3}-5_{1,4}, E2, v_t=0}$&101.12685&60.73&8.37&23.87&98.88&0.01&{208.4$\pm$18.8}\\
&$\rm{7_{-2,5}-7_{1,6}, E2, v_t=0}$&101.29341&90.91&7.72&45.71&97.38&0.05&{373.2$\pm$18.0}\\
&$\rm{8_{-2,6}-8_{1,7}, E2, v_t=0}$&101.46980&109.49&7.25&55.31&97.98&0.09&{426.9$\pm$16.5}\\
&&&&&&&&\\
CH$_3$NCO&$\rm{10_{0,0}-9_{0,0}}$&86.68655&34.97&7.70&14.63&97.83&83.11&{120.0$\pm$6.8}\\
&$\rm{10_{0,10}-9_{0,9}}$&86.68019&22.88&8.28&13.50&97.90&83.06&{119.1$\pm$11.5}\\
&$\rm{10_{2,9}-9_{2,8}}$&86.78077&46.73&6.80&10.45&97.32&79.73&{75.8$\pm$4.5}\\
&$\rm{10_{2,8}-9_{2,7}}$&86.80502&46.74&5.82&11.16&97.10&79.74&{69.2$\pm$3.8}\\
&$\rm{10_{-3,0}-9_{-3,0}}$&86.86674&88.58&10.14&15.80&97.00&75.32&{170.6$\pm$16.6}\\
&$\rm{10_{2,0}-9_{2,0}}$&87.01607&58.88&6.22&8.91&97.12&79.18&{59.0$\pm$6.2}\\
&$\rm{10_{0,10}-9_{0,9}}$&86.67492&285.03&--&--&--&83.07&--\\
&&&&&&&&\\
CH$_3$OCHO&$\rm{17_{3,14}-17_{3,15}}$&86.91761&99.72&6.41&7.09&98.03&1.87&{48.5$\pm$7.6}\\
&$\rm{11_{3,9}-11_{2,10}}$ E&88.68688&44.97&5.41&12.90&97.86&2.44&{74.3$\pm$9.2}\\
&$\rm{11_{3,9}-11_{2,10}}$ A&88.72326&44.96&7.84&17.11&98.32&2.43&{142.8$\pm$29.0}\\
&$\rm{7_{1,6}-6_{1,5}}$ E&88.84318&17.96&8.07&45.87&98.18&18.04&{394.3$\pm$58.9}\\
&$\rm{7_{1,6}-6_{1,5}}$ A&88.85160&17.94&6.64&49.17&97.50&18.04&{348.0$\pm$22.4}\\
&$\rm{8_{1,8}-7_{1,7}}$ E&88.86241&207.10&5.83&23.49&97.64&20.89&{145.9$\pm$10.1}\\
&$\rm{13_{3,11}-13_{2,12}}$ E&101.37050&59.64&5.16&15.90&97.93&2.61&{87.4$\pm$6.7}\\
&$\rm{13_{3,11}-13_{2,12}}$ A&101.41474&59.63&5.93&14.56&97.81&2.60&{92.0$\pm$13.5}\\
&&&&&&&&\\
SiO& (2-1)&86.84696&6.25&--&--&94.1&2.82&--\\
\hline
\hline
\end{tabular}}}
NOTES: $^{1}$ Spectral profile 1, $^{2}$ Spectral profile 2, $^{3}$ Spectral profile 3, $^{i}$ velocity with respect to 101.13915 GHz
transition, $^{ii}$ velocity w.r.t 101.15933 GHz transition, $^{iii}$ velocity w.r.t 101.16715 GHz transition.
\end{table*}

\begin{table}
\centering{
\scriptsize{
\caption{Observed molecular transitions in absorption. \label{table:absorption}}
\begin{tabular}{|c|c|c|c|c}
\hline
\hline
Species&(${\rm J^{'}_{K_a^{'}K_c^{'}}}$-${\rm J^{''}_{K_a^{''}K_c^{''}}}$)&Frequency in GHz&E$_u$ (K)\\
\hline\hline
H$^{13}$CO$^{+}$&1-0&86.75430&4.16\\
HCN&1-0(F=1-1)&88.63041&4.25\\
HCN&1-0(F=2-1)&88.63184&4.25\\
HCN&1-0(F=0-1)&88.63393&4.25\\
\hline
\hline
\end{tabular}}}
\end{table}

\begin{table}
\centering{
\caption{Input parameters of LTE model \label{table:lte-para}}
\begin{tabular}{|c|c|c|}
\hline
Species&   Source size ($''$)& FWHM (km/s)\\
\hline
SO$_2$ &0.7 &8.4\\
H$_2$CO & 1.3 &10.2\\
CH$_3$SH &0.8 & 5.5 \\
CH$_3$OH &1.3 & 8.5\\
CH$_3$NCO &0.8 &6.5\\
CH$_3$OCHO &1.35& 6.0\\
\hline
\end{tabular}
\\
H$_2$ column density = $\rm{1.53\times10^{25} \ cm^{-2}}$\\
T$_{ex}$ =150 K\\
}
\end{table}

\begin{table*}[t]
{ 
\scriptsize
\begin{center}
\caption{Comparison of observed  molecular column density, fractional abundances, and molecular ratios between G31 and other 
sources.\label{table:frac-abun}}
\begin{tabular}{|c|cc|c|c|c|c|}
\hline
Species&\multicolumn{2}{c|}{G31.41+0.31,}& Sgr B2 (N) & Orion KL &G10.47+0.03& Chemical Model\\
&\multicolumn{2}{c|}{Column density}&Column density&Column density&Column density&Column density\\
&\multicolumn{2}{c|}{[Abundance]}&[Abundance]&[Abundance]&[Abundance]&[Abundance]\\
\cline{2-3}
&\multicolumn{2}{c|}{LTE \hskip 1.2 cm Rotation Diagram}&&&&\\

\hline
CH$_3$OH&1.84(19)[1.2(-6)]$^{*}$&2.94(19)[1.92(-6)]&4.00(19)[1.14(-5)]$^{a}$&3.24(18)[1.62(-6]$^{f}$&9.0(18)[1.8(-6)]$^{m}$&7.56(19)[2.10(-5)]$^{n}$\\
CH$_3$OCHO&6.50(18){[4.25(-7)]}&4.30(18){[2.81(-7)]}&1.20(18)[3.33(-7)]$^{b}$&1.04(17)[5.2(-8)]$^{g}$&7.0(17)[1.4(-7)]$^{m}$&3.31(17)[9.20(-8)]$^{n}$\\
CH$_3$NCO&7.22(17){[4.72(-8)]}$^{*}$&1.58(16){[1.04(-9)]}&2.20(17)[6.11(-8)]$^{c}$&7.0(15)[3.5(-9)]$^{h}$&1.2(17)[8.88(-9)]$^{o}$&4.68(16)[1.30(-8)]$^{b}$\\
CH$_3$SH&4.13(17){[2.70(-8)]}&2.85(16)[1.86(-9)]&3.40(17)[9.44(-8)]$^{a}$&2.0(15)[1.0(-9)]$^{i}$&--&1.33(16)[3.70(-9)]$^{a}$\\
SO$_2$&2.88(18){[1.88(-7)]}&&8.28(16)[2.30(-8)]$^{d}$&1.0(17)[5.0(-8)]$^{j}$&3.0(17)[6.0(-8)]$^{m}$&\\
H$_2$CO&2.90(18){[1.90(-7)]}&&1.82(17)[5.06(-8)]$^{e}$&1.5(17)[7.5(-8)]$^{k}$&3.0(18)[6.0(-7)]$^{m}$&\\
SiO&1.46(15)[9.54(-11)]$^{**}$&&4.32(15)[1.20(-9)]$^{d}$&5.4(14)[2.7(-10)]$^{l}$&--&\\
CH$_3$SH/CH$_3$OH&0.0225&0.0010&0.0082&0.0006&&0.0002\\
CH$_3$OCHO/CH$_3$OH&0.3541&0.1463&0.0292&0.0321&0.0778&0.0044\\
CH$_3$NCO/CH$_3$OH&0.0393&0.0005&0.0054&0.0022&0.0049&0.0006\\
\hline
\end{tabular}\\
$^{a}$\cite{mull16}, $^{b}$\cite{bell16}, $^{c}$\cite{bell17}, $^{d}$\cite{higu15}, $^{e}$\cite{bell13}, $^{f}$\cite{favr15}, $^{g}$\cite{favr11}, 
$^{h}$\cite{cern16}, $^{i}$\cite{kole14},  $^{j}$\cite{terc18}, $^{k}$\cite{espl13}, $^{l}$\cite{schi01}, $^{m}$\cite{rolf11},$^{n}$\cite{garr13},
$^{o}$\cite{gora20}\\\

Orion KL ($\rm{N_{H_2}}$)= $\rm{2.0\times10^{24}}$cm$^{-2}$, Sgr B2(N) ($\rm{N_{H_2}}$)= $\rm{3.6\times10^{24}}$cm$^{-2}$, 
G10.47+0.03 ($\rm{N_{H_2}}$)= $\rm{5.0\times10^{24}}$cm$^{-2}$, G31.41+0.31 ($\rm{N_{H_2}}$)= $\rm{1.53\times10^{25}}$cm$^{-2}$\\
$^{*}$ For hot component of methanol, column density is derived using higher value of excitation temperature T = 300 K.
Using LTE, we have found opacities $\geqslant$1 for a few transitions of CH$_3$OH (86.68655 GHz, 86.68019 GHz, 86.78077 GHz, and 86.80502 GHz) 
and CH$_3$NCO (86.68019 GHz and 86.68655 GHz). Thus, the column density estimated for CH$_3$OH and 
CH$_3$NCO by using LTE should be considered as lower limit.\\
$^{**}$ SiO column density and fractional abundance calculated in the outflowing gas measured at V$_{LSR}$ =94.1 km s$^{-1}$ by averaging 
 over the box (7${''}$$\times$${7''}$) centered at ($\alpha$(J2000), $\delta$ (J2000))  = [18$^h$47$^m$34.121$^s$, -1$^{o}$12$'$48.975$''$].
\end{center}}
\end{table*}

\subsection{Radiative transfer modeling}
\begin{figure*}
\includegraphics[width=18cm, height=6.5cm]{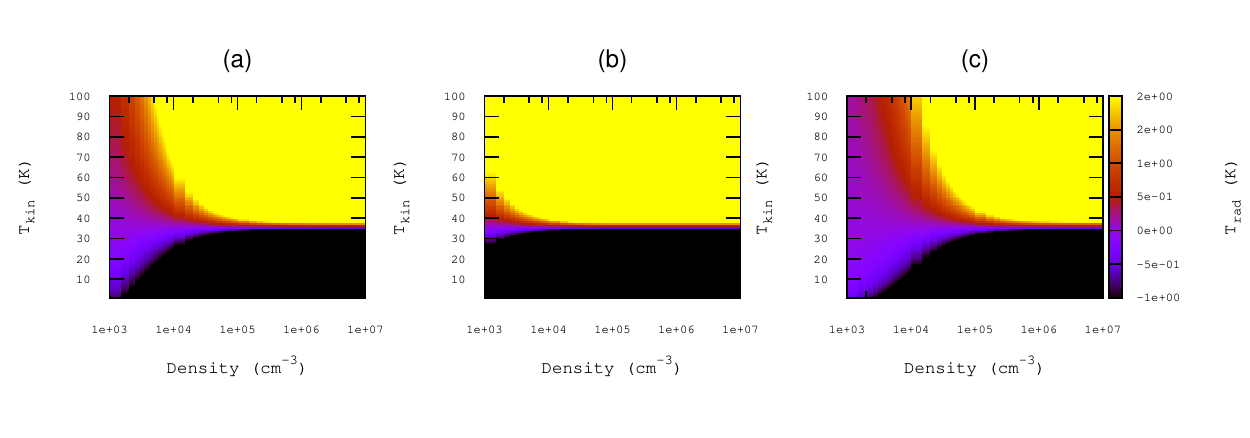}
\caption{Variation of the radiation temperature of (a) HCN(1-0),
(b) H$^{13}$CO$^{+}$ (1-0), and (c) SiO (2-1) respectively for a wide range of parameter space. The vertical axis is the kinetic temperature, labeled as 
$\rm{T_{kin}}$ (K). The units of the radiation temperature (T$_{rad}$) used in the color bar is in Kelvin.}
\label{contour-radiative-transfer}
\end{figure*}

Absorption studies are best-suited to trace the physical conditions (e.g., density, kinetic temperature) of the ISM because it is not induced by beam dilution
as long as the absorbing layer is larger than the continuum source. Here, we used non-LTE modeling to derive excitation conditions and physical properties 
from these absorption features. For the non-LTE modeling, we use RADEX program \citep{vand07}.  
The collisional rates were taken from LAMDA database and the spectroscopic constants were taken from JPL/CDMS catalog.
In RADEX, we used the continuum background temperature (T$_c$) of 36 K, which corresponds to the peak brightness temperature of the 3.2 mm 
continuum by considering the beam dilution. For this modeling, we have used colum density $\rm{3.55 \times 10^{15} \ cm^{-2}}$, 
$\rm{8.92 \times 10^{15} \ cm^{-2}}$, and $\rm{1.18 \times 10^{15} \times cm^{-2}}$ for H$^{13}$CO$^{+}$, HCN, and SiO respectively. 
Figures \ref{contour-radiative-transfer}(a)-(c) shows the variation of radiation temperature of HCN (1-0), $\rm{H^{13}CO^{+}}$ (1-0), and SiO (2-1) 
respectively for a wide range of parameter space. Our parameter space consists of the 
number density of hydrogen (varies in between $\rm{10^3-10^7\ cm^{-3})}$ and kinetic temperature (varies in between 1-100 K). {Figures 
\ref{contour-radiative-transfer}(a)-(c) shows the parameter space for the absorption features of HCN, $\rm{H^{13}CO^{+}}$, and SiO respectively. 
We have estimated the critical density of HCN (1-0), $\rm{H^{13}CO^{+}}$ (1-0), and SiO (2-1), which are 1.51$\times$10$^{6}$ cm$^{-3}$, 
1.83$\times$10$^{5}$ cm$^{-3}$, and 2.26$\times$10$^{5}$ cm$^{-3}$ respectively. We have shown that for hydrogen density
$\geq$ 1.0$\times$10$^{5}$ cm$^{-3}$, all these transitions are showing absorption features.  If H$_2$ density is below the critical density, level 
populations are mostly governed by collisional excitation and de-excitation rather than Boltzmann distribution. . Thus these molecules might show 
emission even at T$\leq$T$_c$. The observed spectral profile of SiO is in line with this results. We found SiO emission at all positions except the 
source's central region (where density is high, which is probably greater than critical density).}

\subsection{Physical properties of the source}
In this source, we obtained a temperature range 33 to 180 K. Hydrogen column density of this source is 
estimated $\sim{\rm1.53\times10^{25}}$ cm$^{-2}$. Various observed molecular emissions are discussed below.

\subsubsection{Infalling envelope}
Infall nature was observed in G31 earlier by various authors \cite[e.g.,]{gira09,gijo14,belt18}. \cite{gira09} observed an inverse 
P-Cygni profile in G31 with low energy lines of C$^{34}$S (7-6). Recently, \cite{belt18} identified red-shifted absorption of H$_2$CO and CH$_3$CN 
and its isotopologues towards the dust continuum of the main core with different upper state energy. \cite{gijo14} observed various inversion transitions (2,2), (3,3), (4,4), (5,5) and (6,6) of ammonia by using Very Large Array (VLA) observations 
to understand the infall motion in G31. To explain the  infall motion in G31, they used a different approach called the central blue spot which can be 
easily identified in the first-order moment maps. The central blue spot consists of blue-shifted emission in the first-order moment towards the 
zeroth-order moment map's peak position. They found the central blue spot infall signature in all observed transitions of ammonia. In this work, we have 
detected one transition of H$^{13}$CO$^{+}$ (1-0) at frequency $86.75430$ GHz, which shows both absorption and emission nature (see Figure \ref{abs-sp}a). 
This is an inverse P-Cygni profile. The infall velocity can be estimated as the difference between systematic velocity and the velocity at 
the observed inverse P-Cygni profile's absorption feature. The infall velocity of the source is obtained to be 4.5 km s$^{-1}$. This suggests that 
matter is infalling from the envelope towards the source's central region, which is consistent with the earlier observations
\citep{gira09,belt18}.

\subsubsection{Molecular outflow}
\begin{figure*}[t]
\centering
\includegraphics[height=14cm,width=16cm]{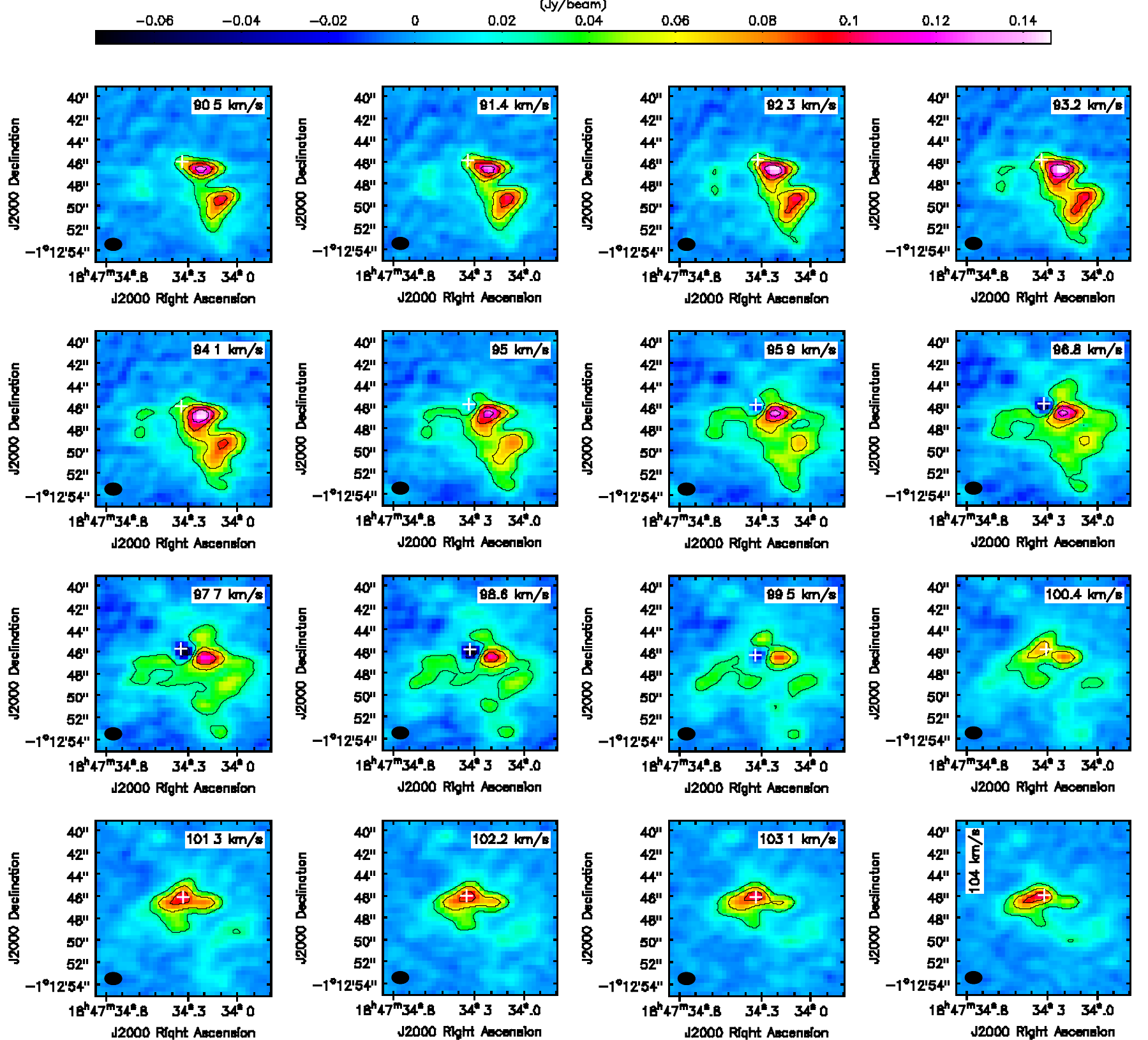}
\caption {Channel maps of SiO (2-1) emission. Ellipse at the below left corner of the image shows the beam size $\rm{1.52'' \times 1.09''}$ 
with PA=88.98$^{o}$. White plus sign indicates the position of the continuum.} 
\label{channel-maps-sio}
\end{figure*}

\begin{figure}
\centering
\includegraphics[height=9cm,width=9cm]{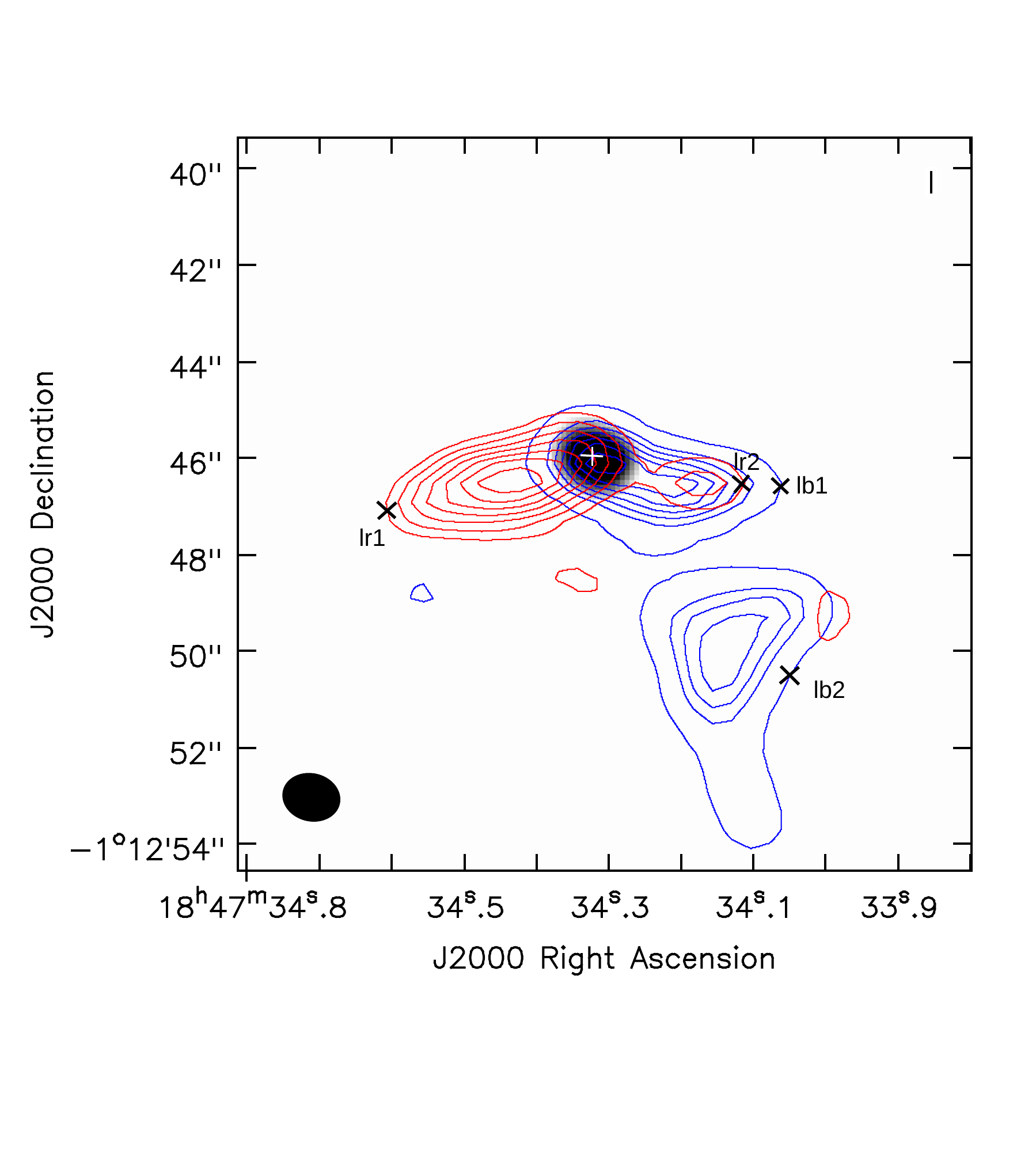}
\caption {Gray scale levels for the continuum at 94 GHz. Blue-shifted emission of SiO integrated over 77-93 km s$^{-1}$ where contour levels are at 
the  20, 30, 40, 50 ,60, 70, 80 $\%$ of the peak intensity.  Red-shifted emission of SiO is integrated over 100-123 km s$^{-1}$. Ellipse at the below 
left corner of the image shows the synthesized beam ($\rm{1.19'' \times 0.98''}$). Plus sign indicates the position of the continuum. Two cross sign at the outer 
contour of red lobes indicates lr1 and lr2 distance another two cross sign on the red lobes indicated lb1 and lb2 distance.} 
\label{outflow-sio}
\end{figure}

\begin{figure}
\centering
\includegraphics[height=9cm,width=9cm]{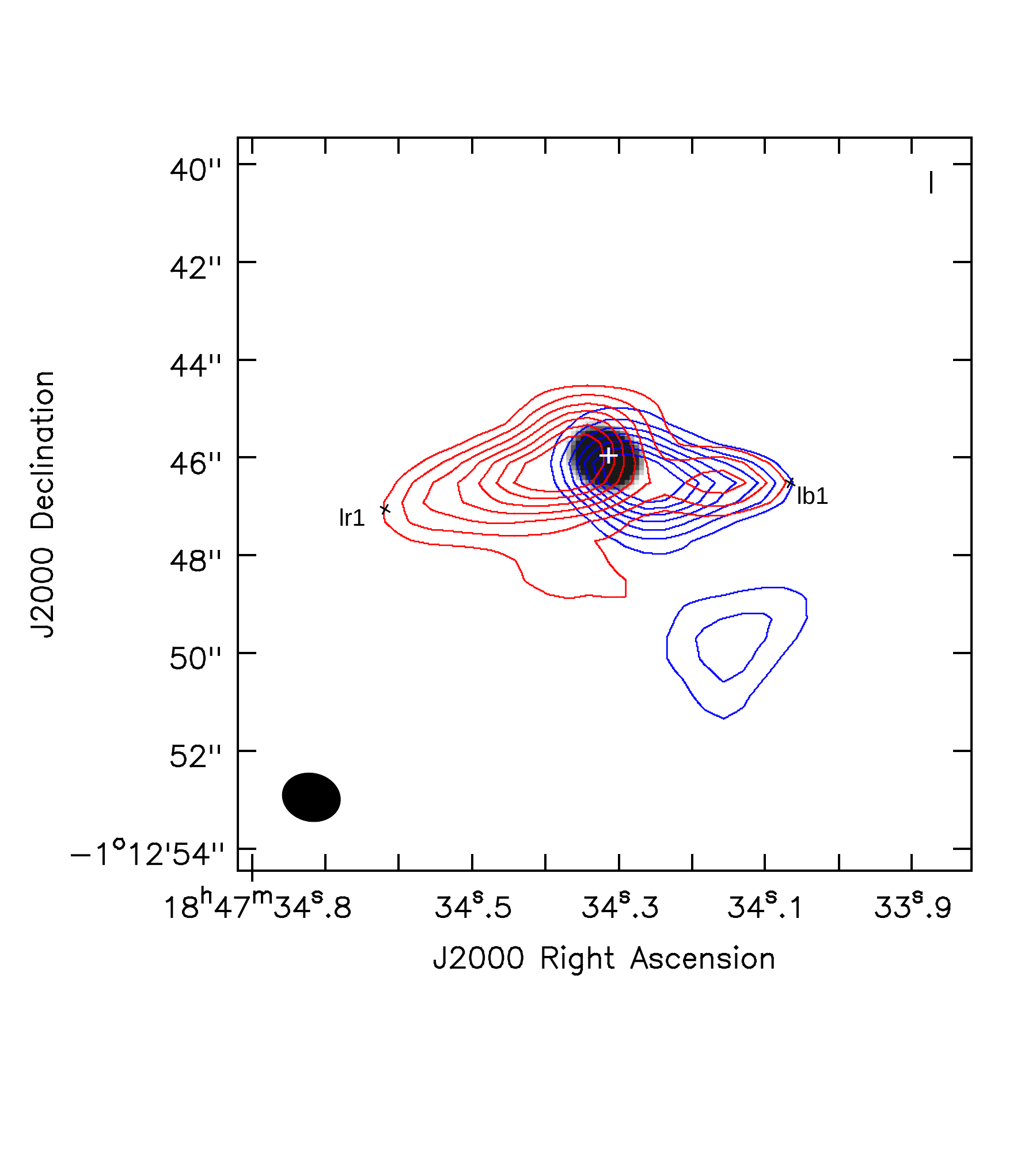}
\caption {Gray scale levels are used for the continuum at 94 GHz. Blue-shifted emission of HCN integrated over 77-90 km s$^{-1}$ contours levels 
are at the  20, 30, 40, 50 ,60, 70, and 80$\%$ of the peak intensity.  Red-shifted emission of HCN integrated over 105-115.7 km s$^{-1}$. 
Ellipse at the below left corner of the image shows the synthesized beam ($\rm{1.19'' \times 0.98''}$) with PA=76.83$^{o}$ } 
\label{outflow-hcn}
\end{figure}

Many authors previously studied molecular outflows associated with G31 \citep[e.g.,][]{cesa11,olmi96,belt18}. SiO and $\rm{^{13}CO}$ revealed the 
presence of two molecular outflows: One in the East-West (E-W) direction and another one in the North-South (N-S) direction. SiO is an excellent 
tracer of shock and molecular outflow associated with the star-forming regions \citep[e.g.,][]{zapa09,leur14}. \cite{belt18} showed average red-shifted 
and blue-shifted wings of SiO emission, which reveals three outflow directions associated with this source: E-W, N-S, and NE-SW. 
. Among these three outflow directions, N-S and E-W components are clearly present in our observation. However, NE-SW component is not very evident 
(see original Figure \ref{outflow-sio} and illustrated in Figure \ref{G31-diagram}). Channel maps of SiO emission is shown in Figure 
\ref{channel-maps-sio}. The blue-shifted and red-shifted integrated emission of SiO along with the continuum, is depicted in Figure \ref{outflow-sio}. 
The presence of HCN is widespread in diverse interstellar environments ranging from dark cloud to star-forming region and circumstellar envelope of a 
carbon-rich star \citep[e.g.,][]{snyd71,ziur86}. HCN also traces molecular outflow in G31. The observed spectrum profile of HCN (see Figure \ref{abs-sp}b) 
is symmetric with respect to the systematic velocity of the source. However, the intensity of emission wings associated with the absorption profile is 
asymmetric. This may be due to the asymmetry in molecular distributions. Figure \ref{outflow-hcn} shows the red-shifted and blue-shifted integrated 
emissions of HCN overlaid on the continuum.

\begin{table*}
\scriptsize
\centering
\caption{Physical parameters of the outflow \label{table:outflow-parameters}}
%  \hskip 0.1cm
\begin{tabular}{|p{4cm}cc|}
\hline
\hline
Parameters&SiO& HCN\\
\hline
\hline
{Distance from continuum position to outer contour (pc)}&&\\
Red lobe (lr1)&0.117  (0.055)&0.118  (0.055)\\
Red lobe (lr2)& 0.087  (0.041)&--\\
Blue lobe1 (lb1)& 0.108  (0.051)&0.09  (0.042)\\
Blue lobe2 (lb2)& 0.135  (0.063)&--\\
\hline
Typical outflow velocity: V$_{outflow}$ (km s$^{-1}$)&&\\
Red lobe & 33.0&40.0\\
Blue lobe& 23.0&30.0\\
\hline
Dynamical time: t$_{dyn}$ (yr)&&\\
Blue lobe &$\rm{4.59\times10^{3}}$  ($\rm{1.62\times10^{3}}$)&$\rm{2.87\times10^{3}}$  ($\rm{1.35\times10^{3}}$)\\
          &$\rm{5.74\times10^{3}}$  ($\rm{1.21\times10^{3}}$)&--\\
Red lobe  &$\rm{3.47\times10^{3}}$  ($\rm{2.16\times10^{3}}$)&$\rm{2.90\times10^{3}}$  ($\rm{1.36\times10^{3}}$)\\
          &$\rm{2.58\times10^{3}}$  ($\rm{2.70\times10^{3}}$)&--\\
Average   &$\rm{4.10\times10^{3}}$  ($\rm{1.92\times10^{3}}$)&$\rm{2.88\times10^{3}}$  ($\rm{1.36\times10^{3}}$)\\

Outflow mass ($\rm{M_\odot}$) &&\\ 
Blue lobe &27  (6)&12  (2.6)\\
Red lobe &24  (5)&55  (12)\\
Total &51  (11) &67  (14.6)\\
Momentum ($\rm{M_\odot}$ km s$^{-1}$)&&\\ 
Blue lobe &619  (136)&352  (77)\\
Red lobe &790  (173) &2213  (485) \\
Total &1409  (309) &2565  (562)\\
Kinetic energy ($\rm{L_\odot}$ yr) &&\\
Blue lobe&$\rm{7.1\times10^{3}}$  ($\rm{1.6\times10^{3}}$)& $\rm{5.3\times10^{3}}$  ($\rm{1.2\times10^{3}}$)\\
Red lobe &$\rm{1.3\times10^{4}}$  ($\rm{8.8\times10^{3}}$)&$\rm{4.4\times10^{4}}$  ($\rm{9.7\times10^{3}}$)\\
Total&$\rm{2.0\times10^{4}}$  ($\rm{1.0\times10^{4}}$)&$\rm{4.9\times10^{4}}$  ($\rm{1.1\times10^{4}}$)\\
Mass loss ($\rm{M_\odot}$ yr$^{-1}$)&&\\
Blue lobe&$\rm{5.2\times10^{-3}}$  ($\rm{4.1\times10^{-3}}$) &$\rm{4.1\times10^{-3}}$  ($\rm{1.9\times10^{-3}}$)\\
Red lobe &$\rm{7.9\times10^{-3}}$  ($\rm{2.1\times10^{-3}}$)&$\rm{1.9\times10^{-2}}$  ($\rm{8.9\times10^{-3}}$)\\
Total&$\rm{1.3\times10^{-2}}$  ($\rm{6.2\times10^{-3}}$)&$\rm{2.3\times10^{-2}}$  ($\rm{10.8\times10^{-3}}$)\\
Momentum loss ($\rm{M_\odot}$ km s$^{-1}$ yr$^{-1}$)&&\\
Blue lobe&0.12  (0.09)&0.12  (0.06)\\
Red lobe &0.26  (0.07)&0.76  (0.35)\\
Total&0.38  (0.16)&0.88  (0.41)\\
Energy loss ($\rm{L_\odot}$)&&\\
Blue lobe&1.38  (1.10)&1.84  (0.85)\\
Red lobe &4.31  (3.63) &15.26  (7.14)\\
Total&5.69  (4.73)& 17.10 (7.99)\\
\hline
\hline
\end{tabular}
\\
Note: To estimate the SiO outflow physical parameters, we considered average value of red lobe (r$_r$=$\frac{lr1+lr2}{2}$) and blue lobe 
(r$_b$=$\frac{lb1+lb2}{2}$) and their corresponding average time scale of blue lobe and red lobe. We have calculated the lobe distance 
in arcsec and then convert into parsec by considering source distance 7.9 kpc. In parenthesis, we have noted the outflows parameters 
by considering the new distance 3.7 kpc of G31.
\end{table*}

To obtain the outflow parameters, we estimated the outflow velocity as the maximum velocity of outflow emission with respect to the systematic LSR 
velocity ($97$ km s$^{-1}$). For 
the blue lobe and red lobe of SiO outflow, velocities are found to be $23$ km s$^{-1}$ and $33$  km s$^{-1}$. We estimated the dynamical 
timescale of outflow by using t$_{dyn}$ =$\frac{l(b,r)}{V_{max}(b,r)}$. For this, we calculate two different lengths (lb1, lb2) of the blue lobe and 
two lengths (lr1, lr2) for red lobes 
(see Figure \ref{outflow-sio}-\ref{outflow-hcn}). We obtained the average dynamical 
time scale of outflow $\sim$ $\rm{4\times10^{3}}$ years by considering the maximum velocity of SiO of the blue lobe and red lobe. The dynamical time 
scale of HCN outflow is obtained to be $\sim$ $\rm{3\times10^{3}}$ years. These values are in good agreement with the dynamical time scale of outflow of 
$^{12}$CO and SiO, as previously reported in G31 by \cite{cesa11} and \cite{belt18}. For all individual positions, the different dynamical time 
scales for both blue and red lobe are presented in Table \ref{table:outflow-parameters}. For HCN, we calculate the outflow parameters for one blue 
lobe (lb1) and one red lobe (lr1). To estimate the other outflow parameters such as mass, momentum, energy, mass loss, momentum loss, and energy
loss per year, we used the formula mentioned in \cite{cabr90}. To estimate the outflow parameters, we assumed same abundance ($\rm{1.0\times10^{-8}}$) 
and excitation temperature ($20$ K) of SiO \citep{code13} and HCN. All the outflow parameters are summarized in Table \ref{table:outflow-parameters}. All these outflow 
parameters are in good agreement with the observed values of \cite{arya08}.

In Figure \ref{channel-maps-sio}, we have shown the channel maps of SiO emission associated with the HMC. This figure depicts that 
blue-shifted channels are distributed in the southwest region and red-shifted channels toward the south-east and north-east regions. Bipolar lobe 
morphology observed from 93.2 km s$^{-1}$ and 103.1 km s$^{-1}$ channels suggest wide-angle bipolar outflow. Integrated emission of SiO and 
HCN transition (Figures \ref{outflow-sio} and \ref{outflow-hcn}) shows that there is a clear outflow direction along the East-West direction. In 
the integrated emission, the red lobe of outflow is slightly tilted to the south-east. This may occur due to another outflow direction along the 
southwest-northeast direction. There is also an outflow in the N-S direction. Outflow of HCN along the N-S direction is wider 
(see Figure \ref{outflow-hcn}) as compared to the SiO outflow (see Figure \ref{outflow-sio}). This may be attributed to the excitation 
conditions and chemical origin of HCN. The estimated outflow is massive, 50 M$_\odot$ for SiO and 67 M$_\odot$ for HCN and the outflow is characterized 
as young (3-4$\rm{\times10^{3}}$ years). However, by considering the new distance measurement of $\sim 3.7$ kpc of this source, we obtained 
the dynamical time scale of outflow around 1-2$\rm{\times10^{3}}$ years. The estimated outflow mass is $\sim$11 M$_\odot$ for SiO and 
$\sim$14.6 M$_\odot$ for HCN.

\subsection{Molecular distributions and abundances}

In this section, we present the molecular abundance of all the observed species. We compare the abundance of 
all detected species between G31 and other sources, which is presented in Table \ref{table:frac-abun}. All the
molecular abundances are reported w.r.t. H$_2$. {Figure \ref{G31-diagram} in the Appendix shows the schematic diagram of kinematics and molecular 
distributions of G31 based on present observation. There are three outflow directions one is along the E-W direction, one is along the N-S direction,
and another is along the NE-SW direction (only for SiO). Outflow components are traced by HCN and SiO emissions. Here, we have obtained different
molecular distributions of COMs. The emitting regions of different transitions of COMs are different. We have separated those into three regions 
such as the emitting region is smaller than the beam size ($\rm{\theta \sim 0.8^{''}}$, see the purple circle in the Figure \ref{G31-diagram}), comparable 
to the beam size ($\rm{\theta \sim 1.1^{''}}$, see green circle outer edge in the Figure \ref{G31-diagram}), and greater than the beam size 
($\rm{\theta \sim 1.4^{''}}$, see cyano circle in the Figure \ref{G31-diagram}). Emitting region of all observed species are 
provided in Table \ref{table:emitting-regions} (see in Appendix). The 
uncertainties of emitting regions listed in Table \ref{table:emitting-regions} are the lower limit of true values as estimated 
emitting regions might be affected by several issues such as poor angular resolution, image noise, and non-Gaussian shape of the source.
Emitting regions of CH$_3$OH and CH$_3$OCHO are comparable or slightly greater as compared to the beam size. Molecular distributions show that 
CH$_3$NCO and CH$_3$SH have a slightly lower-emitting region as compared to beam size (marginally resolved). The data presented in 
this paper suggests possibly different emitting regions for different COMs, but these are not conclusive 
because of comparable beam and source size. High sensitivity data with high angular and spatial 
resolution is required to understand the spatial distributions of these molecules.}

\subsubsection{Sulfur dioxide, SO$_2$}
SO$_2$ is a good tracer of dense and hot gas of typical hot core \citep[e.g.,][]{char97,leur07}. In this source, we detected a 
strong transition ($\rm{8_{3,5}-9_{2,8}}$) of SO$_2$. Using LTE model, we estimated the column density and the fractional abundance of SO$_2$,  
which are $\sim 2.88 \rm{\times10^{18}}$  cm$^{-2}$ and 1.88$\rm{\times10^{-7}}$. The observed abundance of SO$_2$ in G31 is one order of magnitude
higher as compared to Srg B2, Orion KL and G10.47+0.03.

\subsubsection {Formaldehyde, H$_2$CO}
The presence of H$_2$CO with its ortho, para, and isotopic form within the frequency range of $218-364$ GHz was previously reported by \cite{isok13}. 
Red-shifted strong absorption of H$_2$CO (E$_{up}$=21 K, 68 K) had recently been observed in G31, which supports the 
existence of infall in the core \citep{belt18}. Previously, \cite{fran01} reported a similar kind of inverse P-Cygni profile and found that H$_2$CO  
traces infall in IRAS 4A and they also detected the outflow from both 4A and 4B position by observing bright wing emission of H$_2$CO. Formaldehyde,
along with its deuterated counterpart, has also been observed in outflows toward other 
sources \citep[e.g.,][]{sahu18}. The presence of H$_2$CS in G31 has previously been reported by \cite{ligt15}. Here, we also have detected one 
strong emission line of H$_2$CS at 101.478 GHz. However, the observed spectrum of H$_2$CS shows that there is a blending with other molecular lines
(methyl formate, confirmed from LTE modeling). “H$_2$CO and H$_2$CS are ubiquitous in the ISM. These two species have similar chemical structures; just
O atom of H$_2$CO is replaced by S atom in H$_2$CS.

\subsubsection{Methanol, CH$_3$OH}
Methanol is a well-known and one of the most abundant COMs in hot cores. Here, we have identified several transitions of methanol. 
LTE model suggests that it has a high fractional abundance $\sim 1.2\rm{\times10^{-6}}$. We have also estimated methanol abundance 
following rotation diagram analysis.The obtained abundance using rotation diagram analysis is $\sim 1.2\rm{\times10^{-6}}$, which 
is similar to the estimated results using the LTE model. A comparison of fractional abundance of methanol between G31 and other hot molecular 
core is presented in Table \ref{table:frac-abun}.

\subsubsection{Methanethiol, CH$_3$SH}
Methanethiol has a similar chemical structure as methanol; just O atom of methanol is replaced by S atom of methanethiol. In Figure \ref{lte-ch3sh}, 
we show the observed and model (LTE) spectra of CH$_3$SH. Best fitted abundance and column density of methanethiol are 
$\sim 2.70\rm{\times10^{-8}}$ and $\sim 4.13\rm{\times10^{-17}}$ respectively. We also estimated the column density and fractional abundance 
of this species using rotation diagram analysis, which is one order of magnitude lower as compared to LTE model fitted results. 
The observed emitting region of methanethiol is comparable to the beam size. The spatial
distribution of its emission is as compact as the dust continuum. Since our beam size is comparable to the source size. 
Therefore, we have uncertainties in determining the emitting regions of this species. For detail knowledge about the distribution 
of this species, we need a high angular and spatial resolution data.

\subsubsection{Methyl formate, CH$_3$OCHO}
CH$_3$OCHO is one of the most abundant COMs, which was observed in both low-mass and high-mass star-forming regions \citep[e.g.,][]{brow95,caza03}. 
Previously, numerous transitions of CH$_3$OCHO were identified in G31 \citep{isok13,rivi17,belt18}. 
Column density, fractional abundance, and a comparison of fractional abundance of methyl formate 
with other hot cores are provided in Table \ref{table:frac-abun}. The estimated column density and fractional abundance of 
methyl formate are similar using both rotation diagram and LTE model fitting analysis. Emitting diameters of various transitions of CH$_3$OCHO 
are noted down in Table \ref{table:emitting-regions}.

\subsubsection{Methyl isocyanate, CH$_3$NCO}
In the case of CH$_3$NCO, we have obtained a lower value of rotational temperature (48 K, see Figure. \ref{rot-diagram}a), which is low as compared to a
typical hot core. Methyl isocyanate transitions are compact and yield lower value rotational temperature.  A similar trend has also been observed for
cold component of methanol. Figure \ref{lte-ch3nco} shows the observed and model (LTE) spectra of CH$_3$NCO. The LTE model suggests 
that for a column density of $\rm{7.22\times10^{17}}$ cm$^{-2}$, we have a good fit with the observed spectra. We have also estimated the column density
of CH$_3$NCO using rotation diagram analysis, which is one order higher as compared to LTE model fitting result. 
CH$_3$NCO emissions are also concentrated towards the center of the source's dust continuum. However, all these transitions are marginally resolved. 
Thus, we can not provide the detailed distributions of this molecule.

\begin{figure}
\includegraphics[width=8.5cm,height=6.0cm]{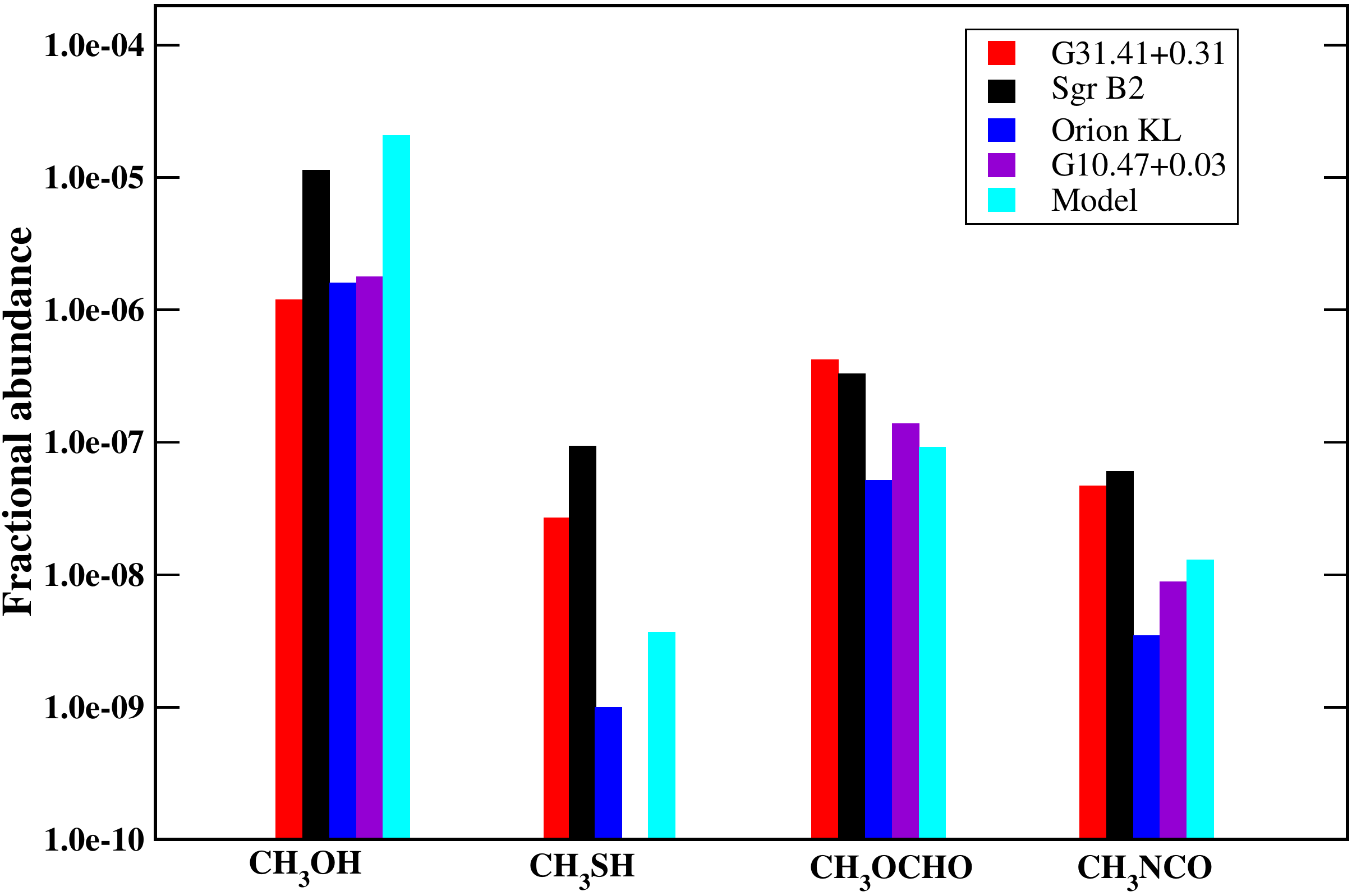}
\caption{Comparison of abundance of CH$_3$OH, CH$_3$SH, CH$_3$CHO and CH$_3$NCO between G31 and other sources (Sgr B2, Orion KL, G10.47+0.03, and model). 
These abundances are given in the Table \ref{table:frac-abun} with their references.}
\label{abundance-comparison}
\end{figure}

\section{Discussions}
Synthesis of many COMs (e.g., ethylene glycol, methyl formate, etc.) in G31 has been discussed by \cite{rivi17}. 
They discussed the formation routes of a few COMs. Thus, we do not concentrate on the synthesis of all the species 
which are identified in the present observation.  Here, we mainly focus on the chemistry of two newly reported molecules, 
CH$_3$NCO and CH$_3$SH and two other COMs such as CH$_3$OH and CH$_3$OCHO that are discussed in the text. Here, we have 
discussed the possible synthesis routes of these species and compared their observed abundance/molecular ratio
between G31 and other hot molecular cores. We have also shown a comparison between observed results and existing astrochemical simulation of 
these species. For the modeling results, we mainly followed \cite{garr13}. They considered an isothermal collapse phase, which is 
followed by a static warm-up phase. In the first phase, the density increases from $\rm{n_H=3\times10^3}$ to $\rm{10^7 \ cm^{-3}}$ under the free-fall 
collapse. At the initial phase, dust temperature falls to 8 K from 16 K. In the second phase; the temperature varies from 8 K to 400 K, where density 
remains fixed at $\rm{10^{7} \ cm^{-3}}$. The temperature of this source is $\sim$150 K, which is typical hot core temperature, and the number density 
($\rm{n_H}$) of this source is $\sim$ $\rm{10^{7}\ cm^{-3}}$ \citep{belt18}. Thus, the hot core model of \cite{garr13} is suitable for 
understanding chemical evolution of COMs in this source. They used three different warm-up models based on the time scale 
(i.e., fast, medium, and slow) and the final temperature of their model is much larger than one that is observed in this source.
The timescale of the fast warm-up model is more suitable for studying the chemical evolution in high-mass star-forming regions. 
We used peak gas-phase abundances of all the species around 100-200 K of fast warm-up timescale model to compare with the observed values. Figure 
\ref{abundance-comparison} compares the observed fractional abundance of CH$_3$OH, CH$_3$SH, CH$_3$OCHO, and CH$_3$NCO between G31 and other sources. 
The chemistry of these COMs is discussed in the following subsections.

\subsection{CH$_3$OH and CH$_3$SH}

CH$_3$OH and CH$_3$SH molecules are the chemical analog of each other and have a similar structure only difference is that one has an oxygen atom 
another sulfur atom. Emitting regions of CH$_3$SH transitions are compact and marginally resolved, whereas the emitting region of methanol is slightly 
extended compared to CH$_3$SH. A large number of previous experimental, theoretical, and observed results suggest that these two species are mainly 
produced on the grain surface via successive hydrogen addition reactions \citep[e.g.,][]{das08,fedo15,mull16,gora17,vida17} with CO and CS, 
respectively. The main route of formation of methanol and methanethiol is given below,

\vskip 0.4 cm
CO $\xrightarrow{\text{H}}$ HCO $\xrightarrow{\text{H}}$ H$_2$CO $\xrightarrow{\text{H}}$ CH$_3$O $\xrightarrow{\text{H}}$ CH$_3$OH  
\vskip 0.3 cm
and
\vskip 0.3 cm
CS $\xrightarrow{\text{H}}$ HCS $\xrightarrow{\text{H}}$ H$_2$CS $\xrightarrow{\text{H}}$ CH$_3$S $\xrightarrow{\text{H}}$ CH$_3$SH.\\

For the above reaction schemes, the first and third steps posses an activation barrier, whereas the second and fourth steps are barrier-less 
\citep[e.g.,][]{hase92,gora17}. At low-temperature regions, methanol and methanethiol are efficiently produced on the grain surface and released 
to the gas phase via non-thermal desorption in the cold phase and during the warm-up stage via thermal desorption process.
From the astrochemical simulation, \cite{mull16} found that peak abundance of methanol arises around $130$ K. Here, our observed results of methanol
predict the gas temperature $\sim 180$ K, which is similar to the typical hot core temperature.
In addition to grain surface reactions, H$_2$CS can also be produced in the gas phase. In the gas phase, H$_2$CS is mainly produced from 
H$_3$CS$^{+}$, where H$_3$CS$^{+}$ is formed via the ion-neutral reaction between S$^{+}$ and CH$_4$. Below $\sim$20 K with a 
density ($\rm{n_H}$) around $\rm{10^{4}-10^{5} \ cm^{-3}}$, H$_2$CS is accreted onto the 
grain surface directly from gas phase, which may also contribute to the CH$_3$SH formation \citep{bonf19}.

The observed abundance of CH$_3$SH in G31 is $\rm{2.7\times10^{-08}}$ and $\rm{1.0\times10^{-09}}$ in Orion KL \citep[][see Table 
\ref{table:frac-abun}]{kole14}. The observed abundance of methanethiol in Sgr B2 is $\sim$3.5 times higher than the G31. Methanol 
abundance is almost similar in G31, Orion KL, G10.47+0.03, and 10 times higher in Sgr B2 (see Figure \ref{abundance-comparison}). 
The observed CH$_3$SH/CH$_3$OH ratio in G31 is similar to Orion KL and it is eight times higher in Sgr B2.
Modeling results of these two species also have comparable abundances and molecular ratio (see Table \ref{table:frac-abun}). 

\subsection{\rm CH$_3$OCHO}
Methyl formate can efficiently be produced on the surface of dust-grains through the reaction between methoxy radical (CH$_3$O) and formyl radical 
(HCO) \citep{garr08,garr13}. Their simulation shows that, around 30-40 K, these radicals are mobile, and the reaction is efficient. It is clear from 
their simulation that gas phase CH$_3$OCHO is mainly coming from the ice phase \citep[see Figure 1 of][]{garr13}. Increased UV photodissociation of 
CH$_3$OH leads to the formation of various radicals CH$_3$, CH$_3$O, and CH$_2$O, around 40 K temperature, these are mobile and form the COMs 
(e.g., CH$_3$OCHO, CH$_3$OCH$_3$). In the gas phase, when T $\sim$ 40 K, protonated methanol and H$_2$CO (when it is abundantly released to the 
gas phase) react and form HC(OH)OCH$_3$$^{+}$. Then electron recombination of HC(OH)OCH$_3$$^{+}$ 
produce CH$_3$OCHO \citep{bonf19}. An efficient gas-phase reaction of methyl formate production was proposed by 
\cite{balu15} for the cold environment, where dimethyl ether is the precursor of methyl formate. Dimethyl ether is also present in this source 
\citep{isok13}. Therefore methyl formate can be produced through both grain surface and gas-phase reactions. The observed abundance of CH$_3$OCHO in G31 is $\rm{4.25\times10^{-7}}$, which is similar to the observed abundance in Sgr B2, G10.47+0.03 and ten 
times higher as compared to the Orion KL and modeling results. In G31, we have observed CH$_3$OCHO/CH$_3$OH abundance ratio $\sim$ 0.14-0.35. For 
other high-mass star-forming regions, this ratio is found to be $0.03$, $0.03$, and $0.08$ in Sgr B2, Orion KL, and G10.47+0.03, respectively 
(see Table \ref{table:frac-abun}). 
%This is well in agreement with the molecular ratio (CH$_3$OCHO/CH$_3$OH) observed in G31. Therefore CH$_3$OH and CH$_3$OCHO may have followed similar formation mechanisms in these sources.

\subsection{CH$_3$NCO}
 In this work, we find emitting regions of CH$_3$NCO is compact. In this source, \cite{rivi17} also observed a similar emitting
region of ethylene glycol. It is primarily produced on the grain surface via recombination of two HCO radicals followed by successive hydrogen addition
reactions \citep{cout17}. In case of CH$_3$NCO formation, \cite{half15} proposed two routes (i) reaction between HNCO(HOCN) and CH$_3$, and 
(ii) reaction between HNCO(HOCN) and $\rm{CH_5^{+}}$. \cite{cern16} proposed that it could be produced on the grain surface. The experiment result 
suggests that CH$_3$NCO can efficiently be produced on the solid phase through the reaction between CH$_3$ and (H)NCO \citep{ligt17}. The hot core 
model suggests that it could be produced via a radical-radical reaction between CH$_3$ and OCN \citep{bell17}. Since the reaction mentioned above is 
exothermic and barrier-less, it can efficiently be produced on the grain surface at a little warmer temperature ($>$30 K), when radical get enough 
mobility. Though CH$_3$NCO is efficiently produced on the grain surface, it could also be destructed by hydrogen addition reactions and produce n-methyl
formamide \citep{bell17}. \cite{quen18} also considered similar radical-radical reactions in their chemical model to study the evolution of CH$_3$NCO 
under hot core conditions. In the hot core, when the temperature reaches around 100 K, CH$_3$ and HNCO can thermally desorb from the grain surface and 
increase the formation of CH$_3$NCO \citep{mart17,gora20}. Our estimated CH$_3$NCO abundance in G31 is $\sim$ $\rm{4.72\times10^{-8}}$. For other hot molecular cores 
(e.g., Sgr B2, Orion KL), we have found a comparable abundance of CH$_3$NCO. The observed fractional abundance of CH$_3$NCO in this work is well 
within the modeled results (model 5) of \cite{bell17}.

The identification of CH$_3$OH, CH$_3$OCHO, CH$_3$NCO, and CH$_3$SH in G31 suggesting both grain surface and gas phase chemistry is efficient in this source for the 
formation of COMs. However, we can not distinguish the contribution from the gas and grain phase separately due to the observational limit. To 
explain the various observed line profile in this work, it requires a combined chemical and radiative transfer model, which will be carried out 
in our follow-up study.

\section{Conclusions}
We analyzed ALMA band 3 data of a hot molecular core, G31.41+0.31, and the following conclusions are made. 

\begin{itemize}
\item {We detected methyl isocyanate (CH$_3$NCO) and methanethiol (CH$_3$SH), for the first time in G31.\\}

\item{We obtained two temperature components by using rotation diagram analysis. The cold component is obtained
by CH$_3$OH and CH$_3$NCO where temperature varies between 33 and 48 K. Hot component is obtained by CH$_3$SH, CH$_3$OH and CH$_3$OCHO 
where the temperature ranges between 143 and 181 K. Estimated hydrogen ($\rm{N_{H_2}}$) column density is $\sim$ 1.53$\times10^{25}$ 
cm$^{-2}$ and dust is optically thin at 94 GHz.\\}

\item{We estimated column density and fractional abundances of various complex species CH$_3$NCO, CH$_3$SH, CH$_3$OH, and CH$_3$OCHO in G31. A 
comparison of column density/fractional abundances of these species between G31 and other hot molecular cores is discussed. We also compared our 
observed results with available chemical modeling under hot core conditions. Our observed abundances of COMs in G31 are consistent with other hot
cores, Sgr B2, Orion KL, and G10.47+0.03, which suggests similar formation mechanisms of these complex species in these sources.\\}

\item{For the kinematics of this source, a low excitation line of $\rm{H^{13}CO^{+}}$ is observed as an inverse P-Cygni profile, which suggests that
there is an infall toward the center of this source, i.e., this source is in the younger stage of its evolution.\\}

\item{Protostellar outflows are traced by SiO and HCN. Their blue-shifted and red-shifted emission clearly trace outflow along East-West direction. We
estimated various outflow parameters. The average dynamical time scale of outflow is $\sim$ $10^{3}$ years.\\}

\end{itemize}

\section{Acknowledgement}
This paper makes use of the following ALMA data: ADS/JAO.ALMA$\#$2015.1.01193.S. ALMA is a partnership of ESO (representing its member states), NSF
(USA), and NINS (Japan), together with NRC (Canada), NSC, and ASIAA (Taiwan), and KASI (Republic of Korea), in cooperation with the Republic of
Chile. The Joint ALMA Observatory is operated by ESO, AUI/NRAO, and NAOJ.  We are grateful to the anonymous referee for numerous comments and 
suggestions, which improved the manuscript. PG acknowledges CSIR extended SRF fellowship (Grant No. 09/904 (0013) 2018
EMR-I) and and Chalmers Cosmic Origins postdoctoral fellowship. AD acknowledges ISRO respond project (Grant No. ISRO/RES/2/402/16-17) for partial financial support. B.B. acknowledges DST-INSPIRE Fellowship 
[IF170046] for providing partial financial assistance. SKM acknowledges CSIR fellowship (Ref no. 18/06/2017(i) EU-V).
This work is partially supported by Consortium for the Development of Human Resources in Science and Technology by Japan Science and Technology 
Agency.  This research was possible in part due to a Grant-In-Aid from the Higher Education Department of the Government of West Bengal.
DS acknowledges ASIAA postdoctoral research fund and travel grant. AD and TS also acknowledge support from DST JSPS grant.

\software{CASA \cite[v4.7.2;][]{mcmu07}, CASSIS (Cassis Team At CESR/IRAP 2014), RADEX \citep{vand07}}

%{  \appendix}
\clearpage
\appendix
\restartappendixnumbering
% \counterwithin{figure}{section}
\section{Fitted spectra, LTE model spectra, moment maps, and schematic diagram of G31 based on present observation.}
%{\hskip 7.5cm \Large   Appendix}\\\\
\vskip 4 cm
\begin{figure}
\centering
\includegraphics[width=\textwidth]{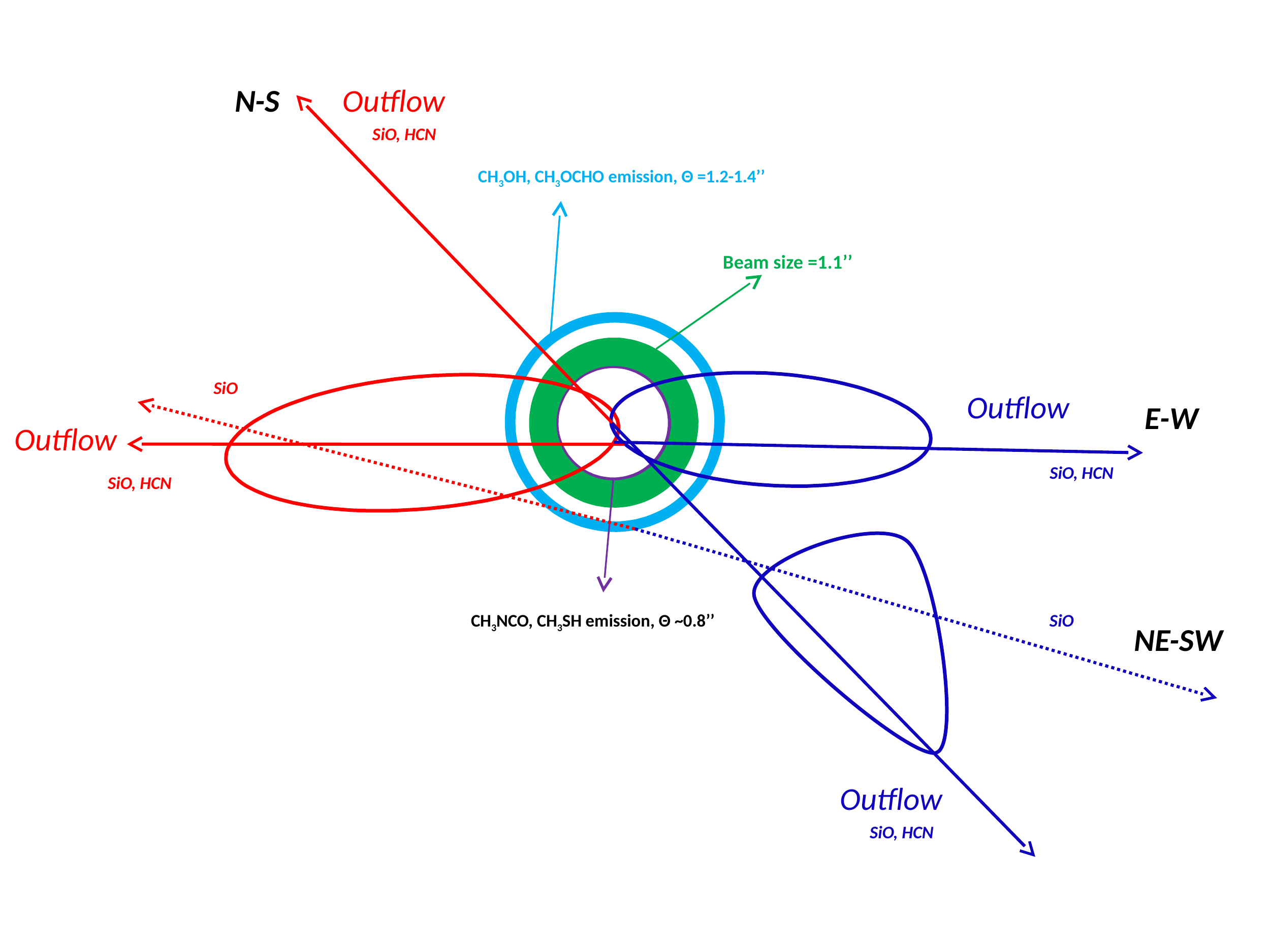}
% \begin{minipage}{0.35\textwidth}
% \includegraphics[width=\textwidth]{ch3sh-LTE.pdf}
% \end{minipage}
\caption{{Schematic diagram (not up to scale) of kinematics and molecular distribution in G31.41+0.31 based on present observational results (see details in 
the Section 3.6.6).}}
\label{G31-diagram}
\end{figure}
%\noindent {{  Figure A1:} Spectral Line Fitting .}
%\clearpage

\begin{figure}
\begin{minipage}{0.35\textwidth}
\includegraphics[width=\textwidth]{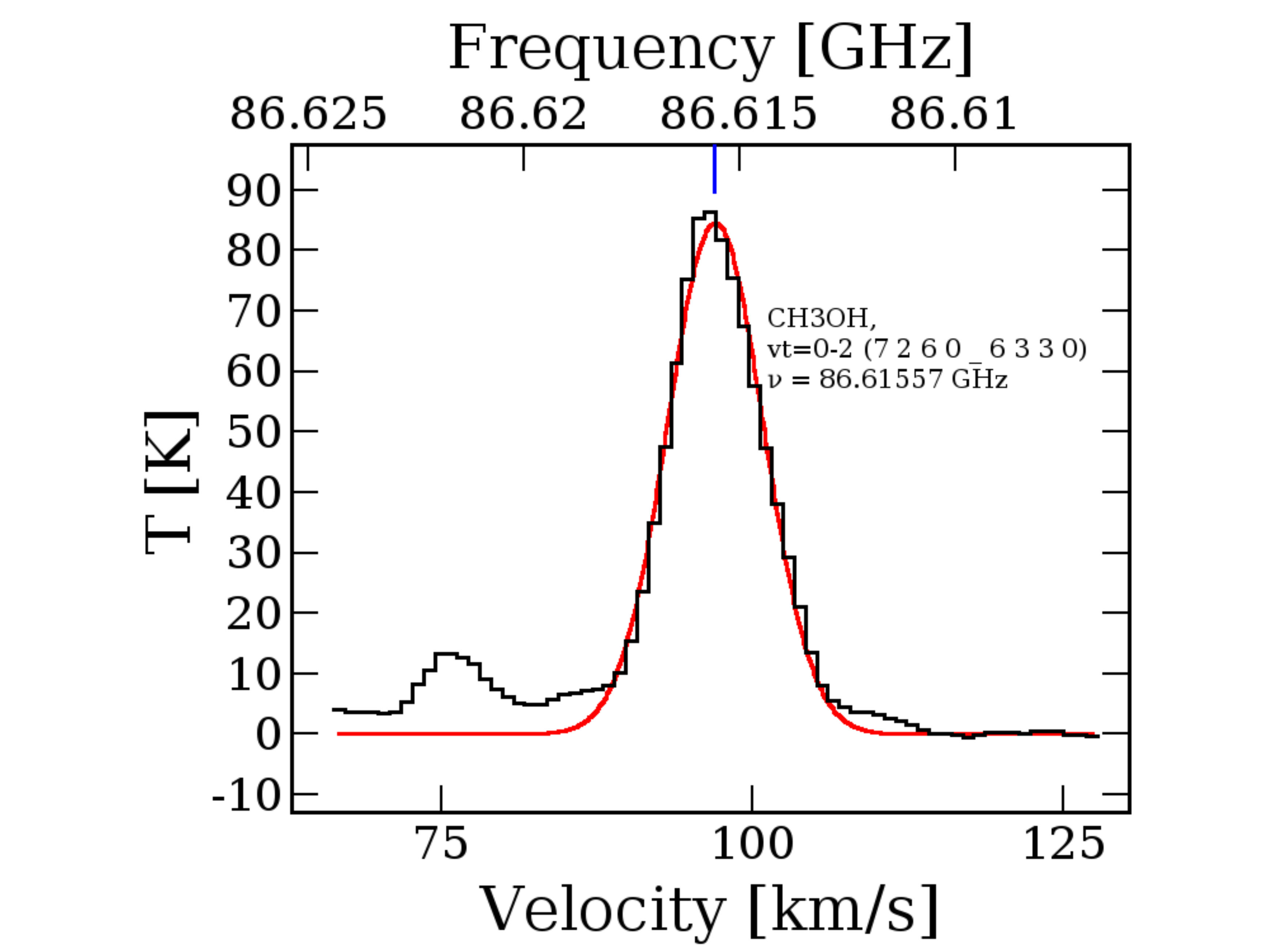}
\end{minipage}
\begin{minipage}{0.35\textwidth}
\includegraphics[width=\textwidth]{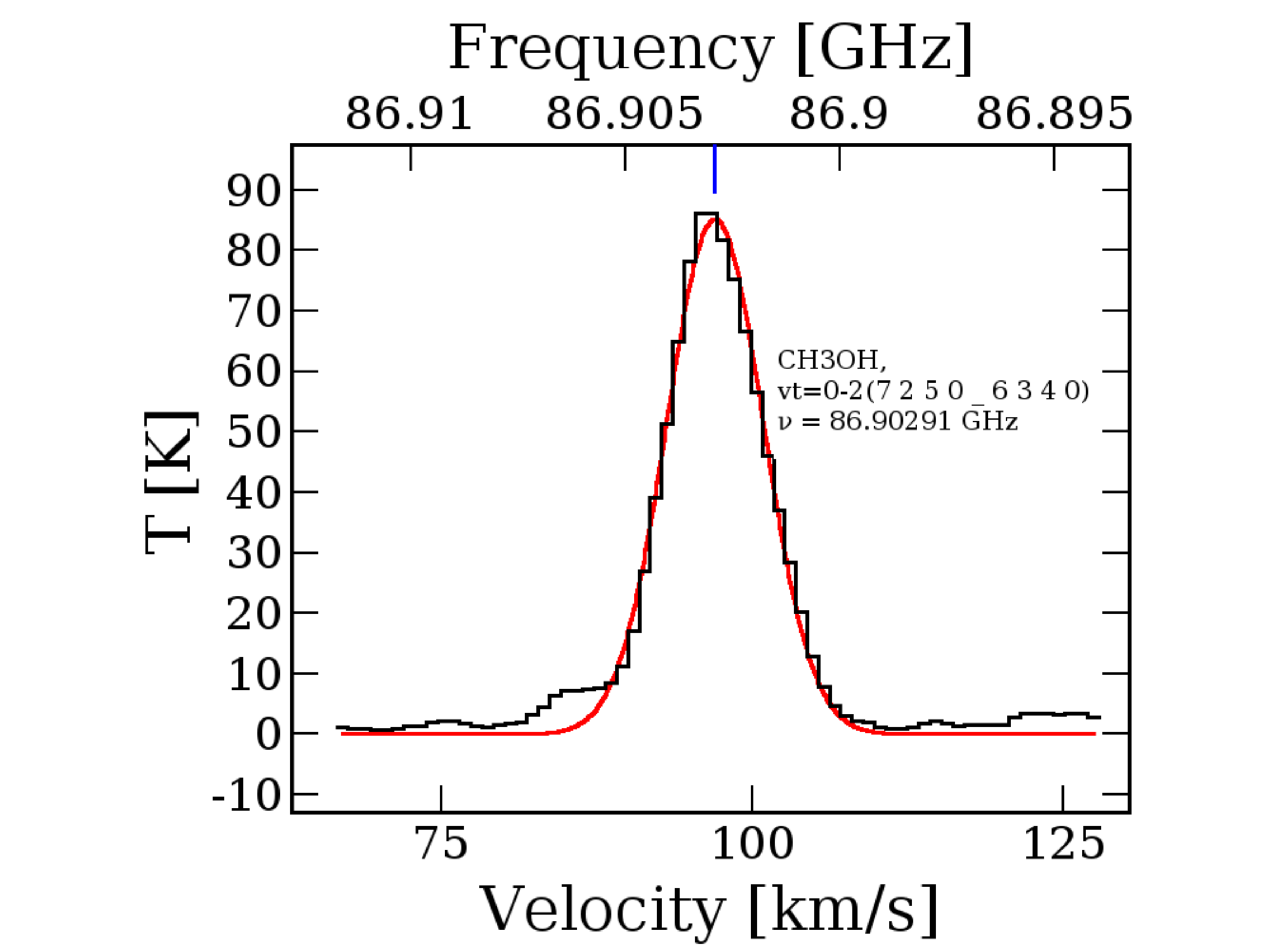}
\end{minipage}
\begin{minipage}{0.35\textwidth}
\includegraphics[width=\textwidth]{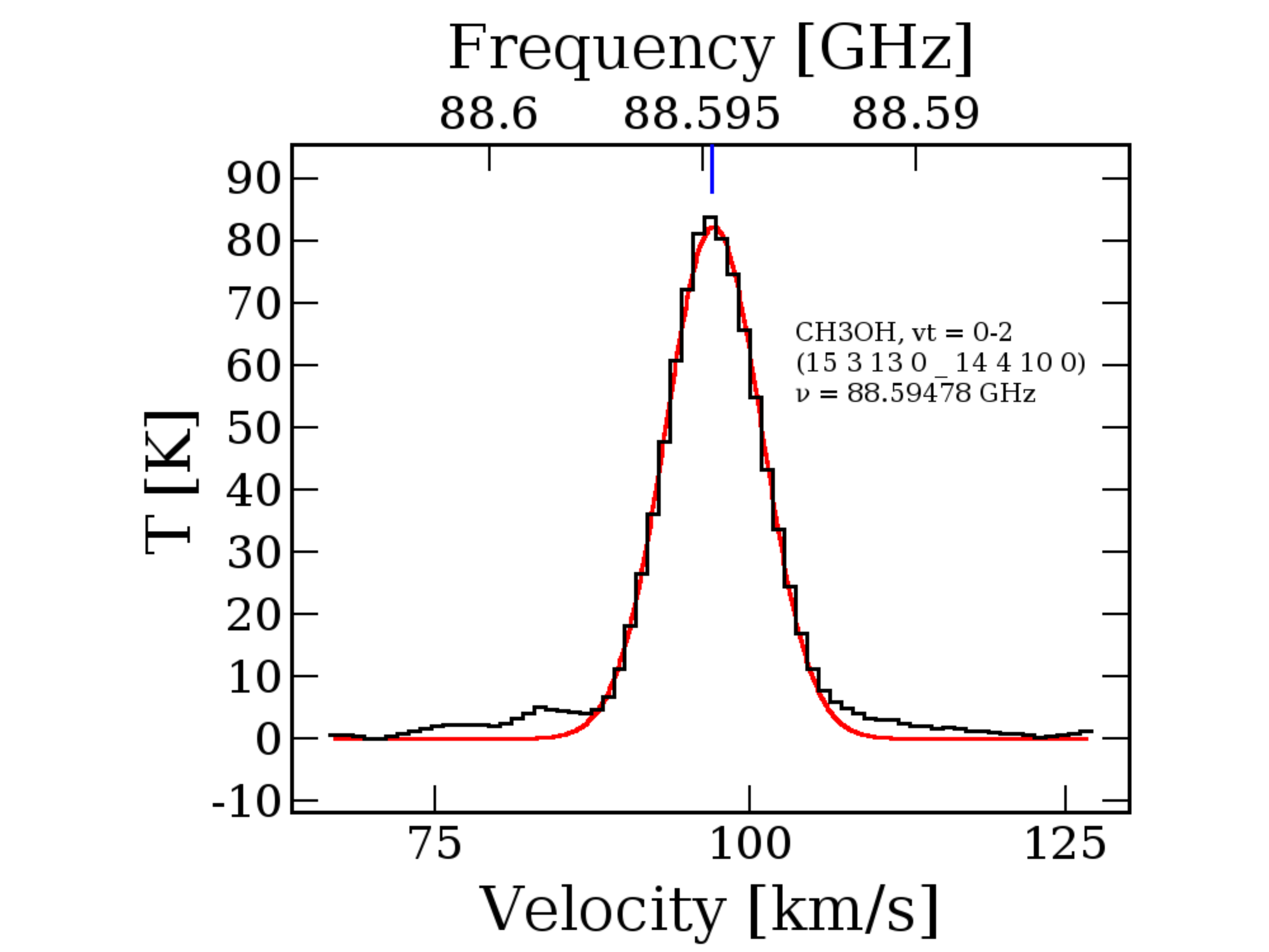}
\end{minipage}
\begin{minipage}{0.35\textwidth}
\includegraphics[width=\textwidth]{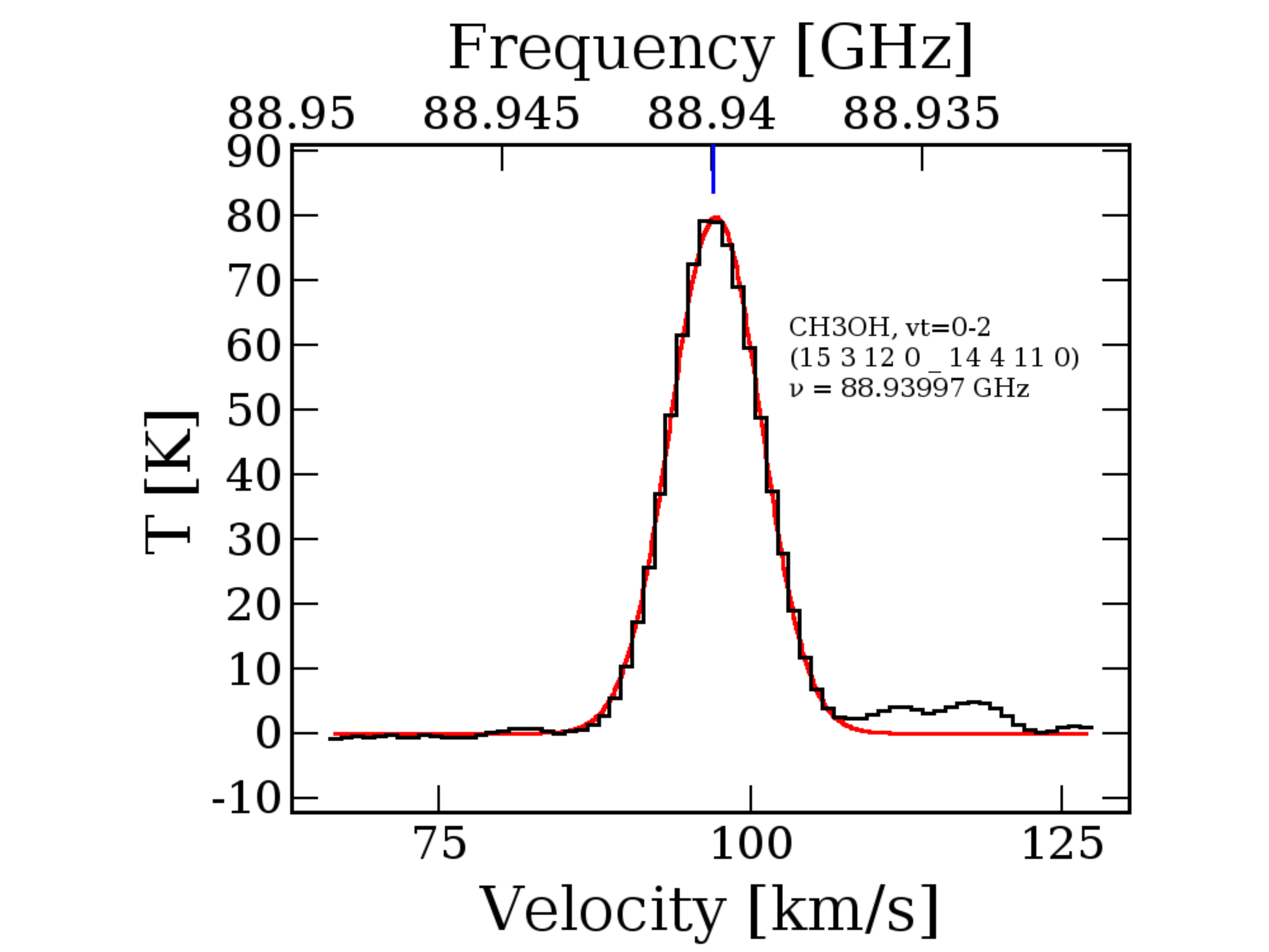}
\end{minipage}
\begin{minipage}{0.35\textwidth}
\includegraphics[width=\textwidth]{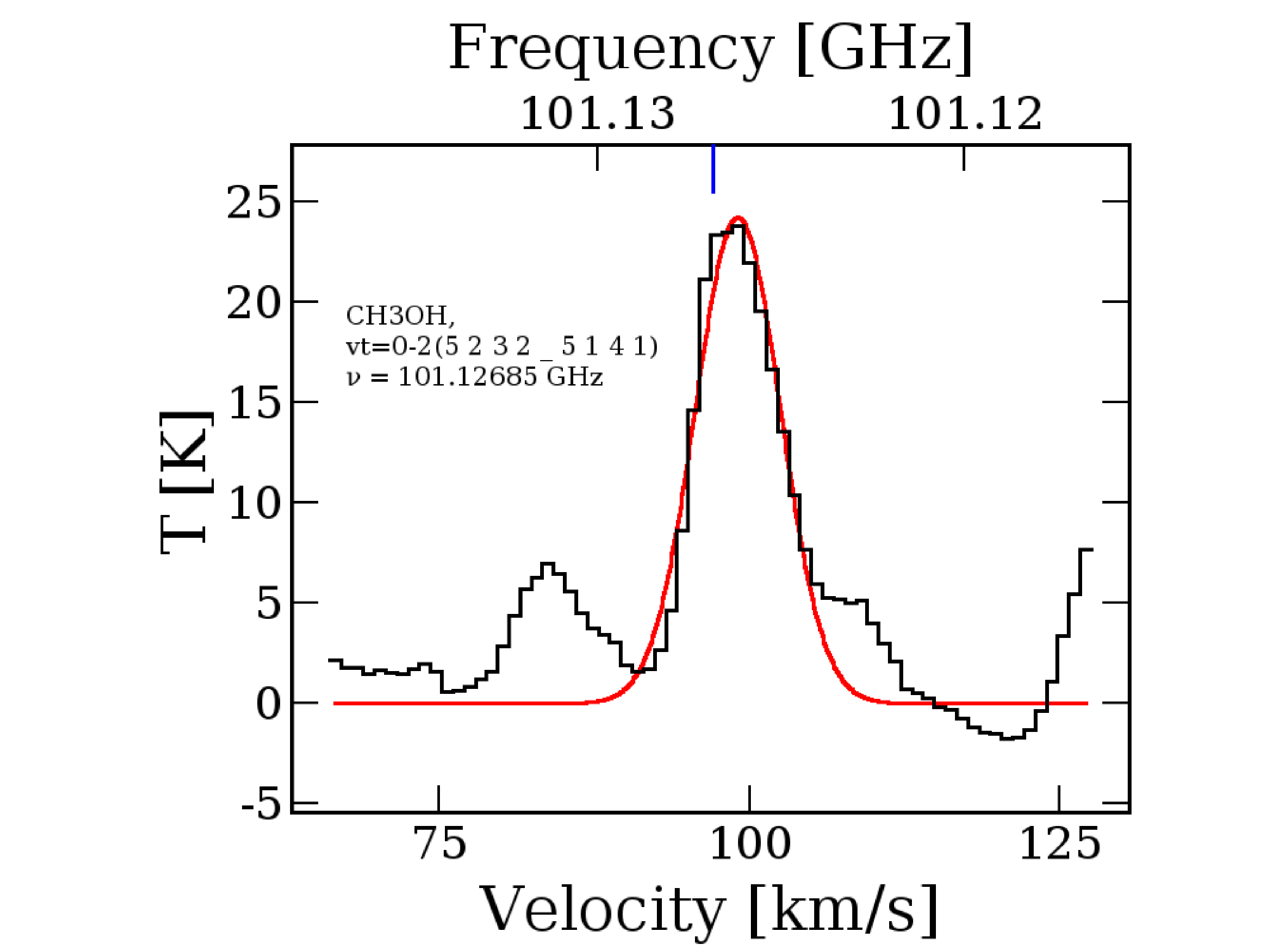}
\end{minipage}
\begin{minipage}{0.35\textwidth}
\includegraphics[width=\textwidth]{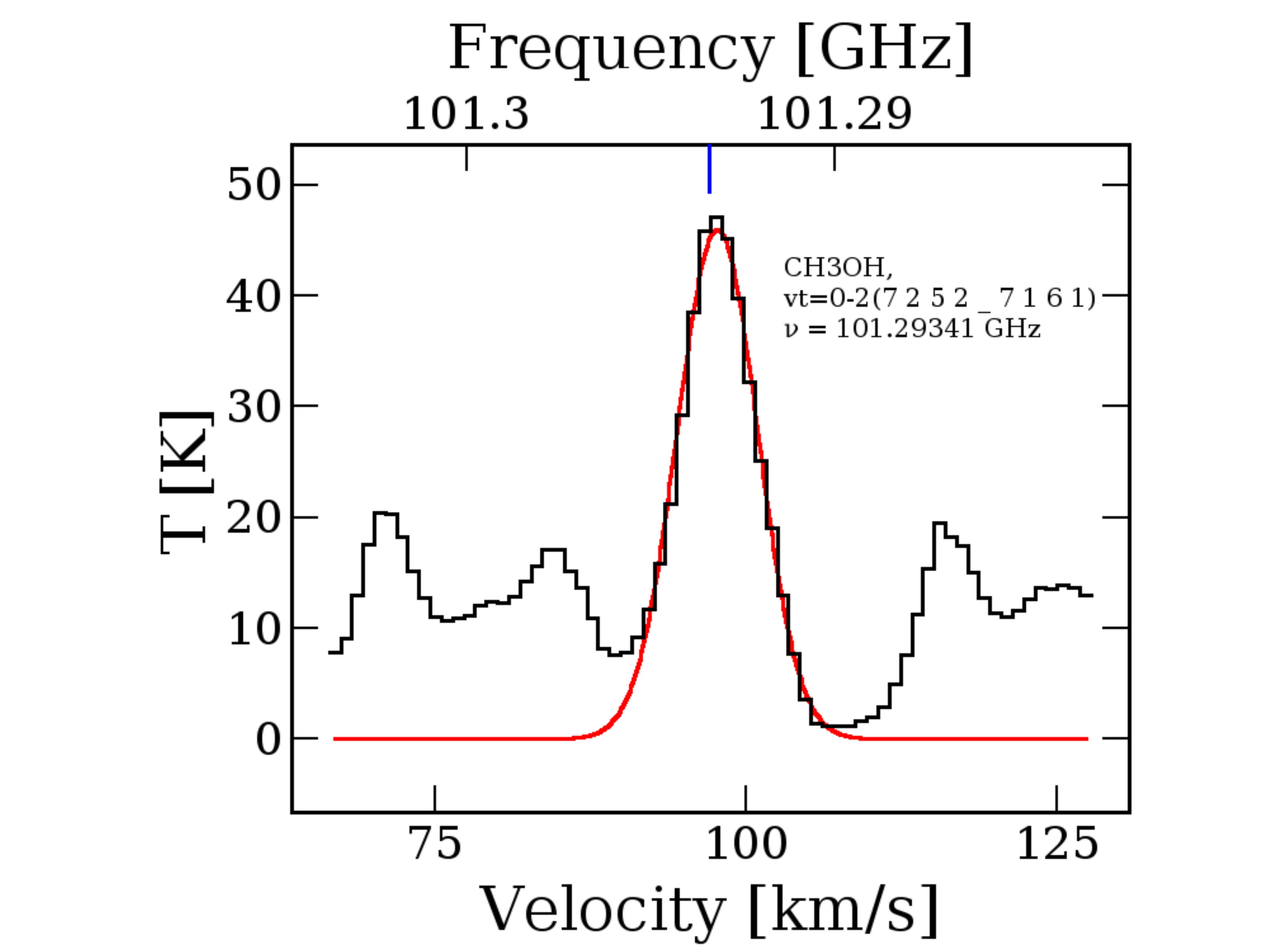}
\end{minipage}
\begin{minipage}{0.35\textwidth}
\includegraphics[width=\textwidth]{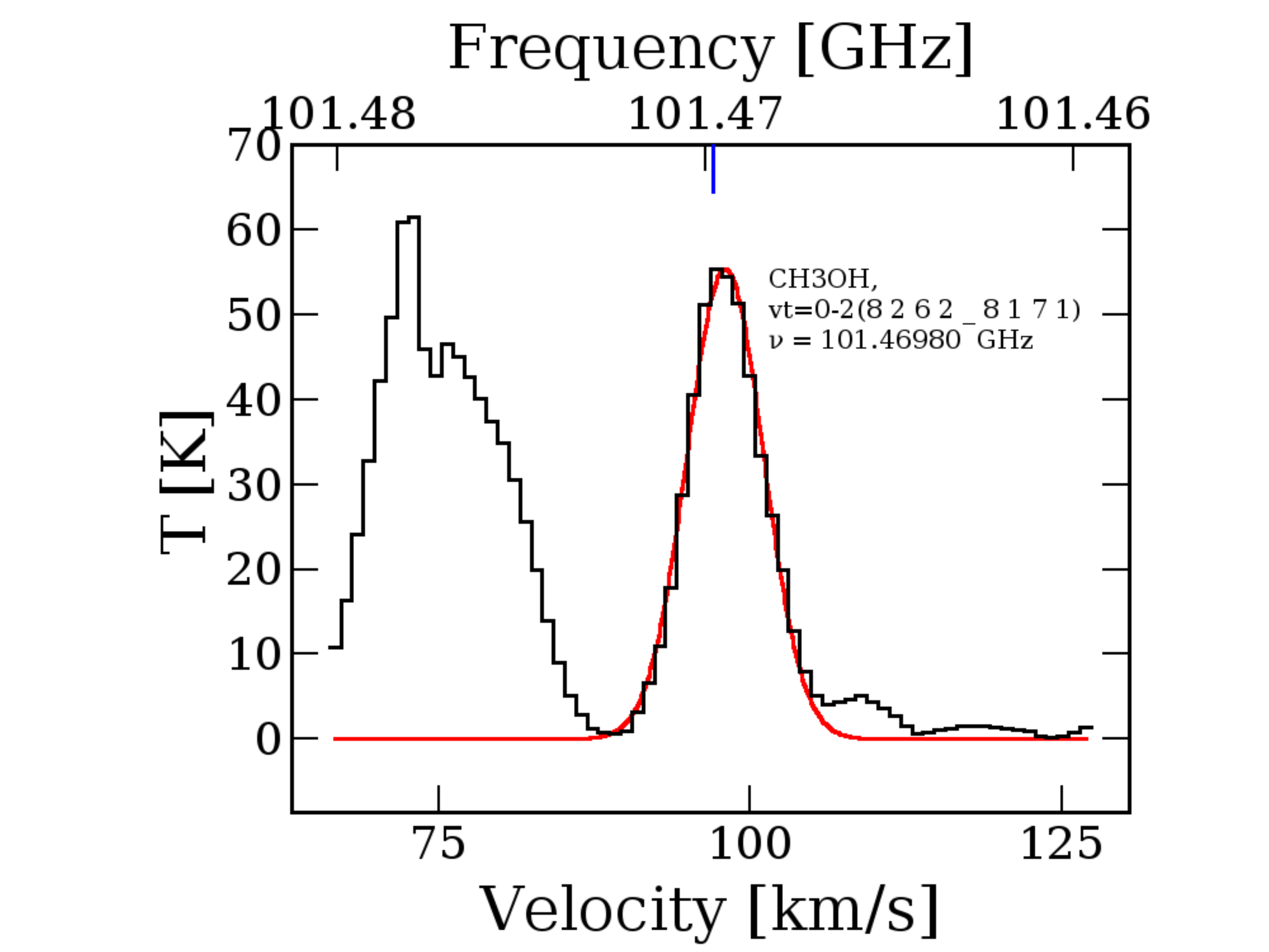}
\end{minipage}
\caption{Black line represents observed emission spectra of methanol (CH$_3$OH) and red line represents a Gaussian profile fitted to the observed 
spectra.}
\label{Gfit-CH3OH}
\end{figure}

\begin{figure}
\begin{minipage}{0.35\textwidth}
\includegraphics[width=\textwidth]{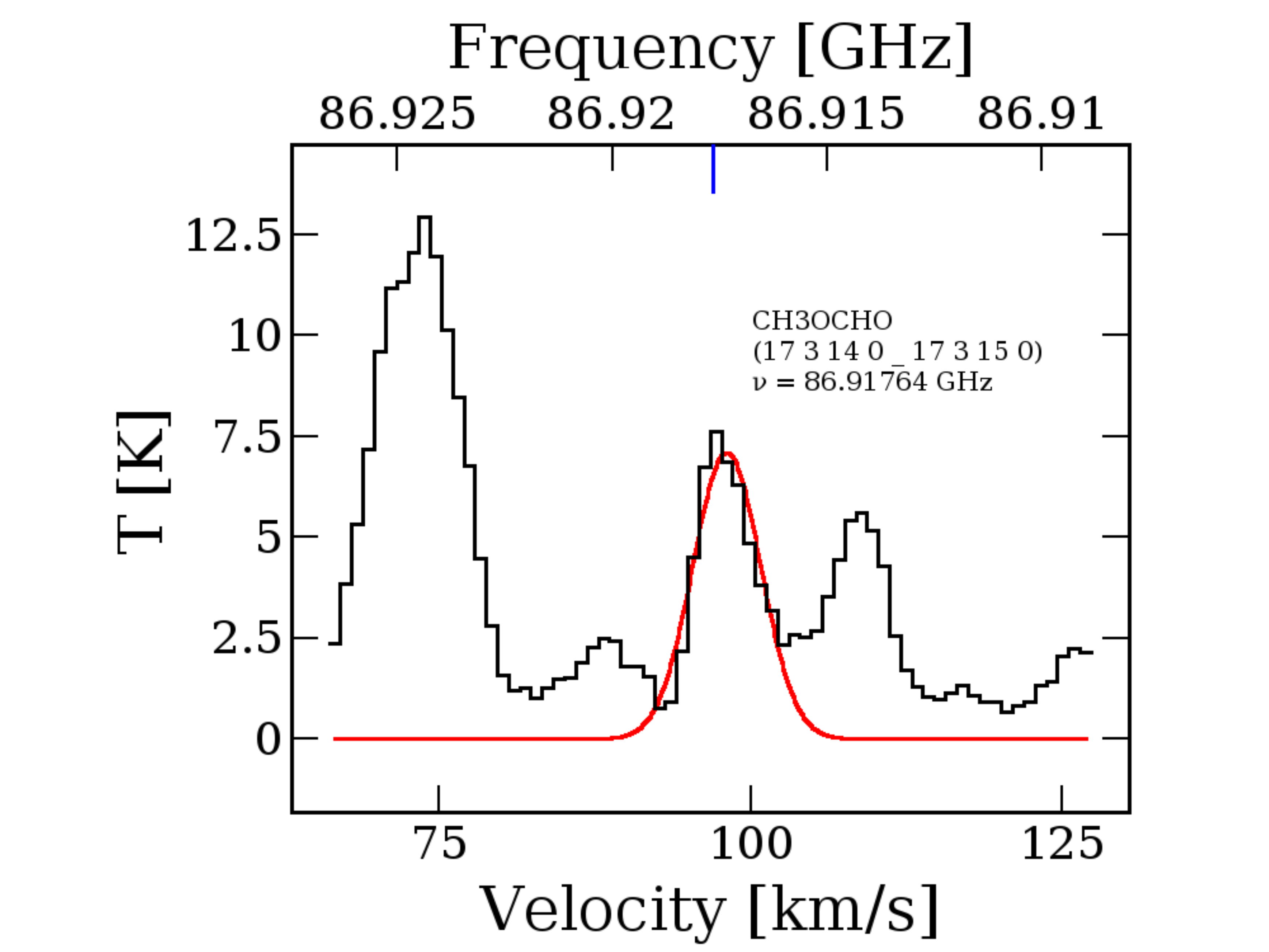}
\end{minipage}
\begin{minipage}{0.35\textwidth}
\includegraphics[width=\textwidth]{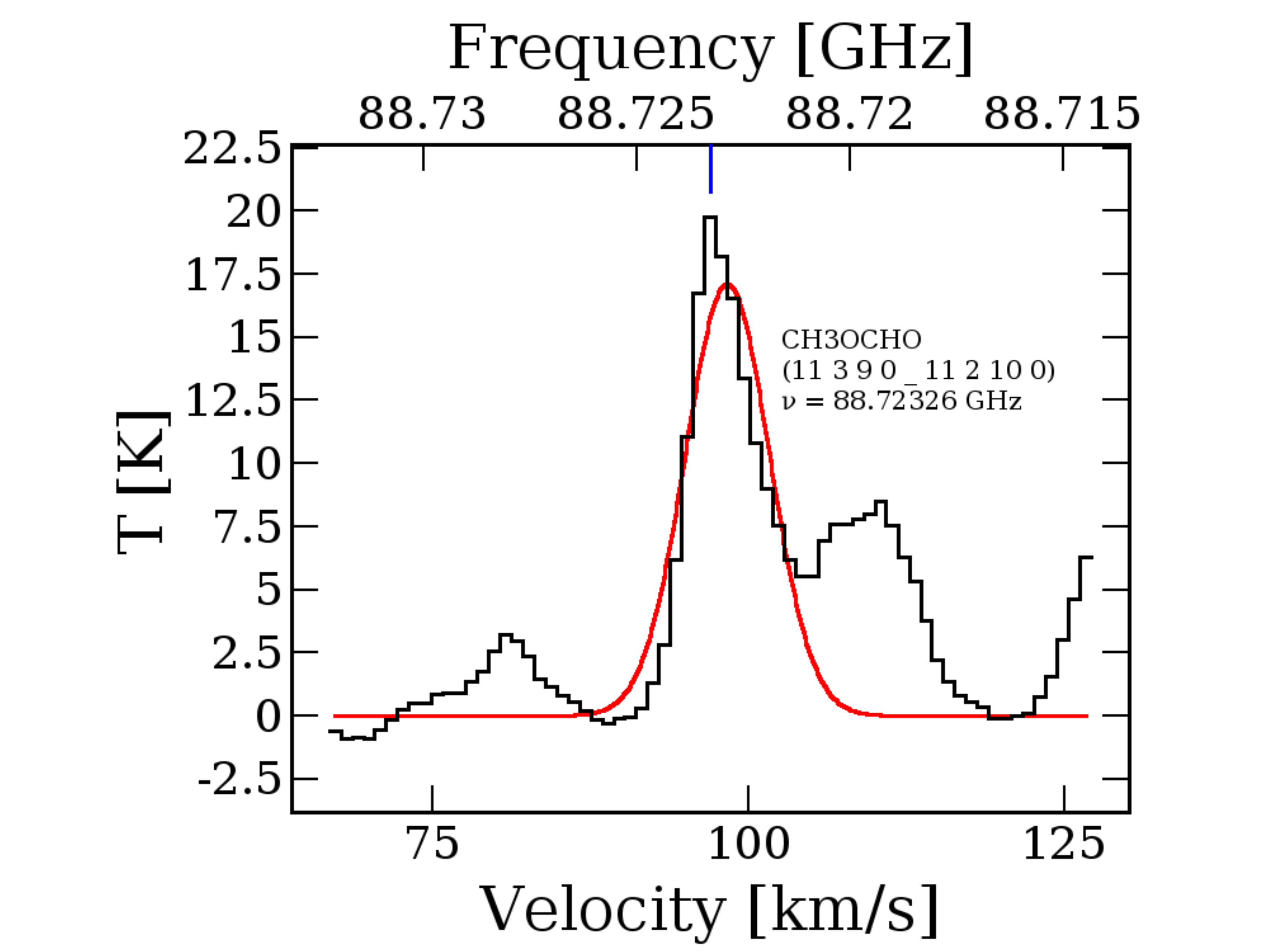}
\end{minipage}
\begin{minipage}{0.35\textwidth}
\includegraphics[width=\textwidth]{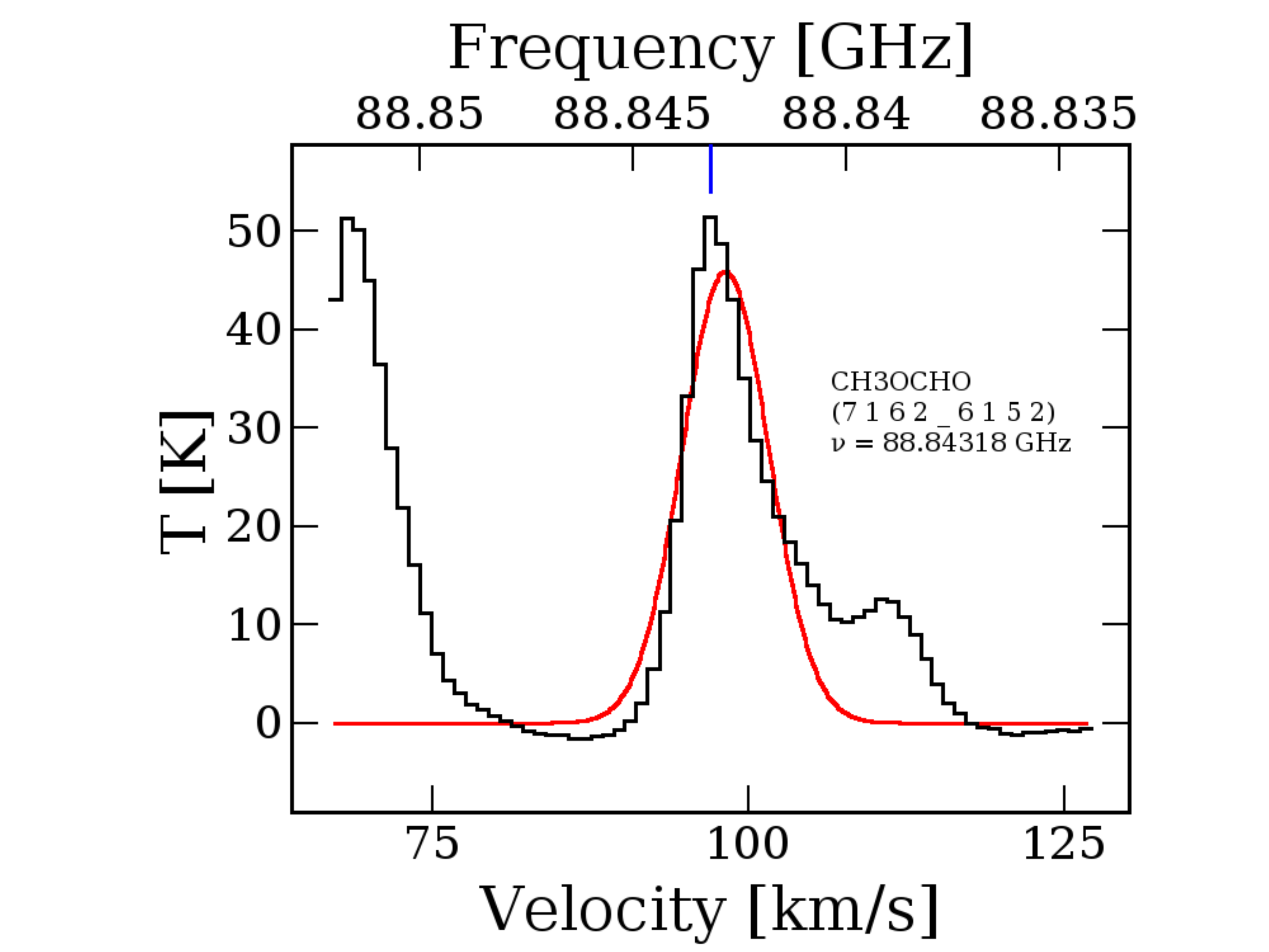}
\end{minipage}
\begin{minipage}{0.35\textwidth}
\includegraphics[width=\textwidth]{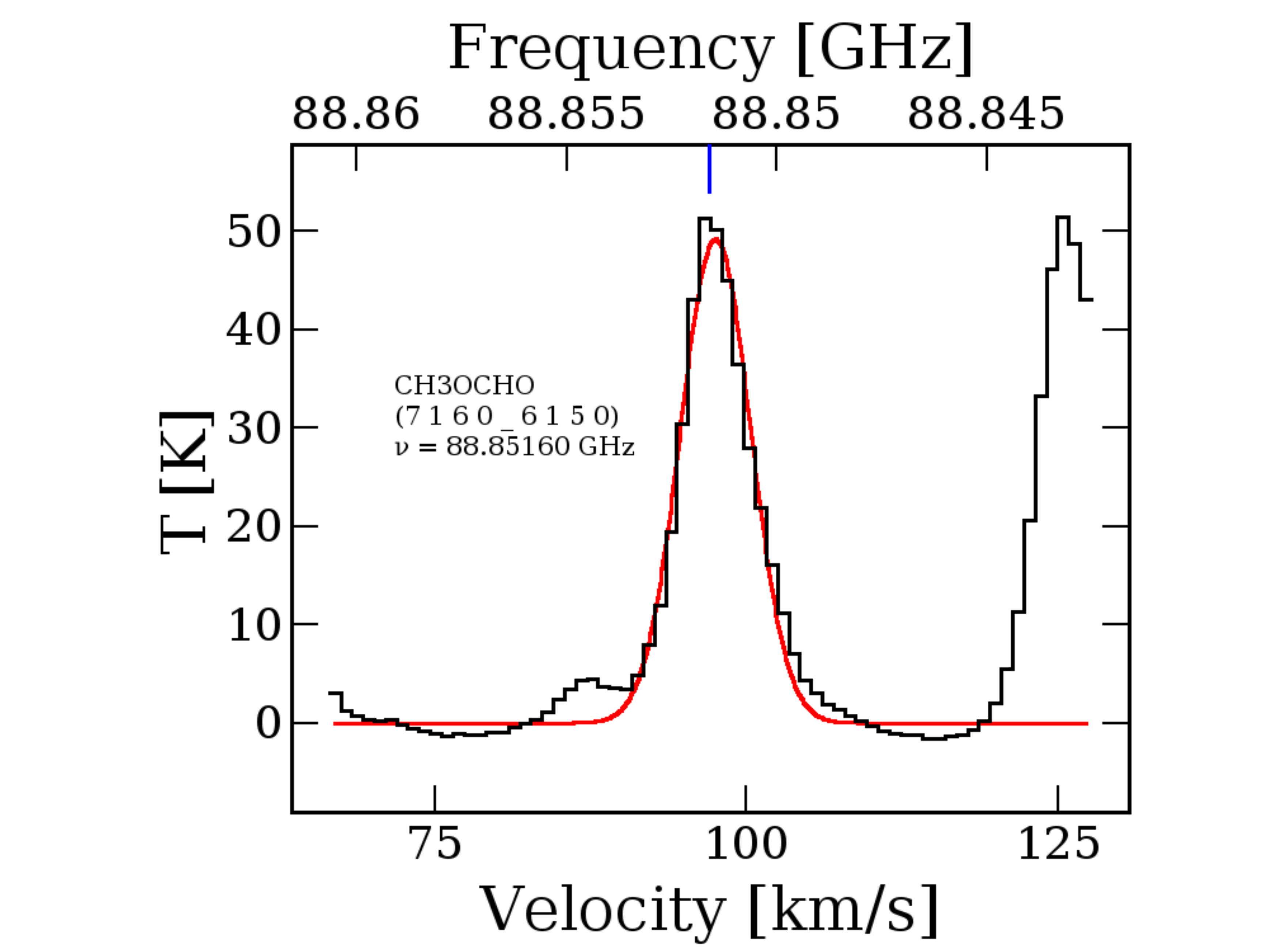}
\end{minipage}
\begin{minipage}{0.35\textwidth}
\includegraphics[width=\textwidth]{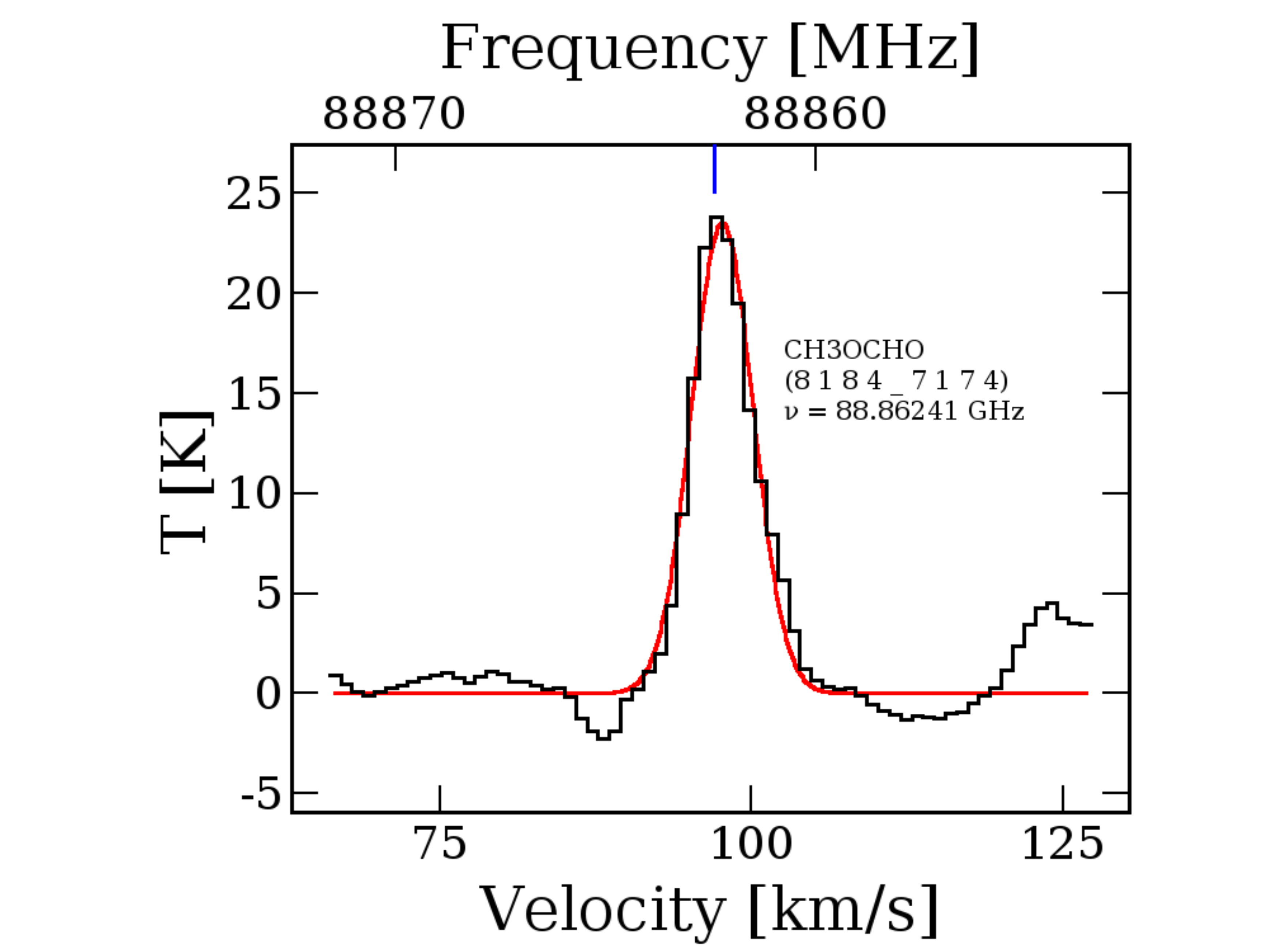}
\end{minipage}
\begin{minipage}{0.35\textwidth}
\includegraphics[width=\textwidth]{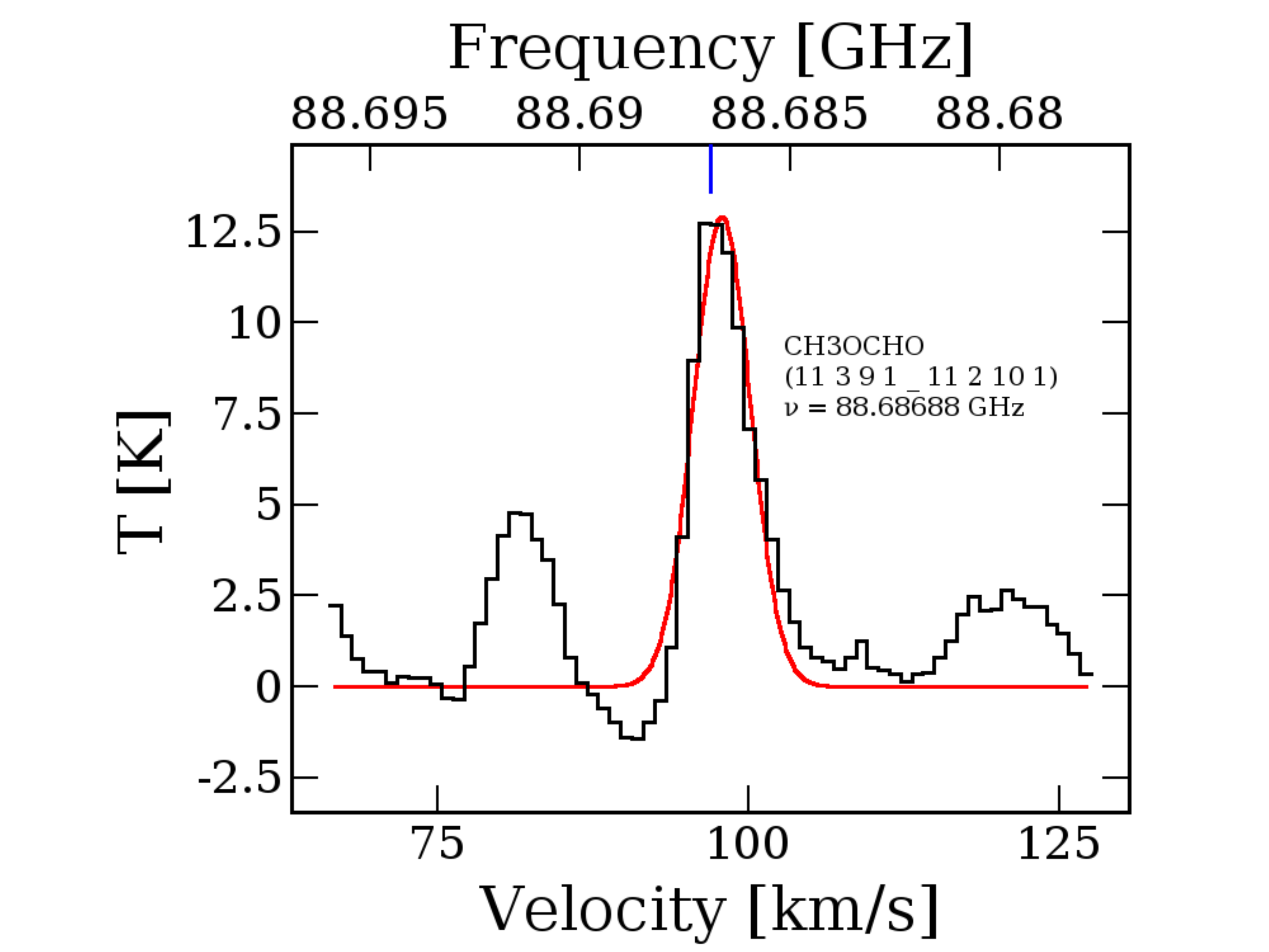}
\end{minipage}
\begin{minipage}{0.35\textwidth}
\includegraphics[width=\textwidth]{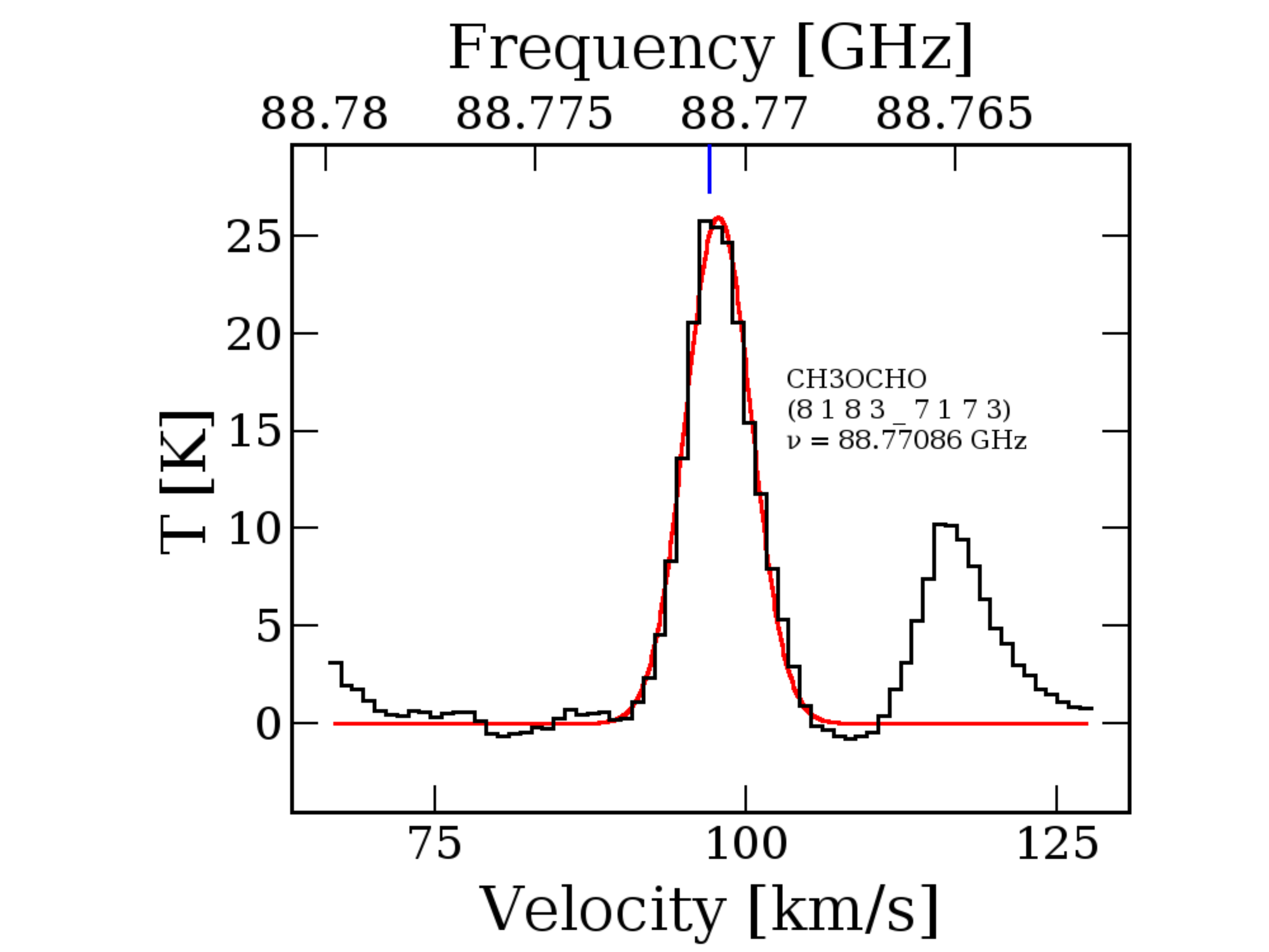}
\end{minipage}
\begin{minipage}{0.35\textwidth}
\includegraphics[width=\textwidth]{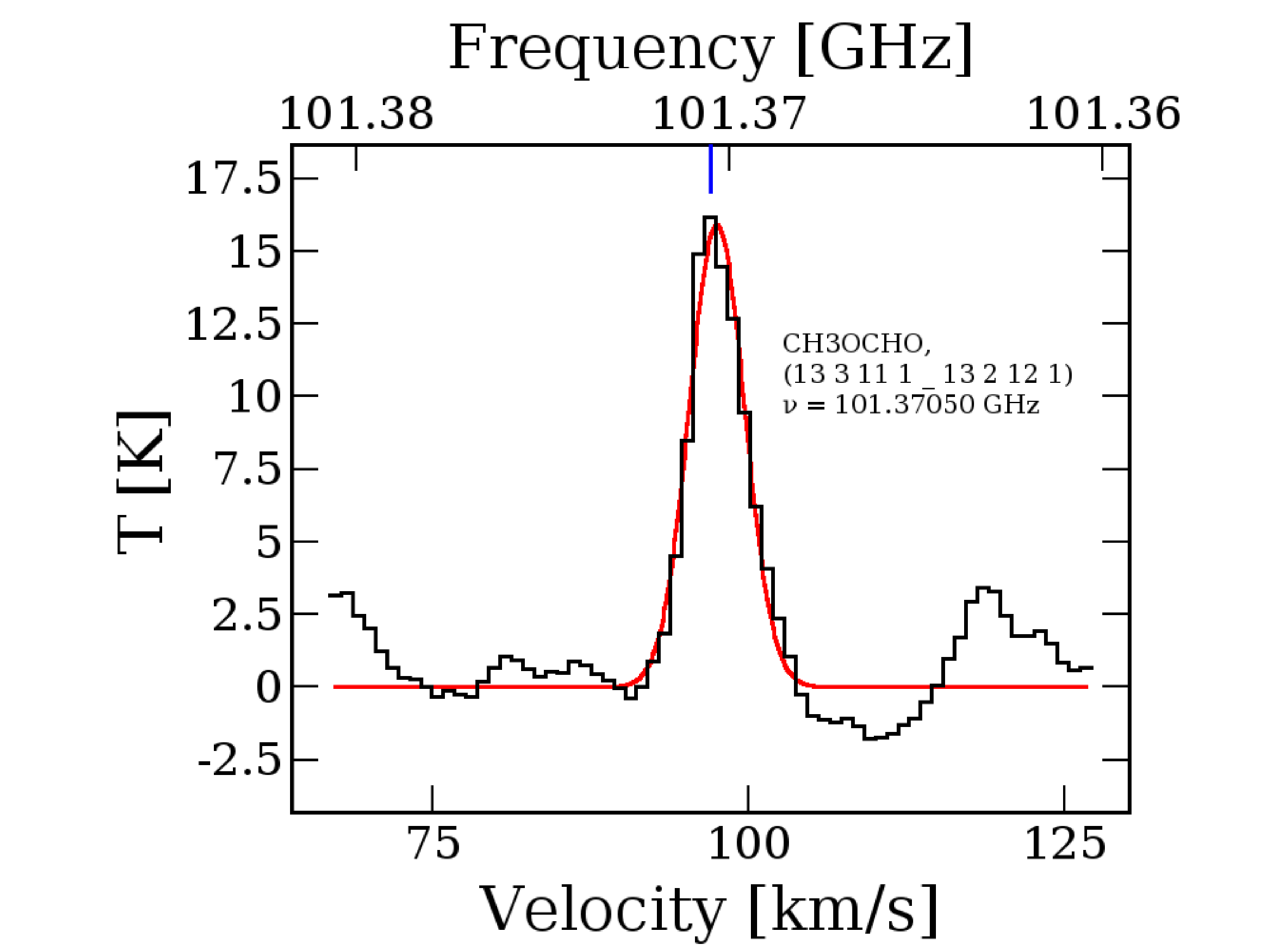}
\end{minipage}
\begin{minipage}{0.35\textwidth}
\includegraphics[width=\textwidth]{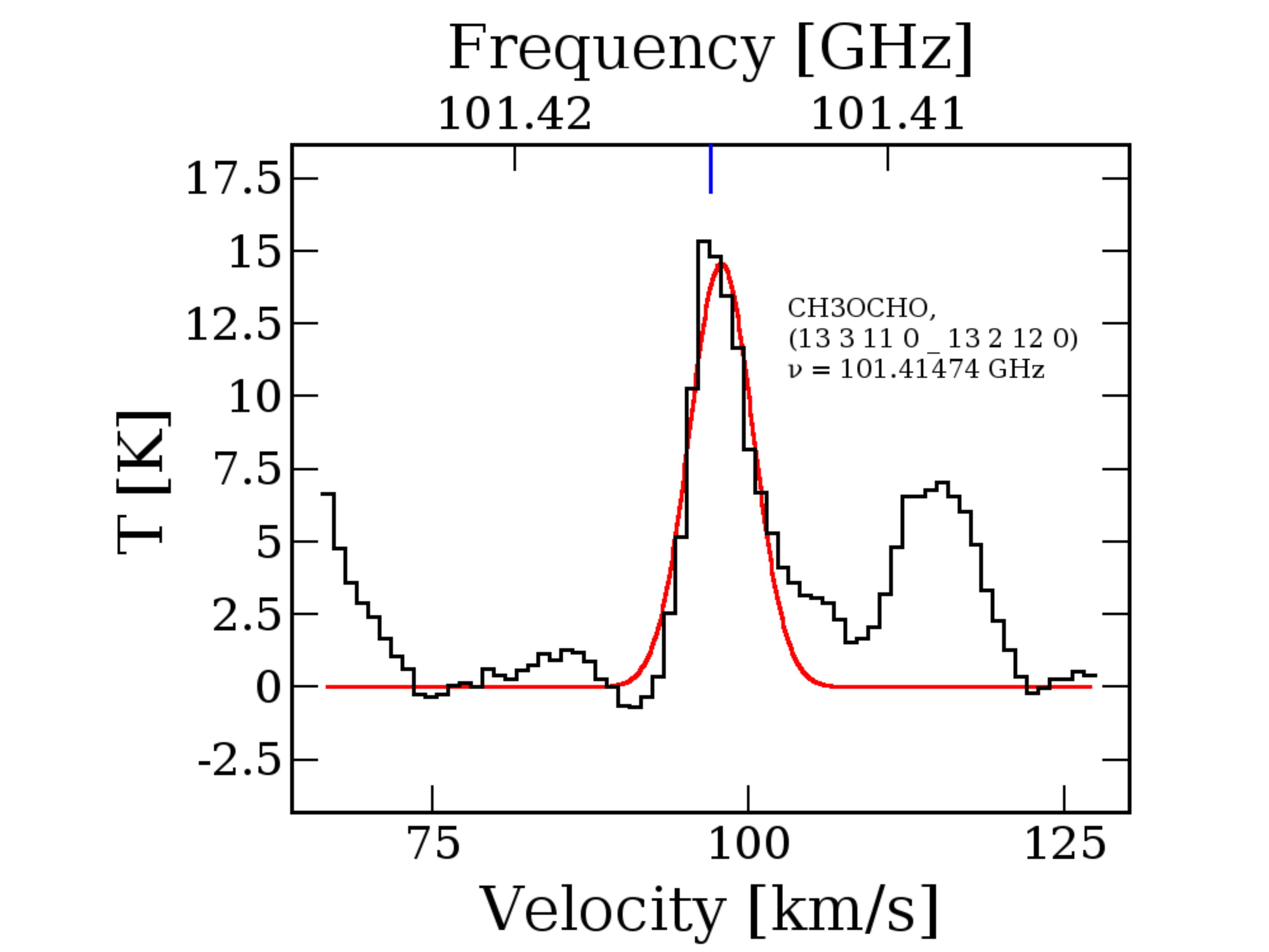}
\end{minipage}

\caption{Black line represents observed emission spectra of methyl formate (CH$_3$OCHO) and red line represents a Gaussian profile fitted to the observed 
spectra. $\rm{11_{39}-11_{2, 10} (A)}$ transition may blend with some other molecular transition.}
\label{Gfit-CH3OCHO}

\end{figure}

\begin{figure}
\begin{minipage}{0.35\textwidth}
\includegraphics[width=\textwidth]{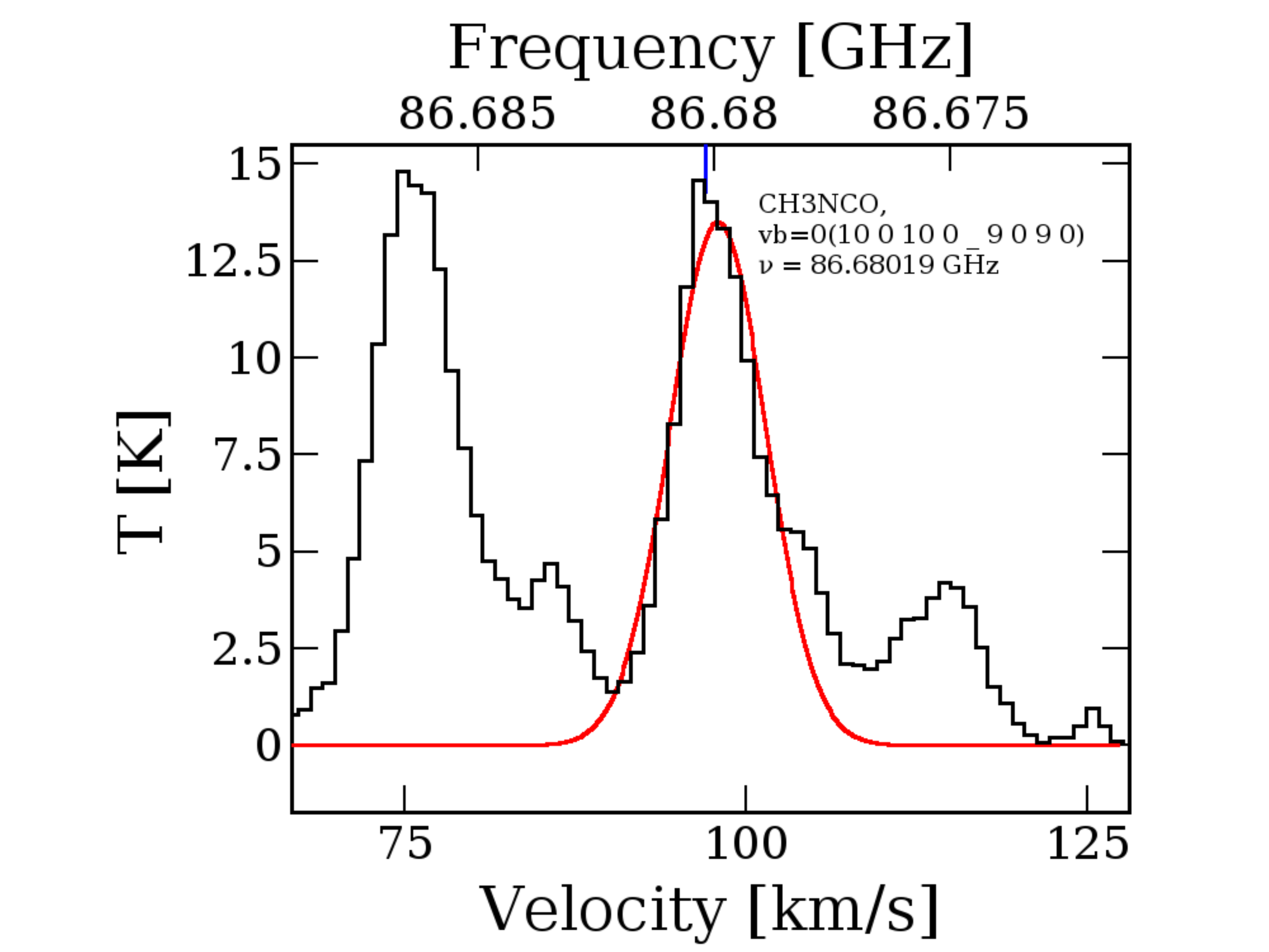}
\end{minipage}
\begin{minipage}{0.35\textwidth}
\includegraphics[width=\textwidth]{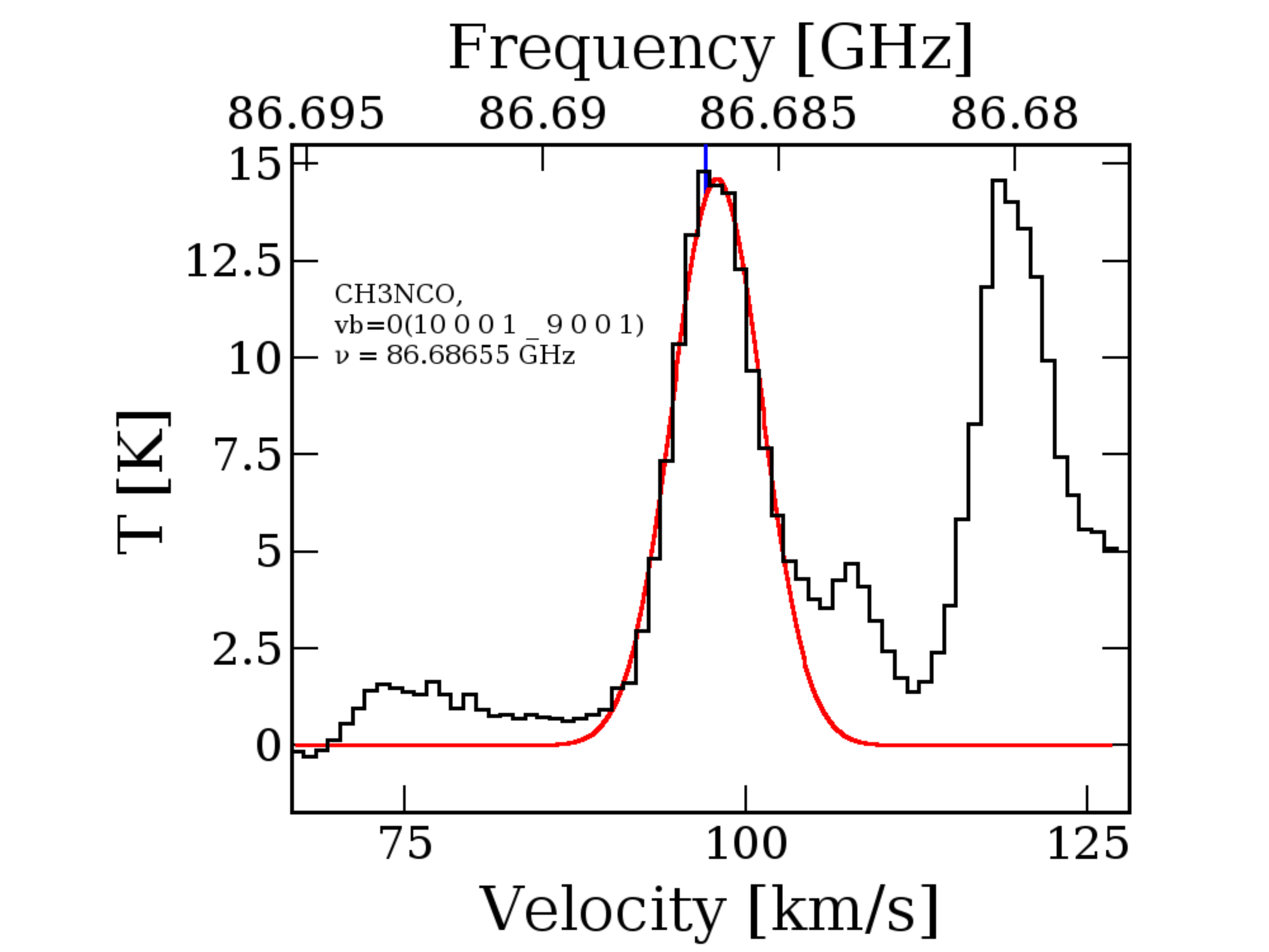}
\end{minipage}
\begin{minipage}{0.35\textwidth}
\includegraphics[width=\textwidth]{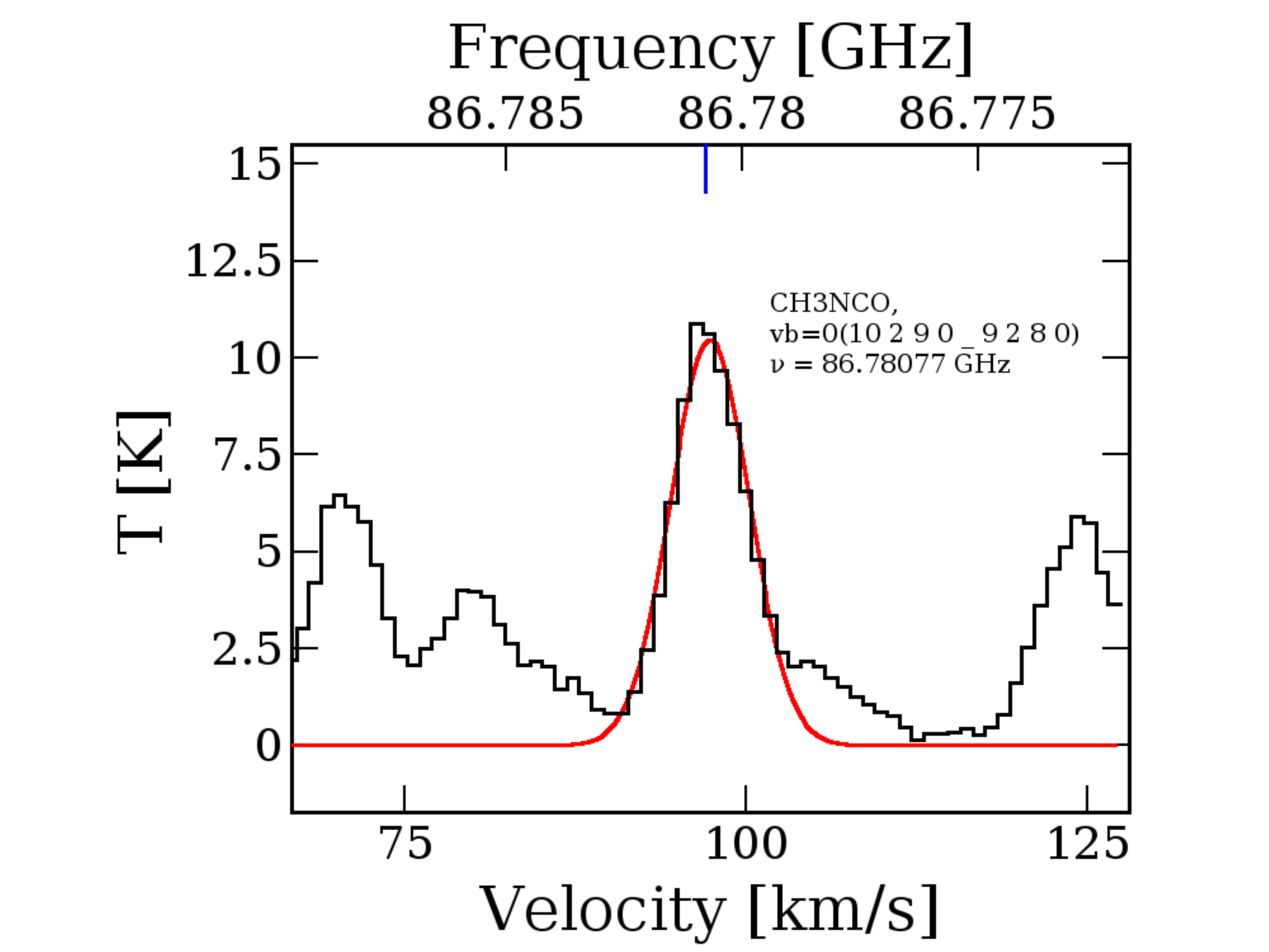}
\end{minipage}
\begin{minipage}{0.35\textwidth}
\includegraphics[width=\textwidth]{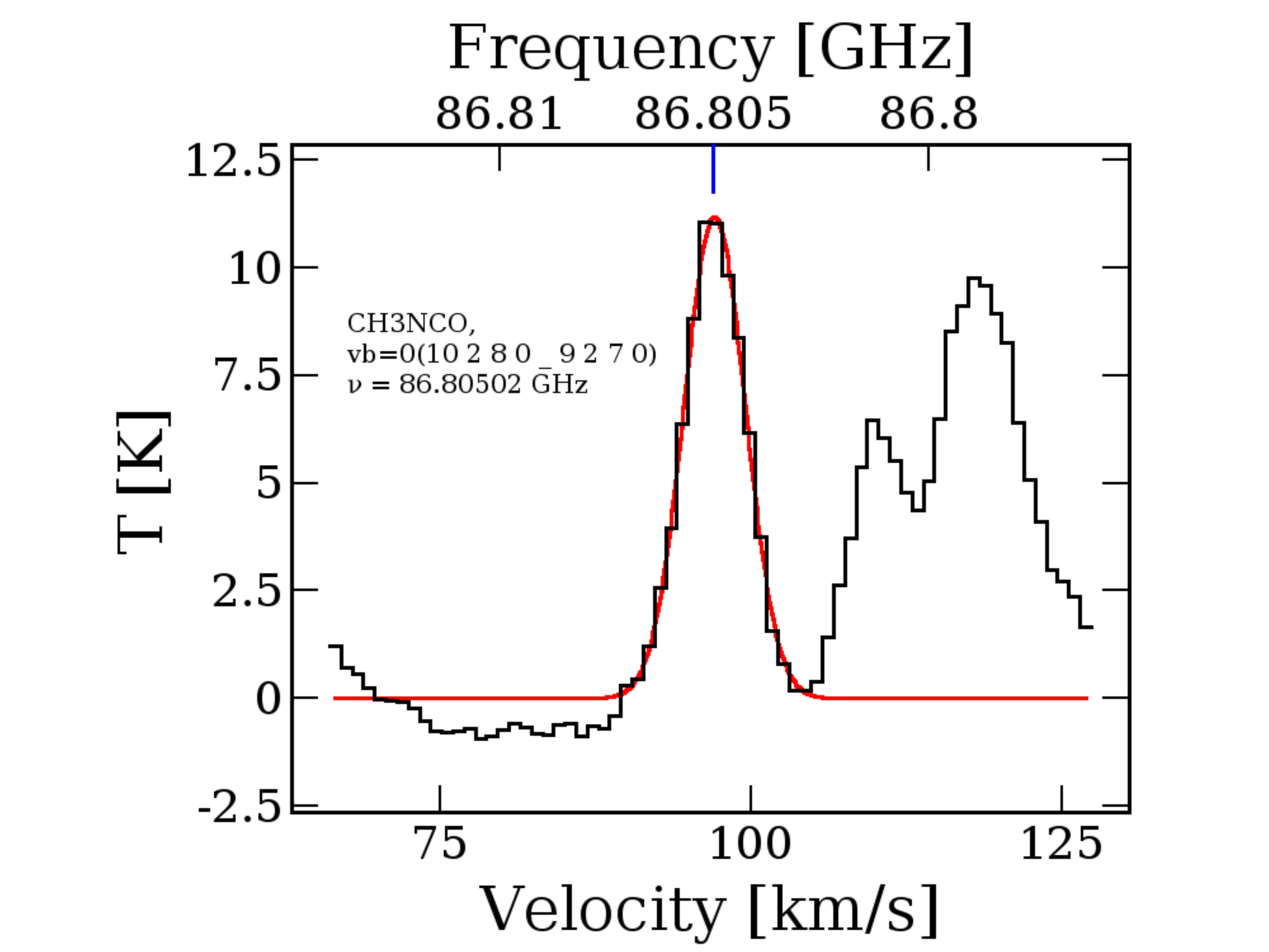}
\end{minipage}
\begin{minipage}{0.35\textwidth}
\includegraphics[width=\textwidth]{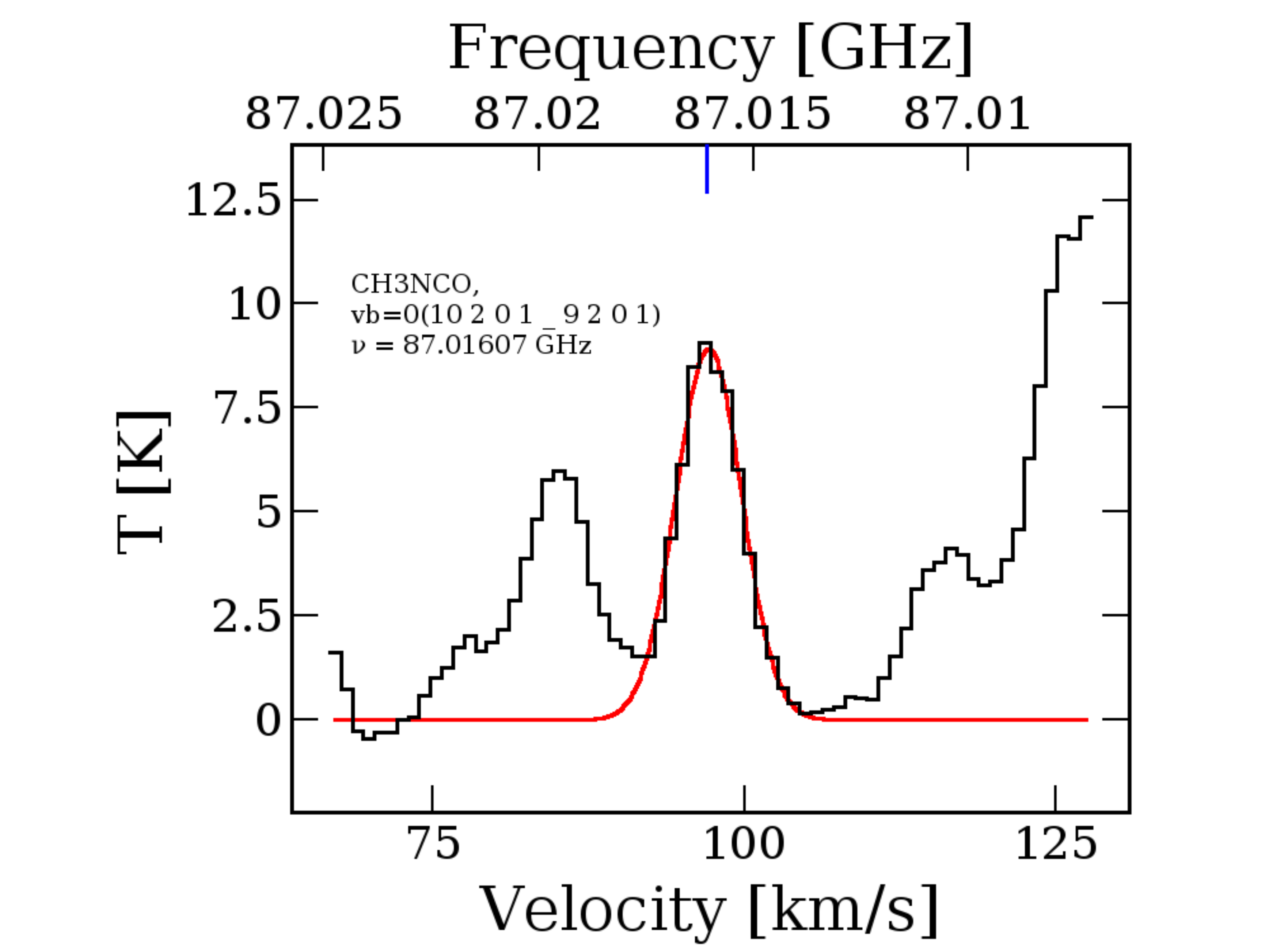}
\end{minipage}
\begin{minipage}{0.35\textwidth}
\includegraphics[width=\textwidth]{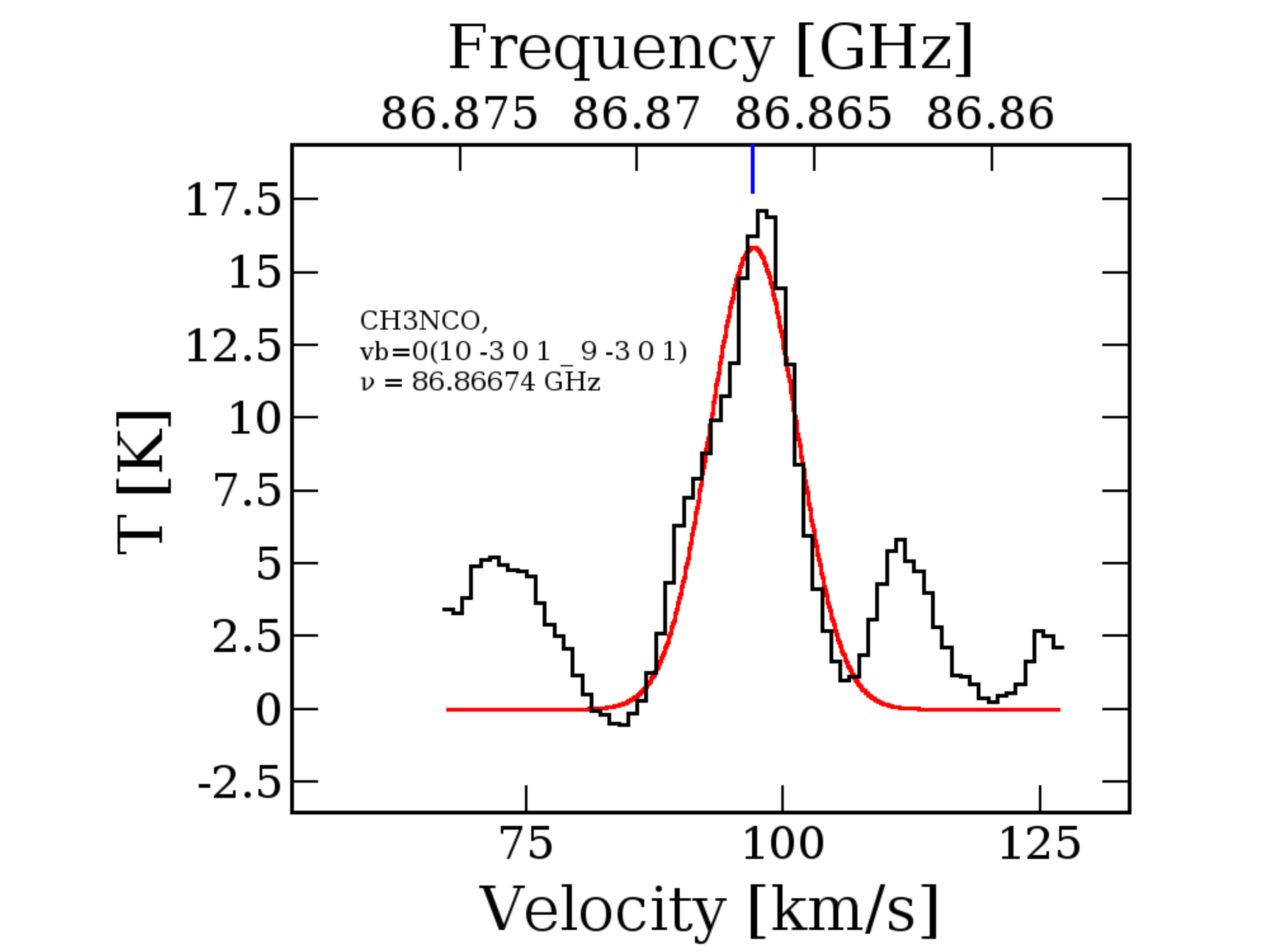}
\end{minipage}
\caption{Black line represents observed emission spectra of methyl isocyanate (CH$_3$NCO) and red line represents a Gaussian profile fitted to the observed 
spectra. $\rm{10_{-30}-9_{-30}}$ transition may strongly blend with some other molecular transitions.}
\label{Gfit-CH3NCO}
\end{figure}

\begin{figure}
\begin{minipage}{0.35\textwidth}
\includegraphics[width=\textwidth]{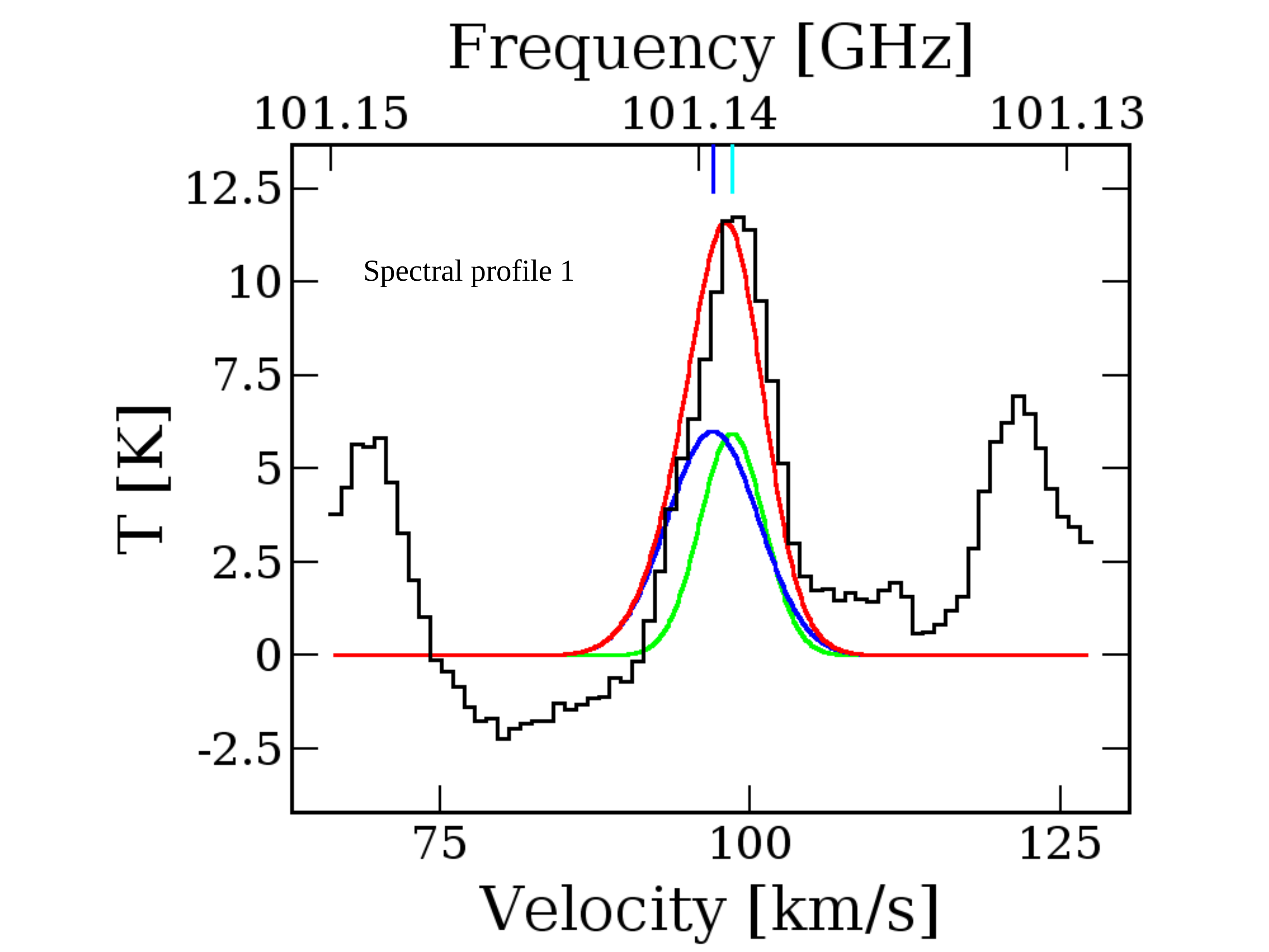}
\end{minipage}
\begin{minipage}{0.35\textwidth}
\includegraphics[width=\textwidth]{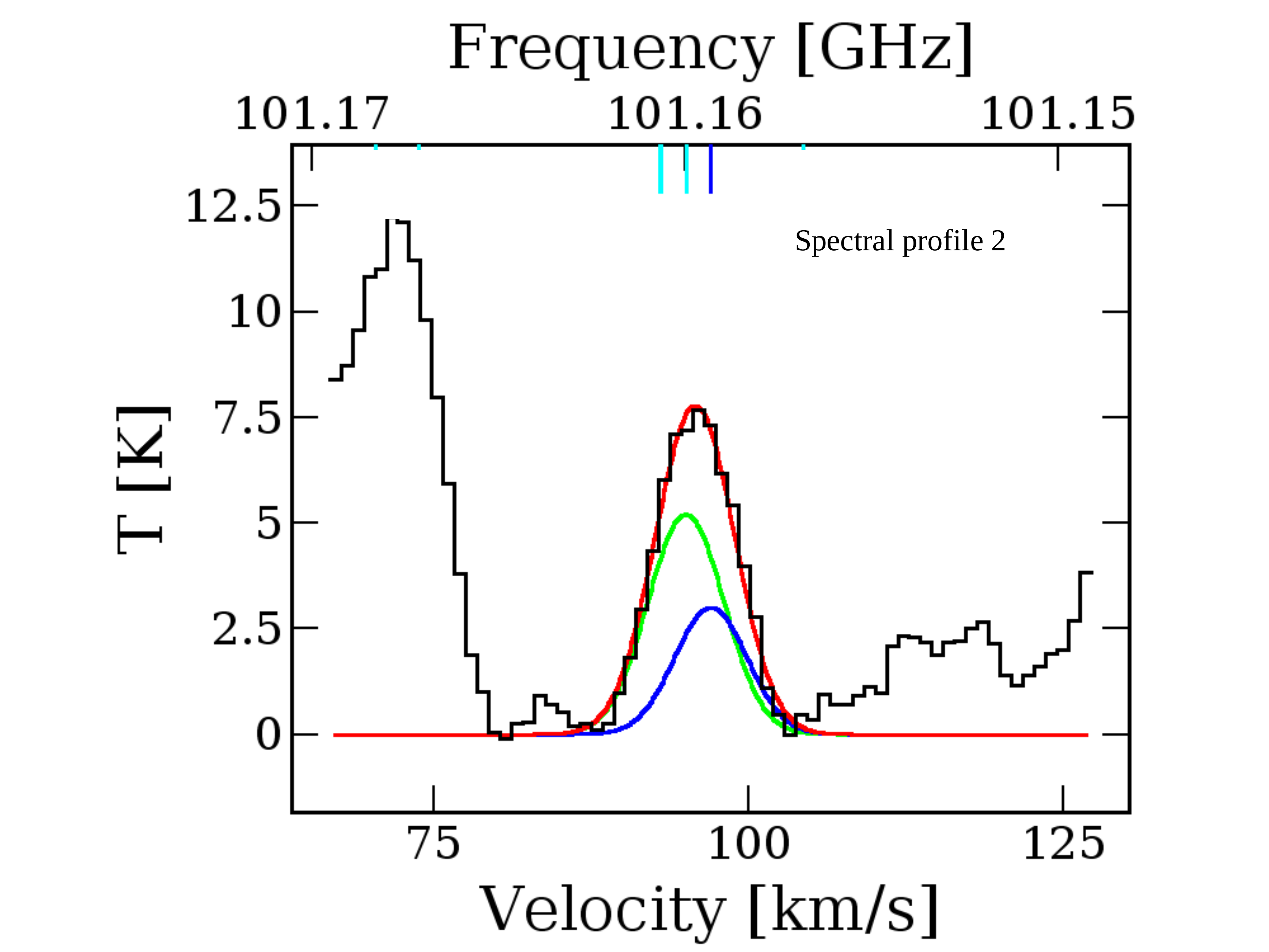}
\end{minipage}
\begin{minipage}{0.35\textwidth}
\includegraphics[width=\textwidth]{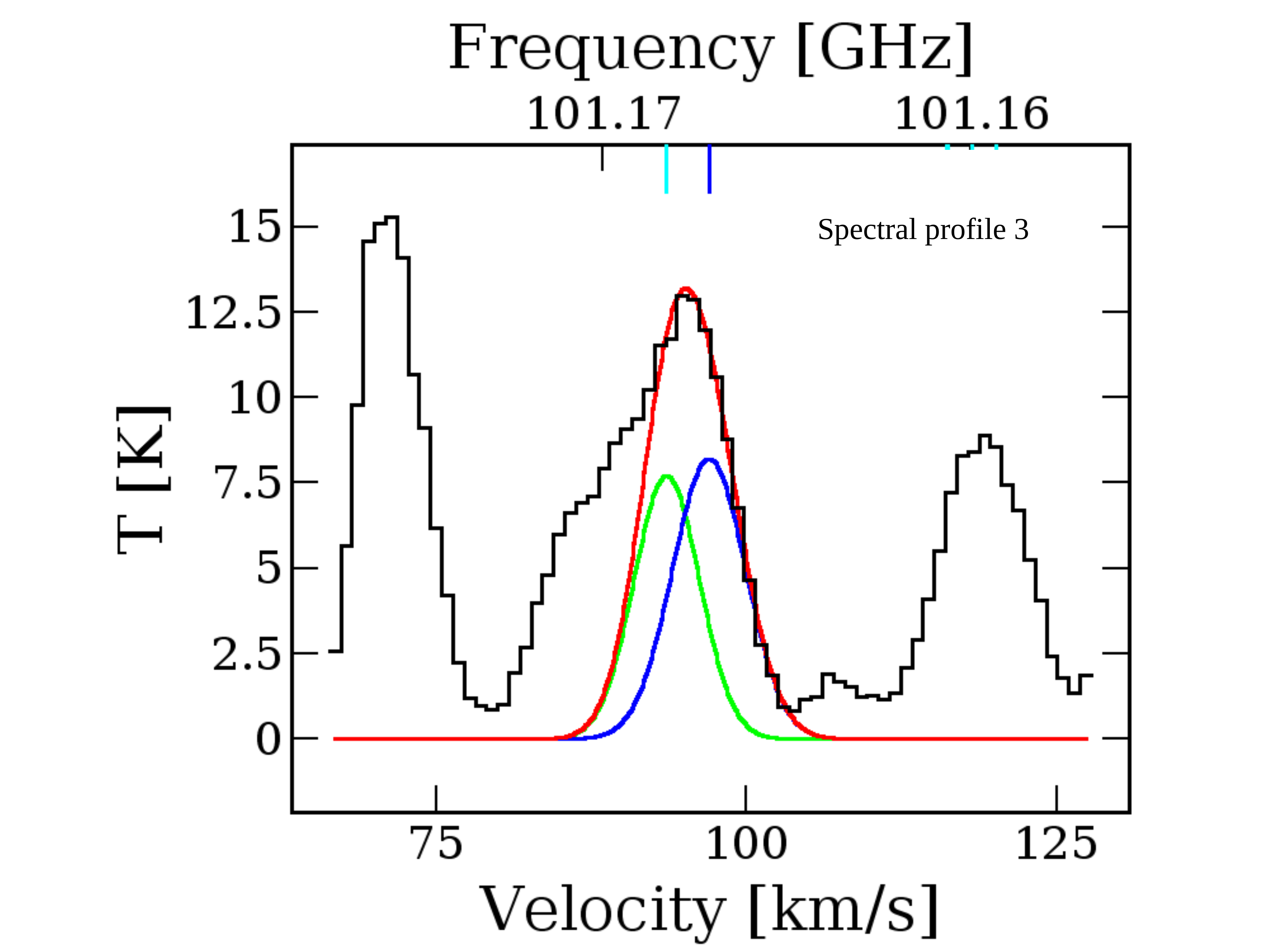}
\end{minipage}
\caption{Black line represents observed emission spectra of methanethiol (CH$_3$SH) and red line represents the resultant Gaussian profile 
of multiple Gaussian components fitted to the observed spectra. In the left panel, blue profile represent the 101.13915 GHz line and green line
represent the 101.13965 GHz transition. In the middle panel, though we observe four transitions, we consider two Gaussian components one (blue profile) 
for 101.15933 GHz and another one (green profile) for other three multiplets (101.15999 GHz, 101.16066 GHz, and 101.16069 GHz). 
In the right panel, blue and green lines represent 101.16715 GHz and 101.16830 GHz transitions respectively.}
\label{Gfit-ch3sh}
\end{figure}

\begin{figure}
\centering
\begin{minipage}{0.35\textwidth}
\includegraphics[width=\textwidth]{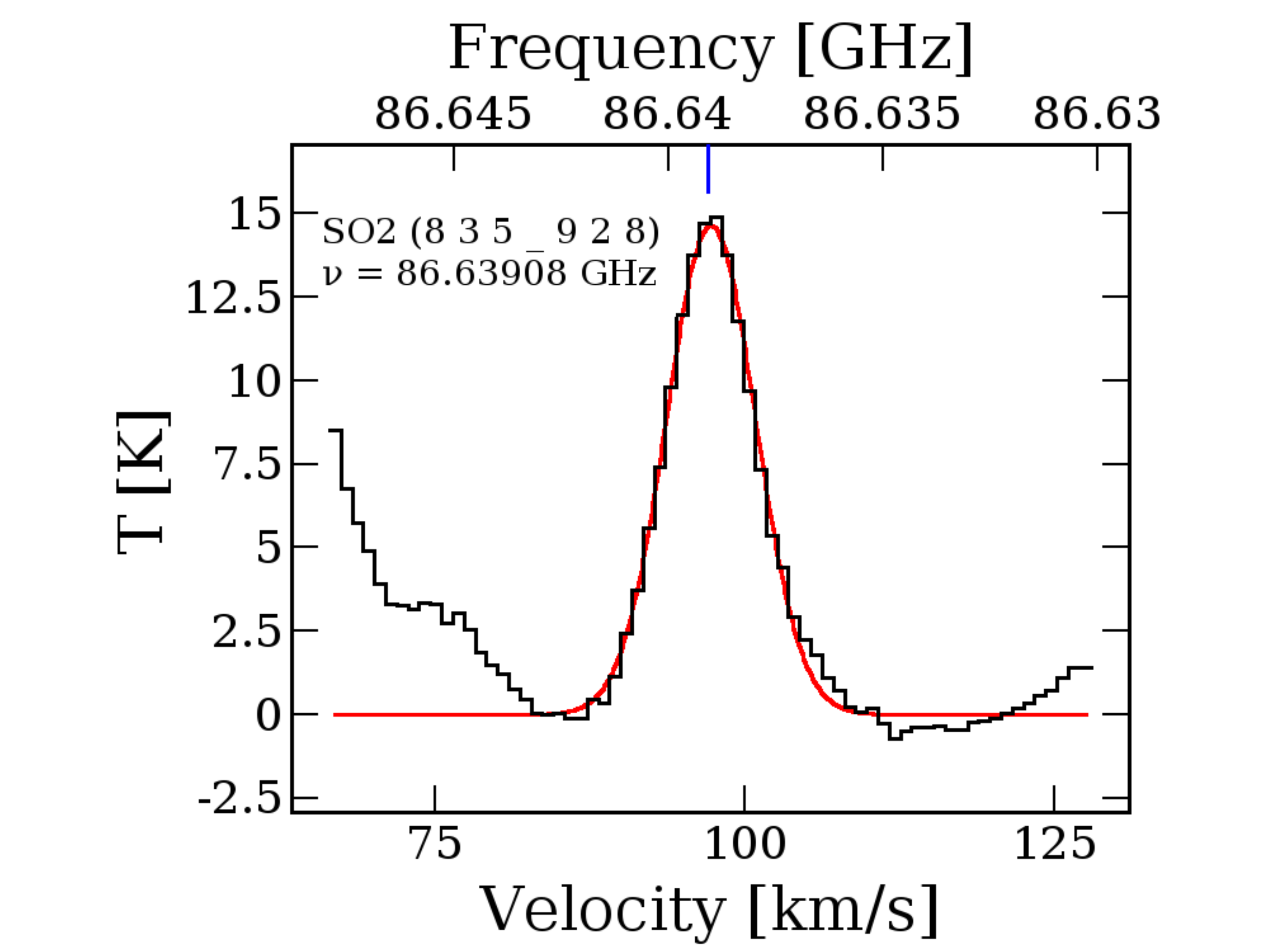}
\end{minipage}
\begin{minipage}{0.35\textwidth}
\includegraphics[width=\textwidth]{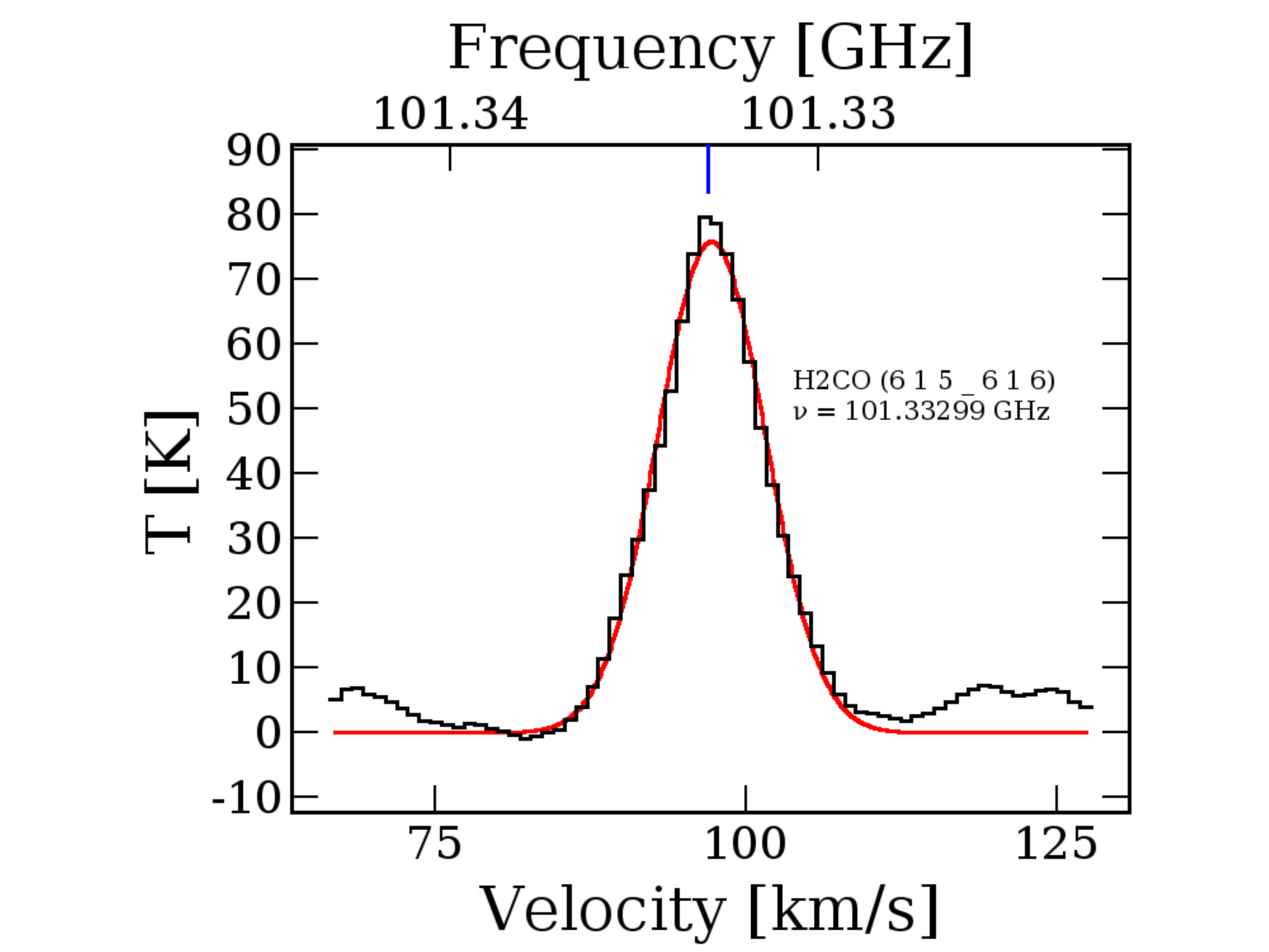}
\end{minipage}
\caption{Black line represents observed emission spectra of SO$_2$, and H$_2$CO and red line represents a Gaussian profile fitted to the observed 
spectra.}
\label{Gfit-other}
\end{figure}

\begin{figure}
\centering
% \begin{minipage}{0.40\textwidth}
\includegraphics[height=7.5cm, width=13cm]{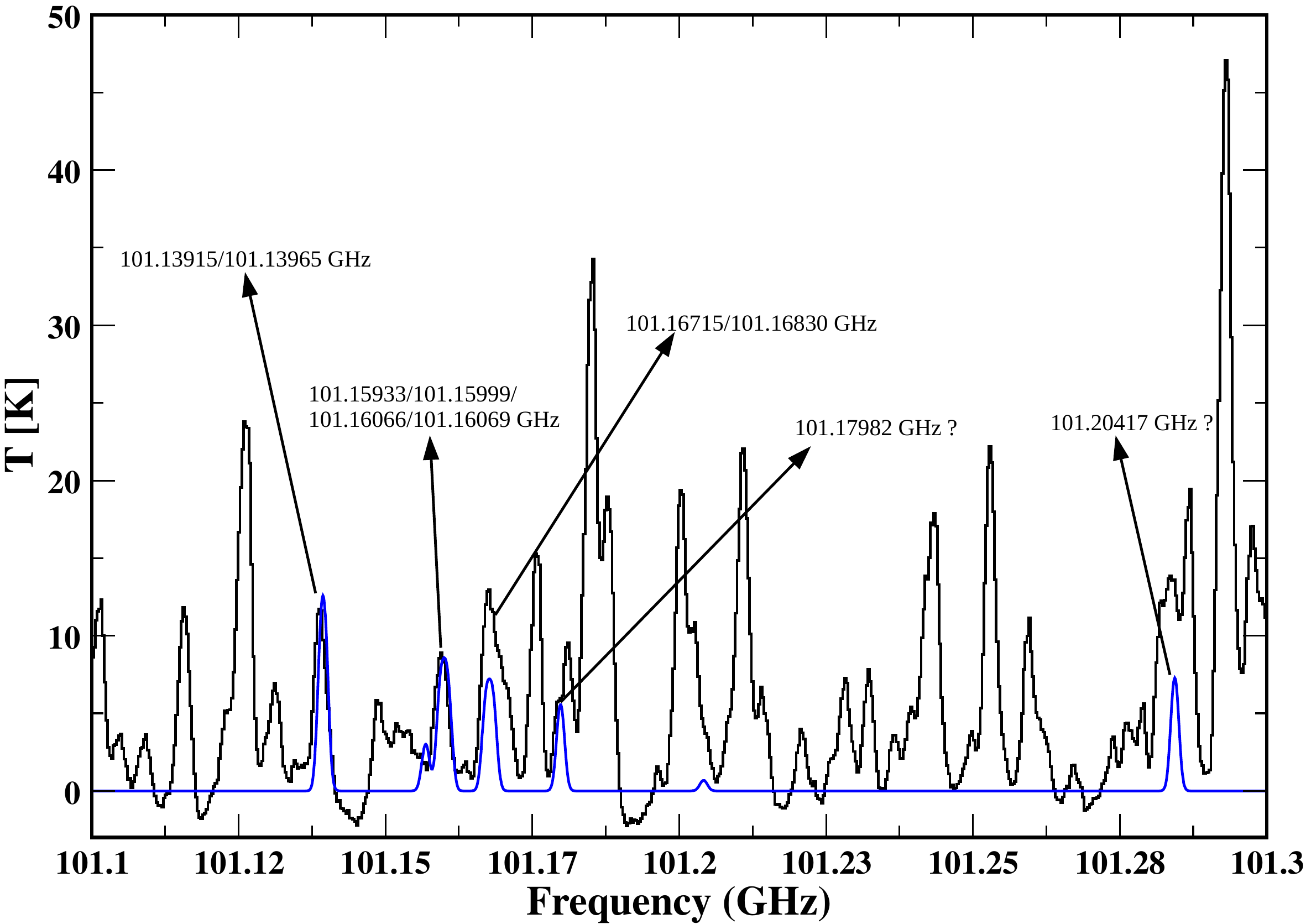}
% \end{minipage}
% \hskip 2cm
% \begin{minipage}{0.40\textwidth}
% \includegraphics[width=\textwidth]{CH3SH-159.pdf}
% \end{minipage}
% \begin{minipage}{0.35\textwidth}
% \includegraphics[width=\textwidth]{ch3sh-LTE.pdf}
% \end{minipage}
\caption{Black line represents observed spectra and blue line represents the (LTE) model  spectra of methanethiol (CH$_3$SH). In addition to the 
observed transitions of CH$_3$SH listed in Table \ref{table:line-parameters}, we have also see signature of two other transitions (101.17982 GHz 
and 101.20417 GHz), which are heavily blended with other species.}

\label{lte-ch3sh}
\end{figure}

\begin{figure*}[t]
\centering
\includegraphics[height=8cm,width=14cm]{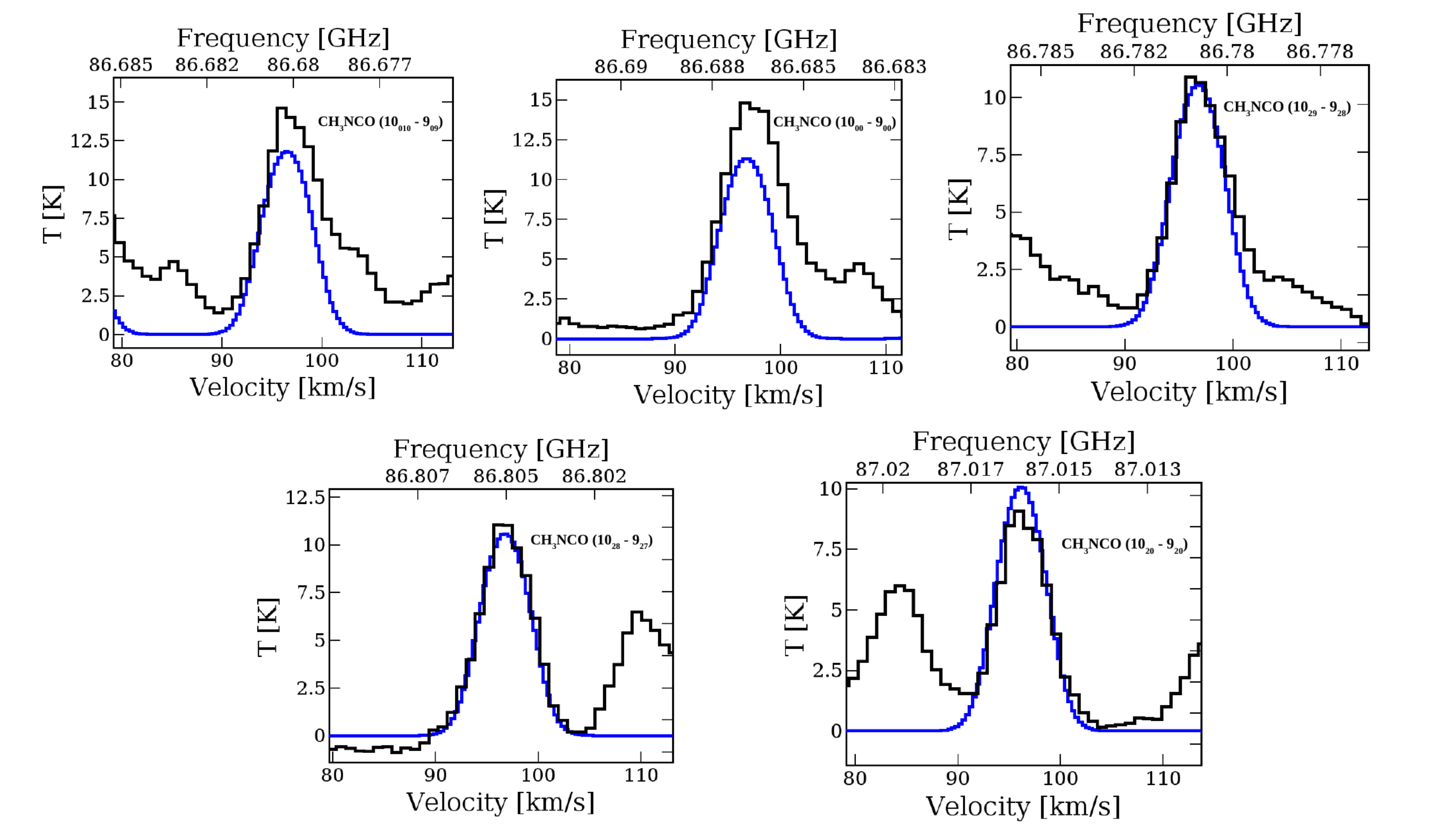}

\caption{Black line represents observed spectra and blue line represents the (LTE) model spectra of methyl isocyanate (CH$_3$NCO).}
\label{lte-ch3nco}
\end{figure*}

\begin{figure}
% \begin{minipage}{0.35\textwidth}
% \includegraphics[width=\textwidth]{ch3oh-726-633.pdf}
% \end{minipage}
\begin{minipage}{0.35\textwidth}
\includegraphics[width=\textwidth]{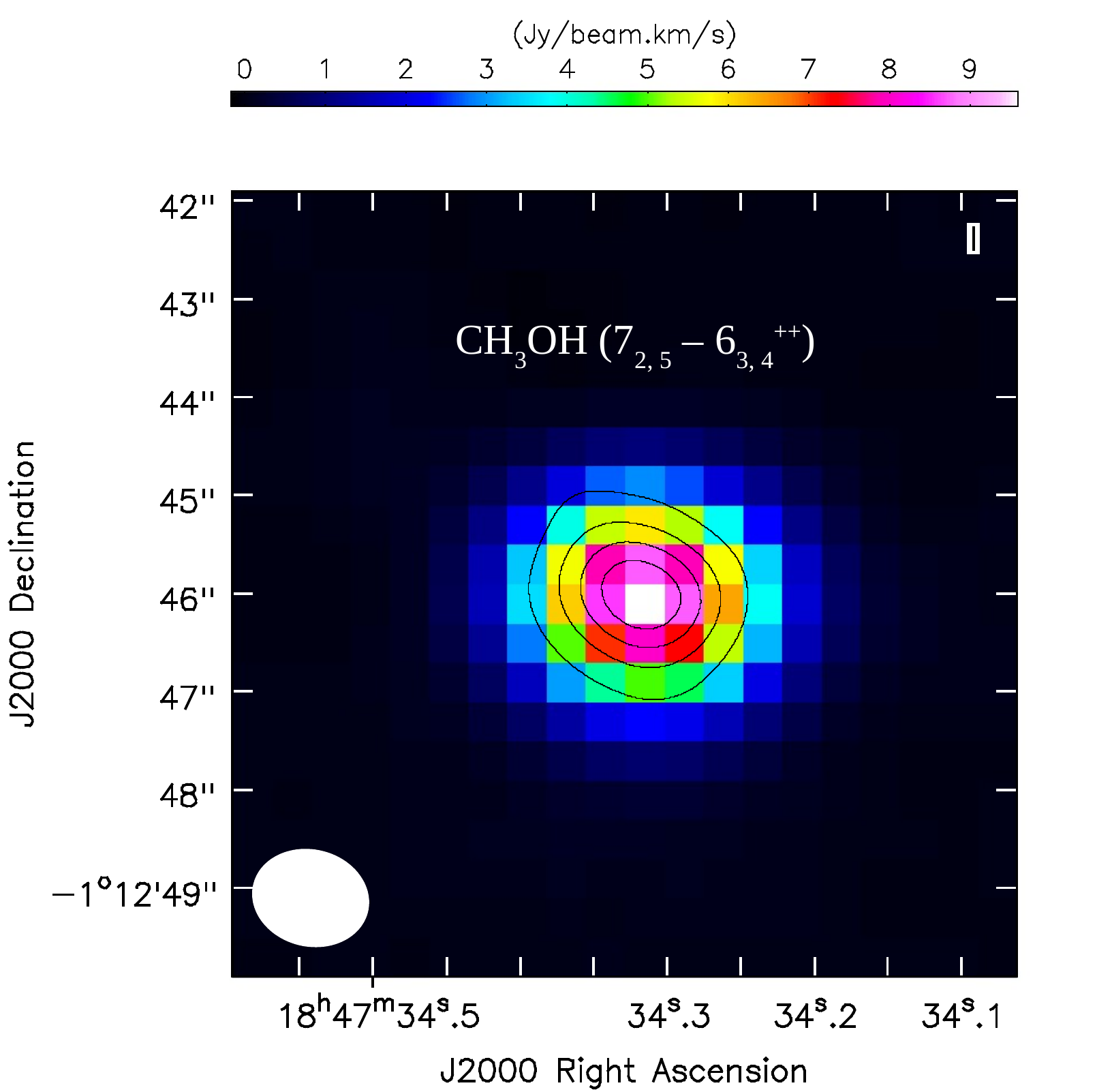}
\end{minipage}
\begin{minipage}{0.35\textwidth}
\includegraphics[width=\textwidth]{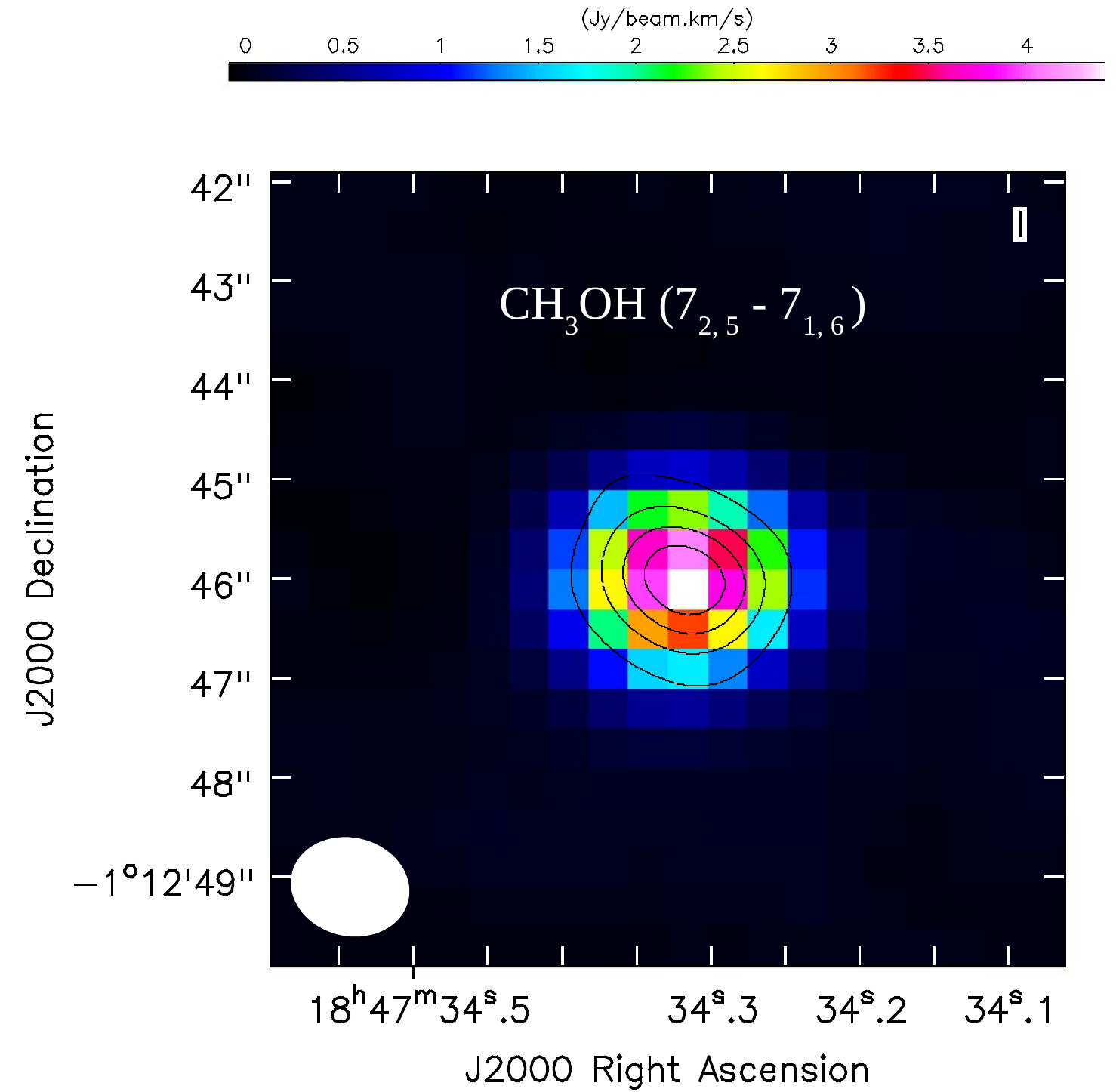}
\end{minipage}
\begin{minipage}{0.35\textwidth}
\includegraphics[width=\textwidth]{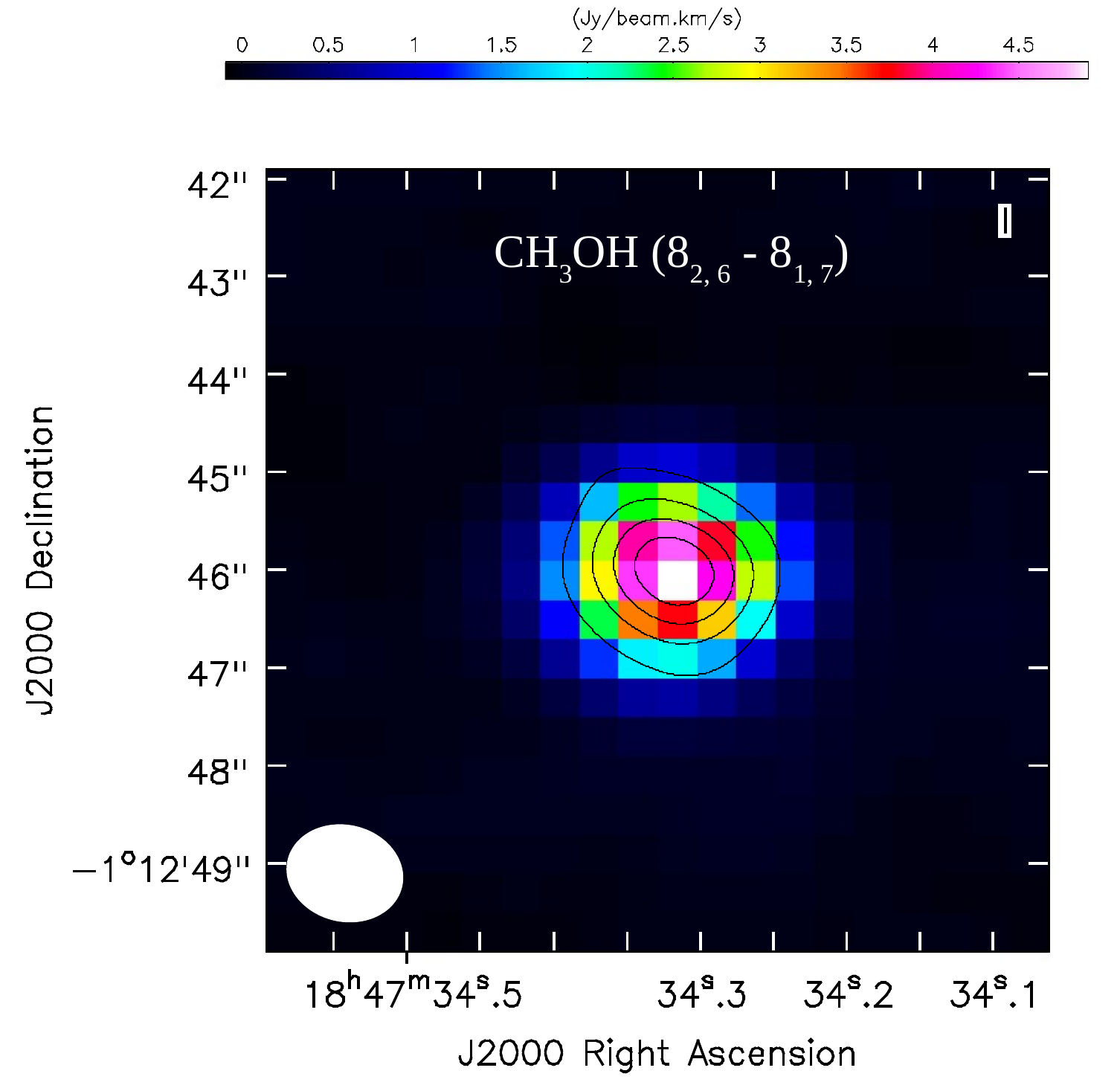}
\end{minipage}
\begin{minipage}{0.35\textwidth}
\includegraphics[width=\textwidth]{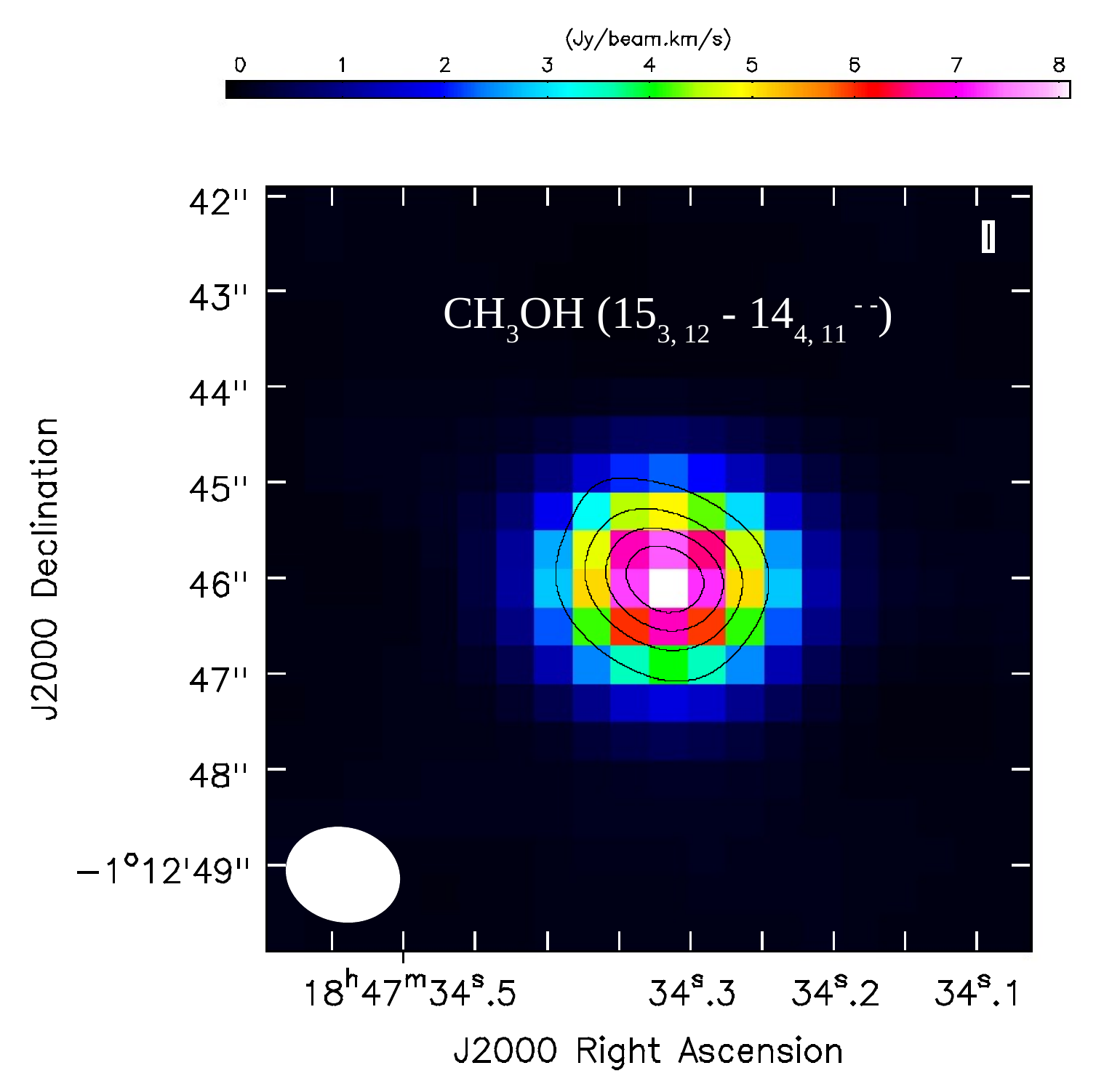}
\end{minipage}
\begin{minipage}{0.35\textwidth}
\includegraphics[width=\textwidth]{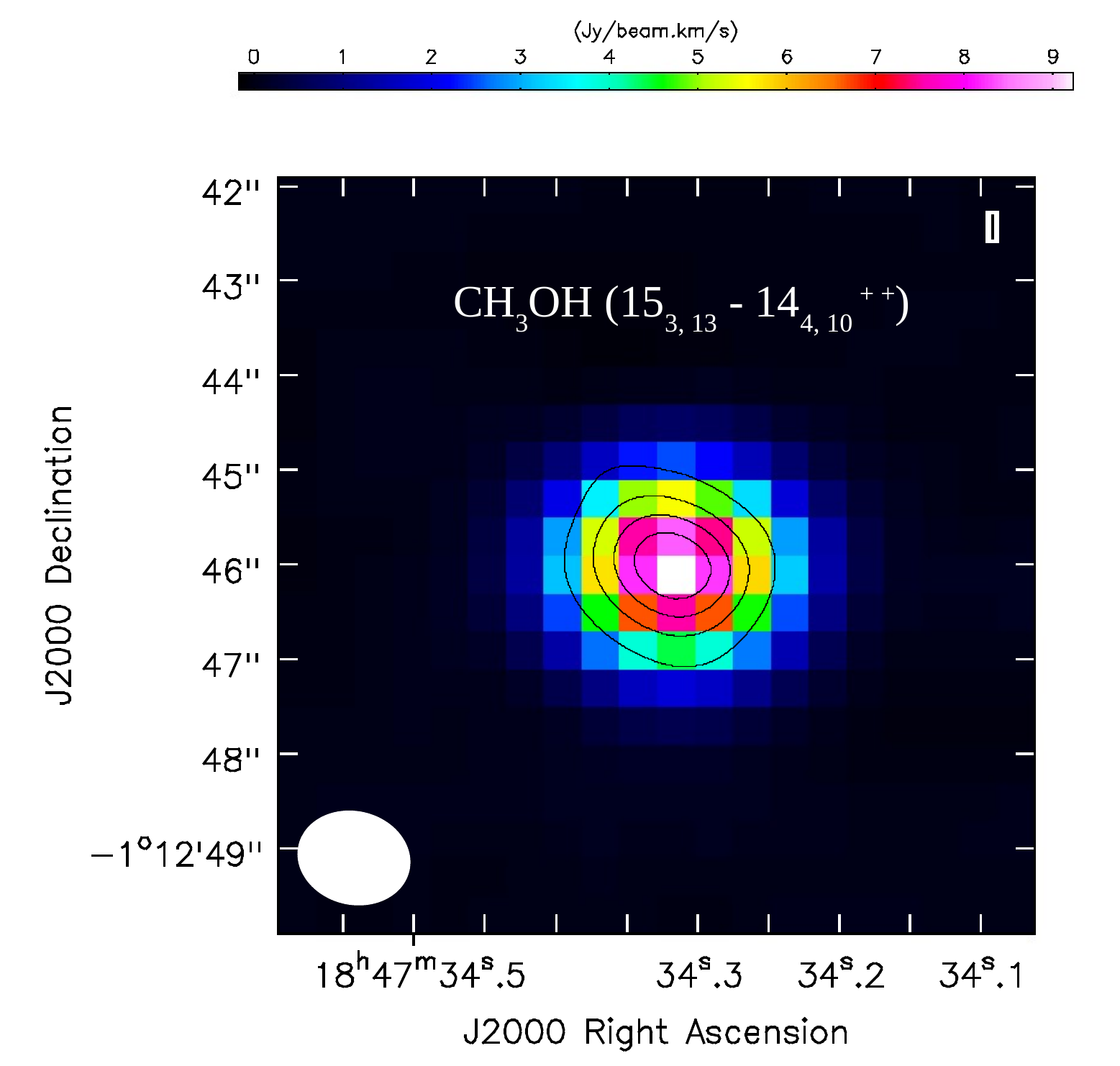}
\end{minipage}
\begin{minipage}{0.35\textwidth}
\includegraphics[width=\textwidth]{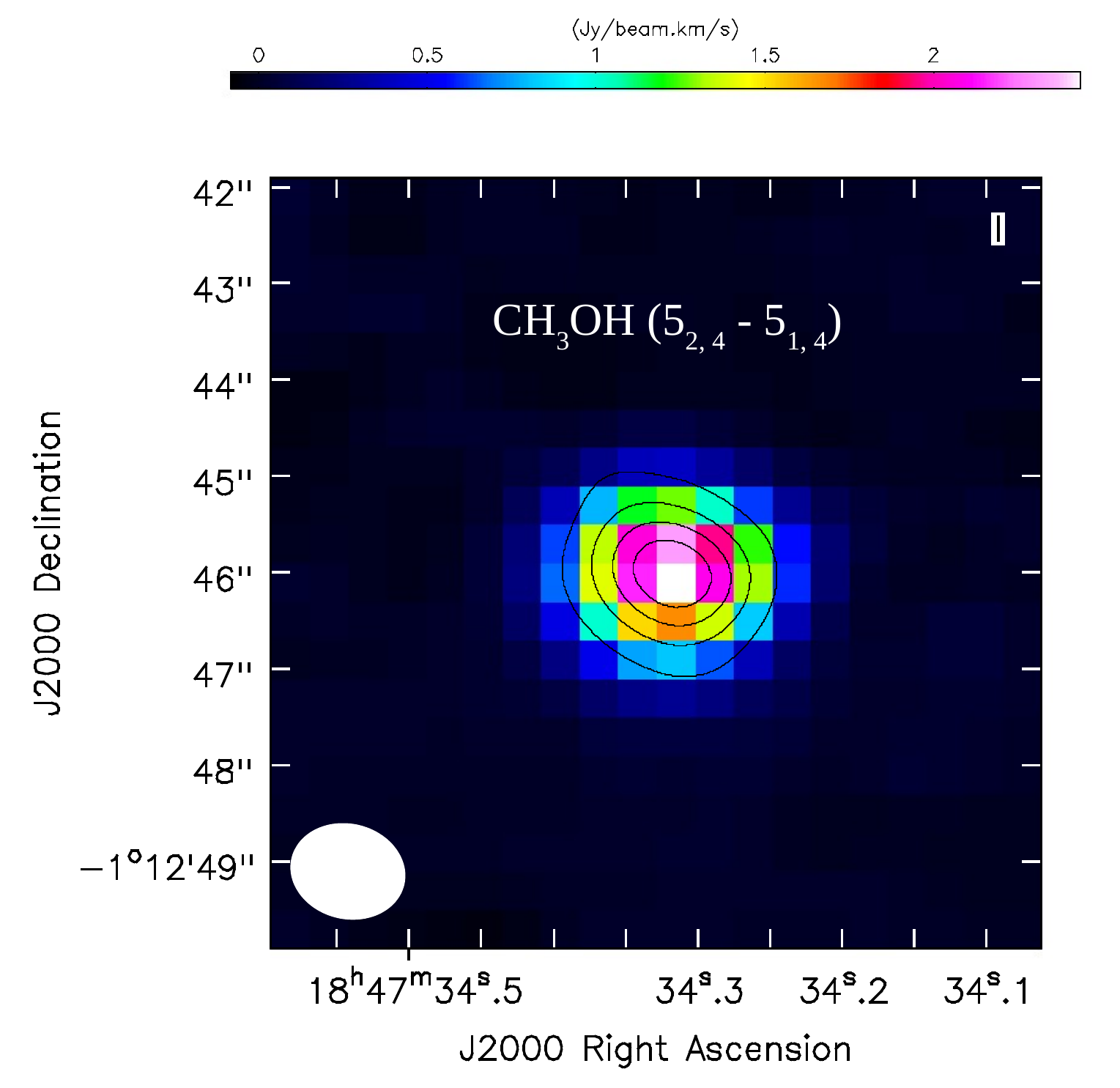}
\end{minipage}
\caption{Moment 0 maps of the integrated intensity distribution (color) of CH$_3$OH transitions is overlaid on the 3.1 mm continuum 
emission (black contours). Contour levels are at 20\%, 40\%, 60 \%, and 80\% of the peak flux of the continuum image. The synthesized 
beam is shown in the lower left-hand corner of each figure.}
\label{int-int-ch3oh}
\end{figure}

\begin{figure}
\begin{minipage}{0.35\textwidth}
\includegraphics[width=\textwidth]{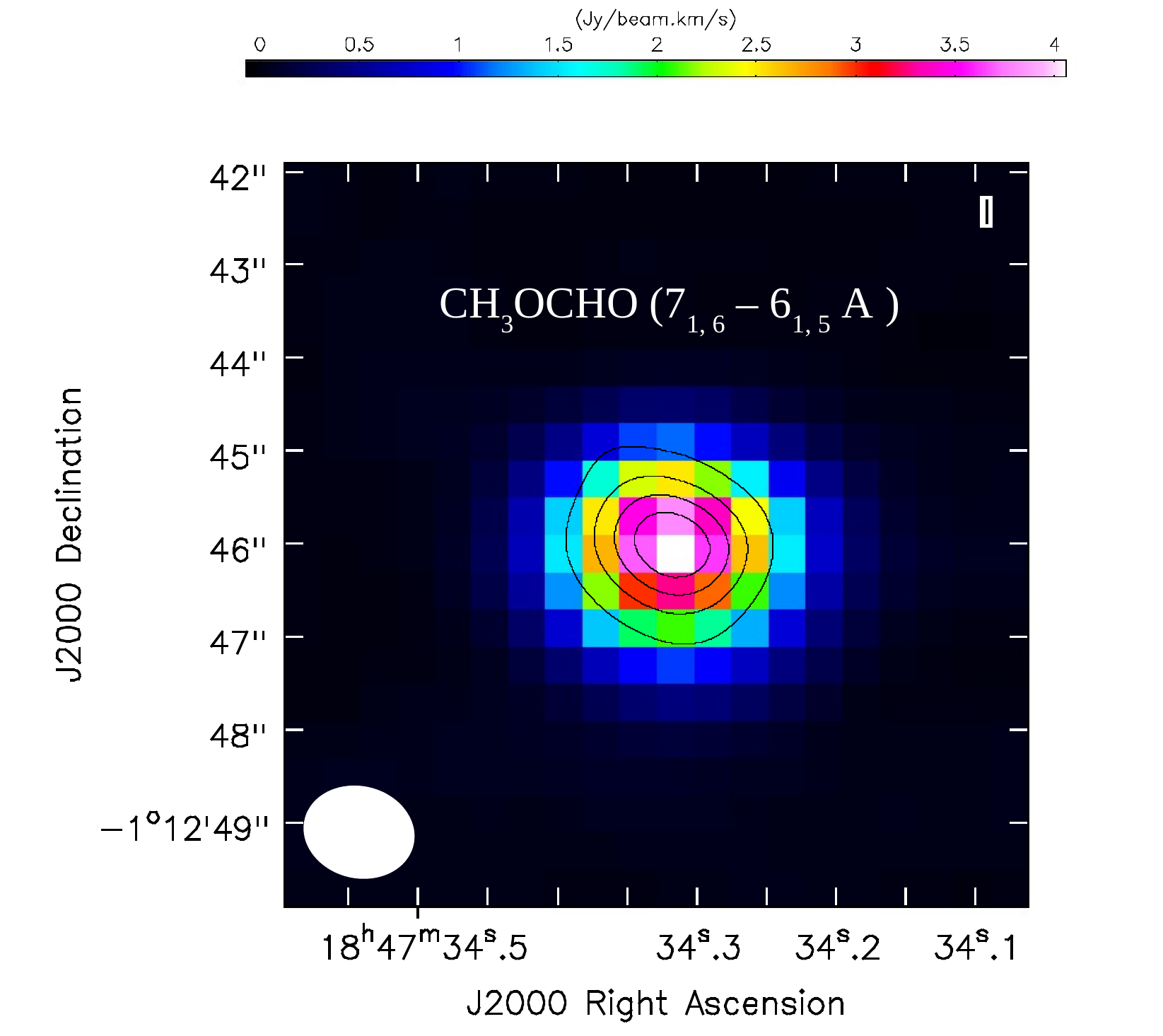}
\end{minipage}
\begin{minipage}{0.35\textwidth}
\includegraphics[width=\textwidth]{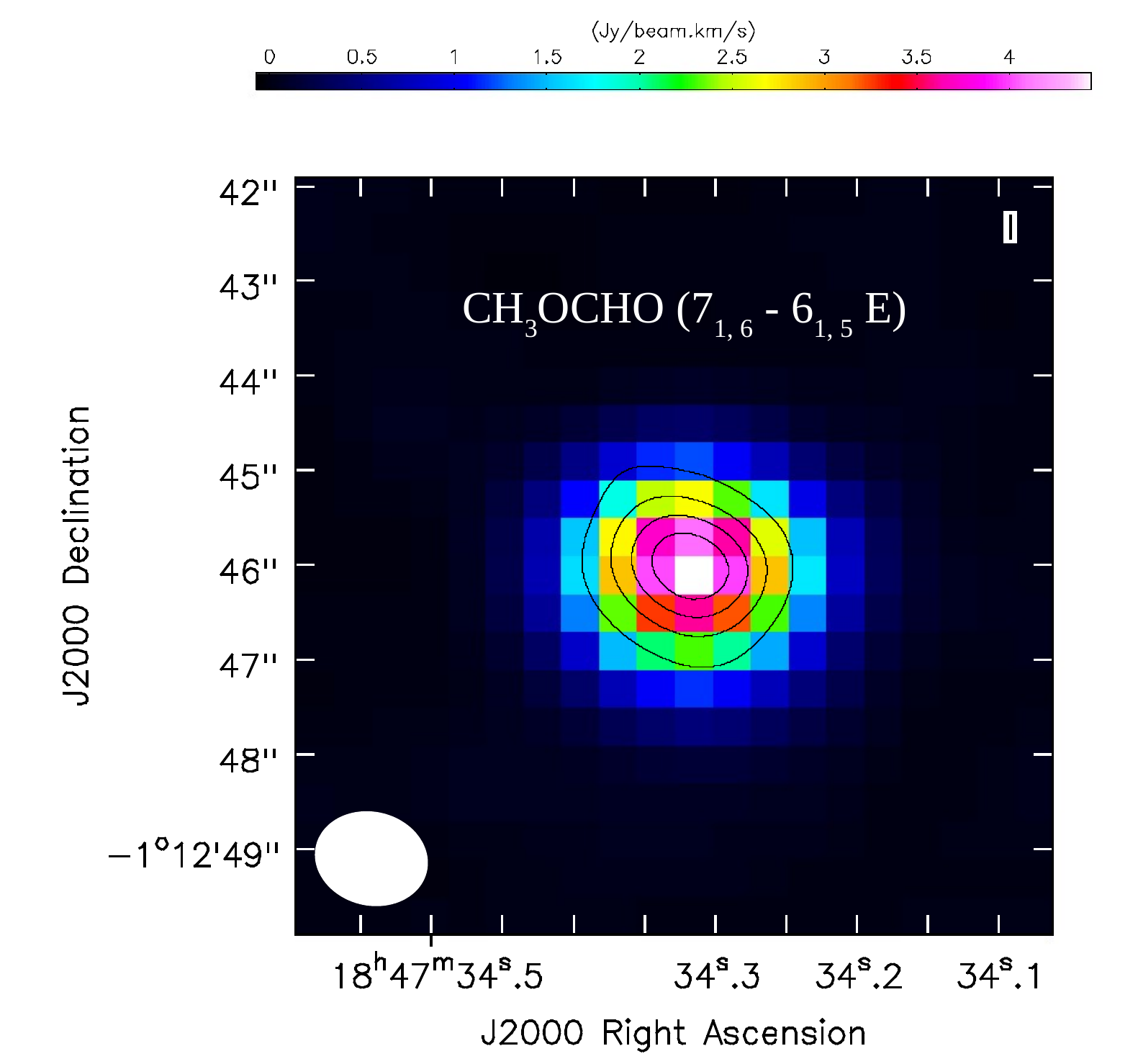}
\end{minipage}
\begin{minipage}{0.35\textwidth}
\includegraphics[width=\textwidth]{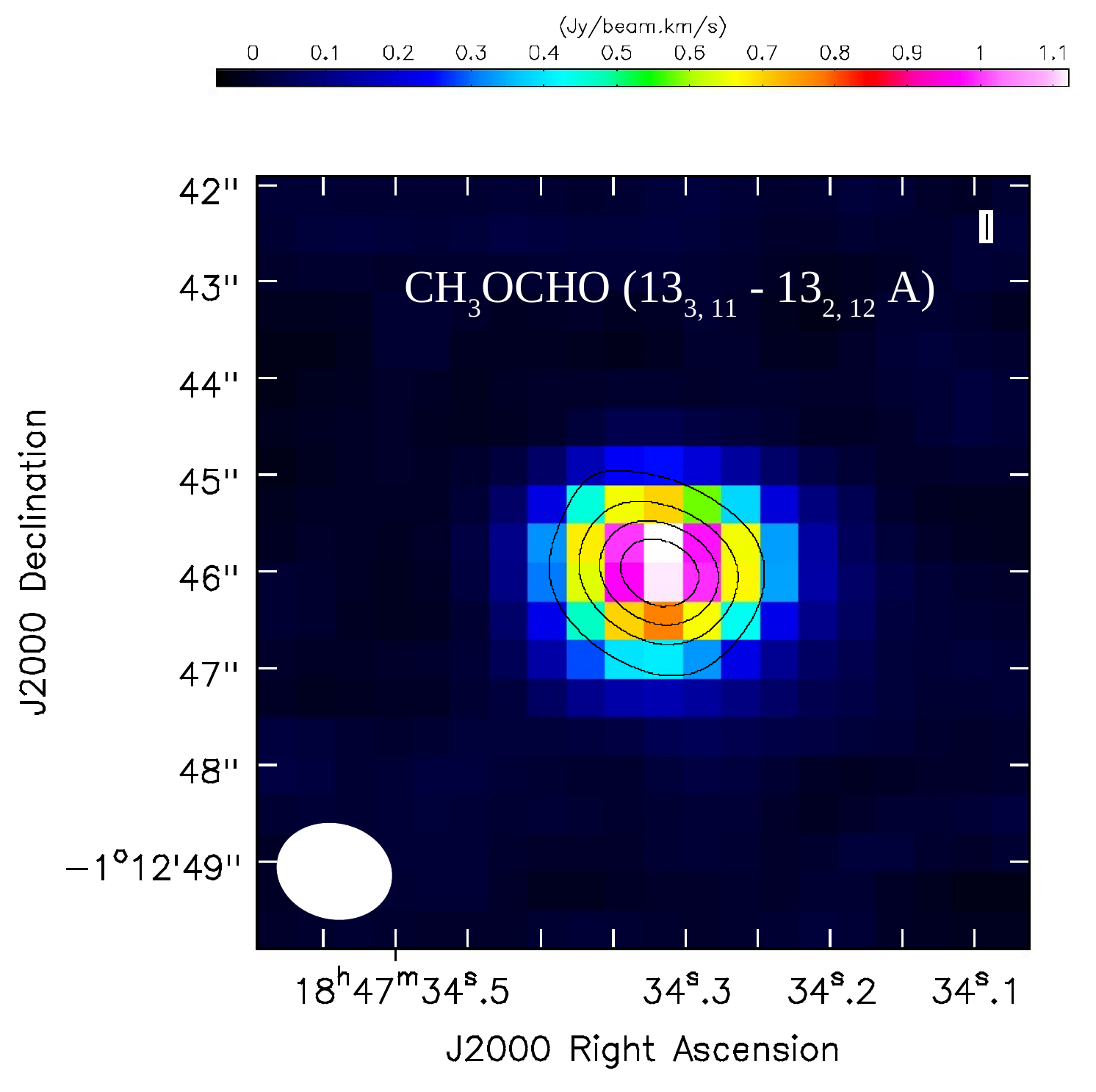}
\end{minipage}
\begin{minipage}{0.35\textwidth}
\includegraphics[width=\textwidth]{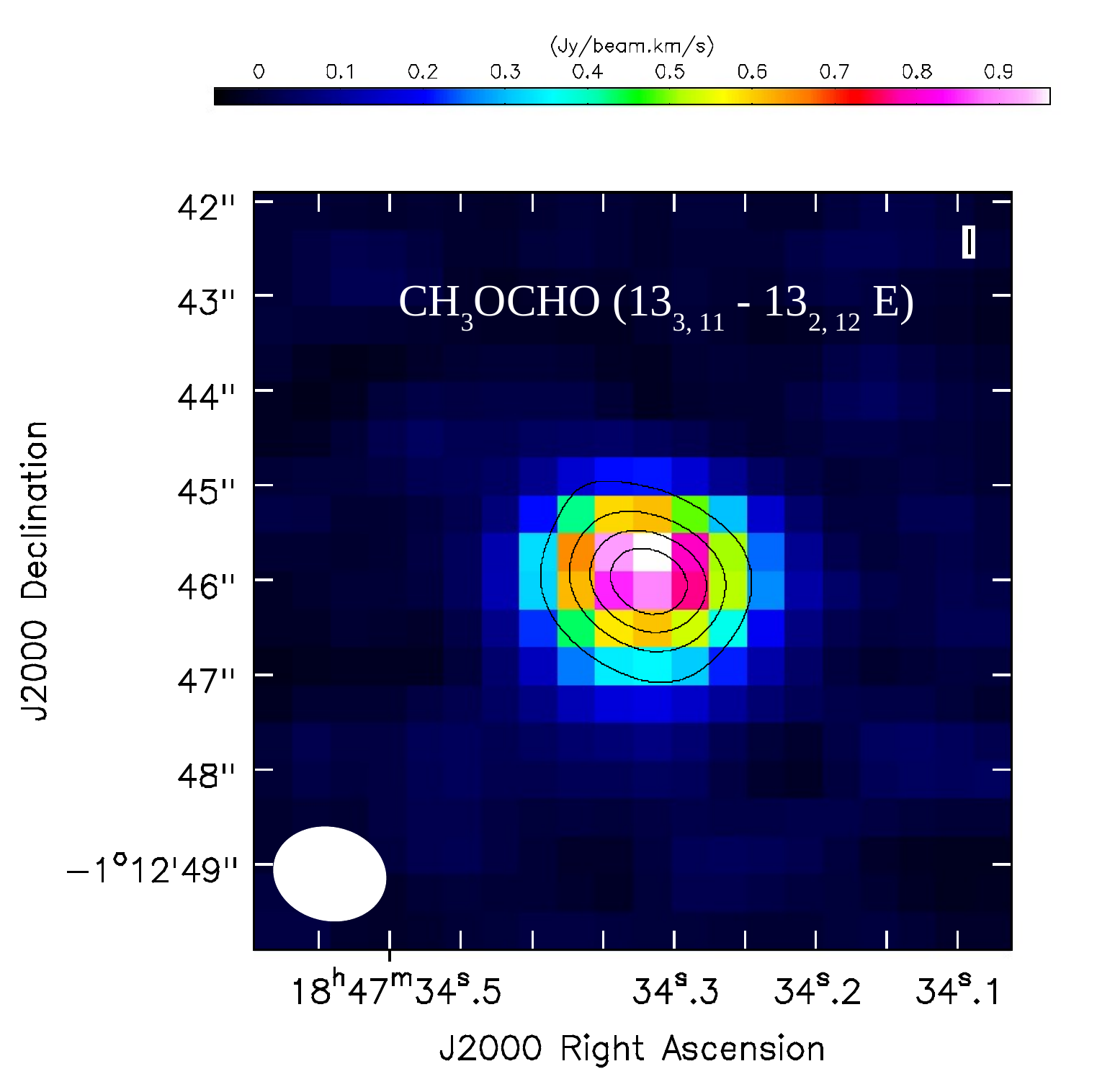}
\end{minipage}
\begin{minipage}{0.35\textwidth}
\includegraphics[width=\textwidth]{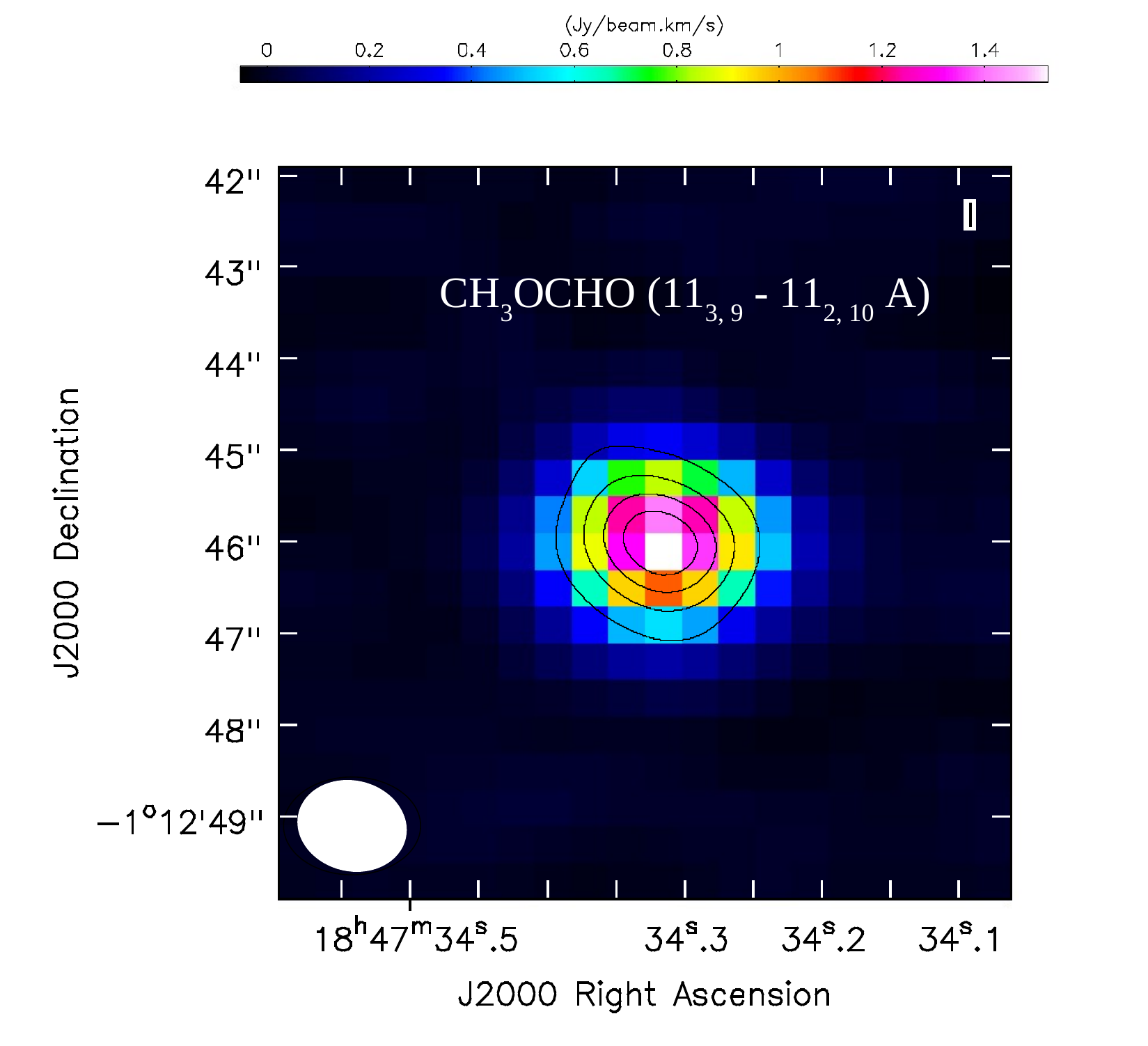}
\end{minipage}
\begin{minipage}{0.35\textwidth}
\includegraphics[width=\textwidth]{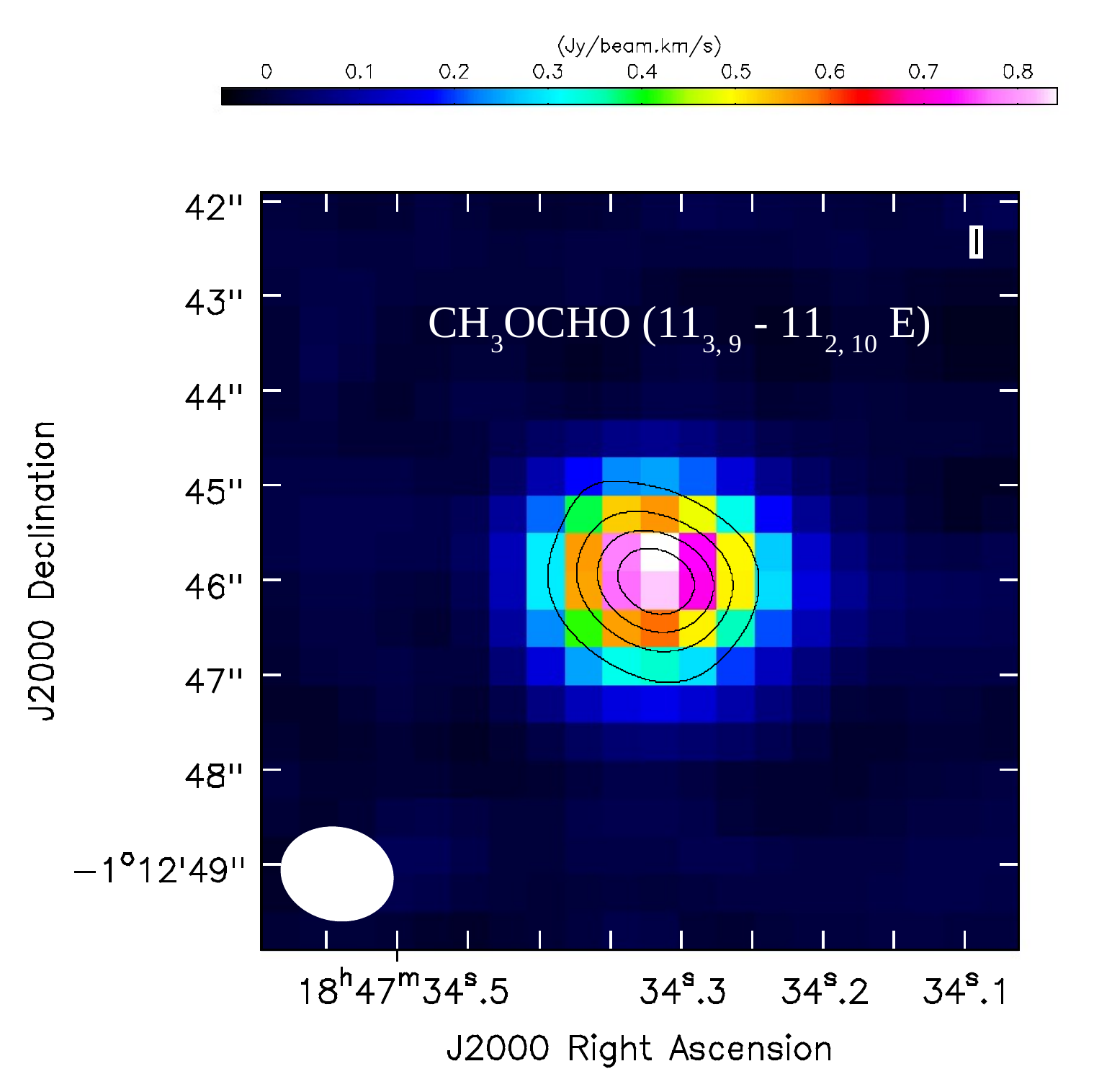}
\end{minipage}
% \begin{minipage}{0.35\textwidth}
% \includegraphics[width=\textwidth]{ch3ocho-818-717E.pdf}
% \end{minipage}
\caption{{Moment 0 of the integrated intensity distribution (color) of CH$_3$OCHO transitions is 
overlaid on the 3.1 mm continuum emission (black contours). Contour levels are at 20\%, 40\%, 60 \%, and 80\% of the peak flux of 
the continuum image. The synthesized beam is shown in the lower left-hand corner of each figure.}}
\label{int-int-ch3ocho}
\end{figure}

\begin{figure}
\begin{minipage}{0.35\textwidth}
\includegraphics[width=\textwidth]{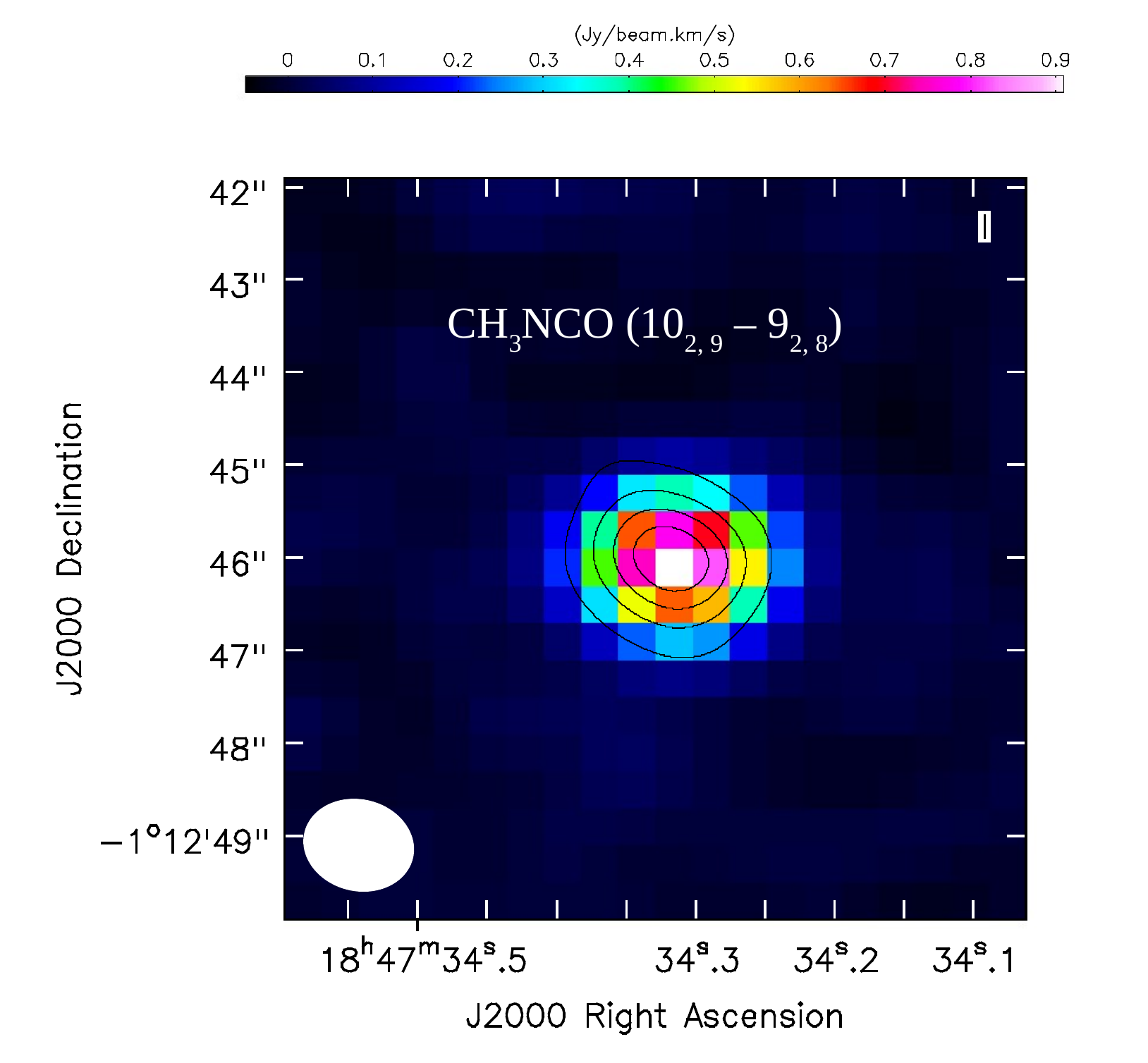}
\end{minipage}
\begin{minipage}{0.35\textwidth}
\includegraphics[width=\textwidth]{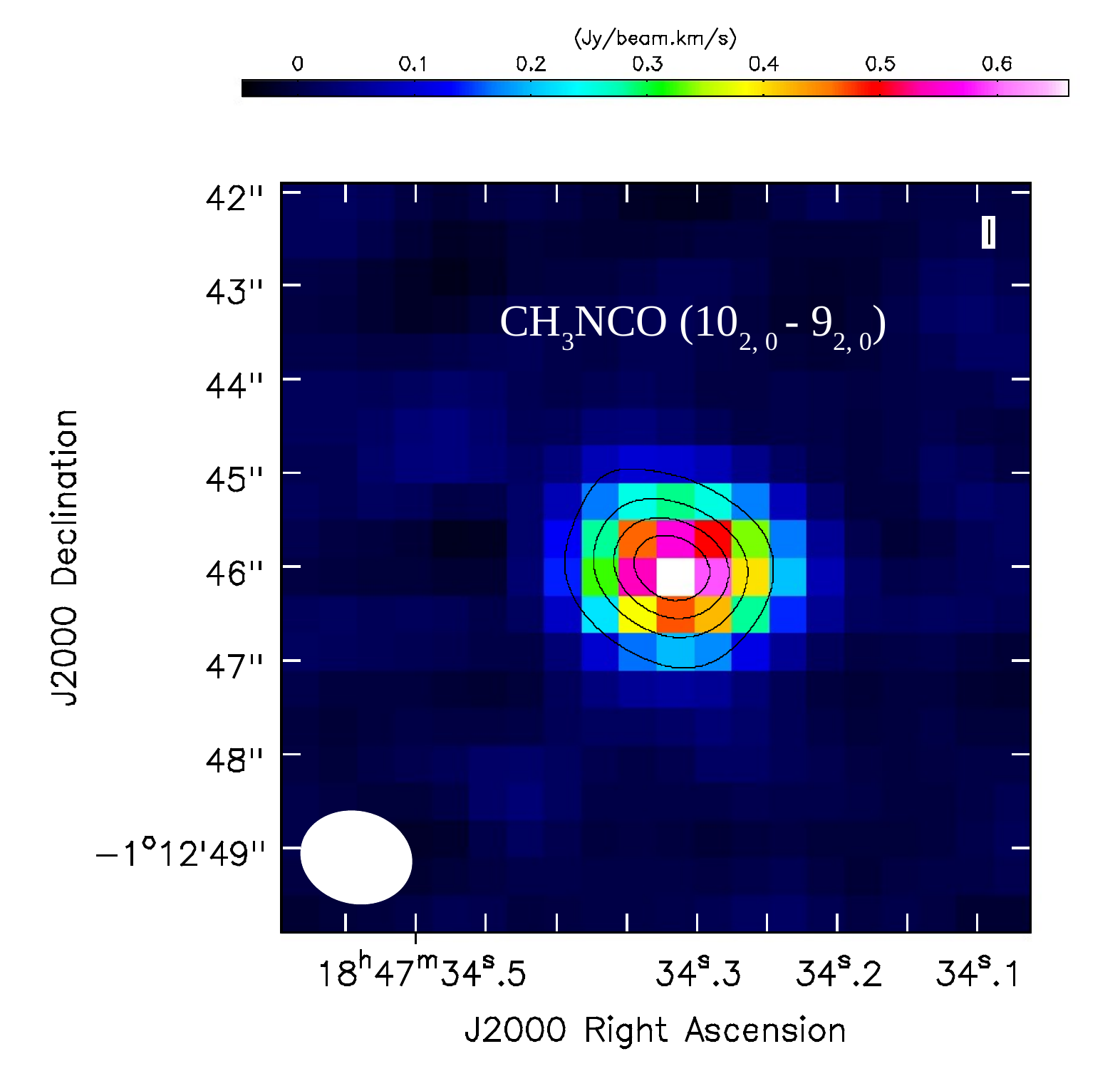}
\end{minipage}
\begin{minipage}{0.35\textwidth}
\includegraphics[width=\textwidth]{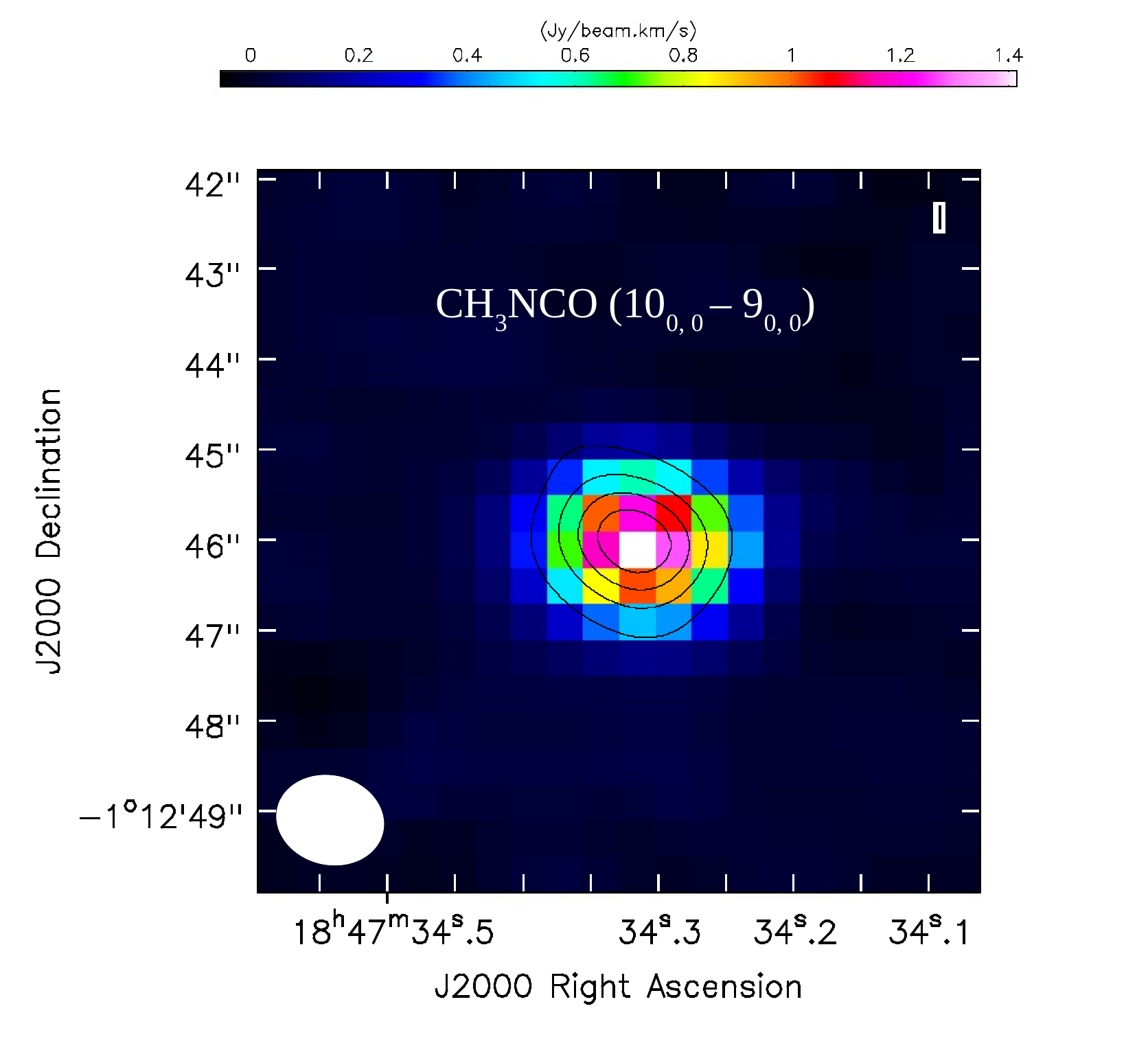}
\end{minipage}
\begin{minipage}{0.35\textwidth}
\includegraphics[width=\textwidth]{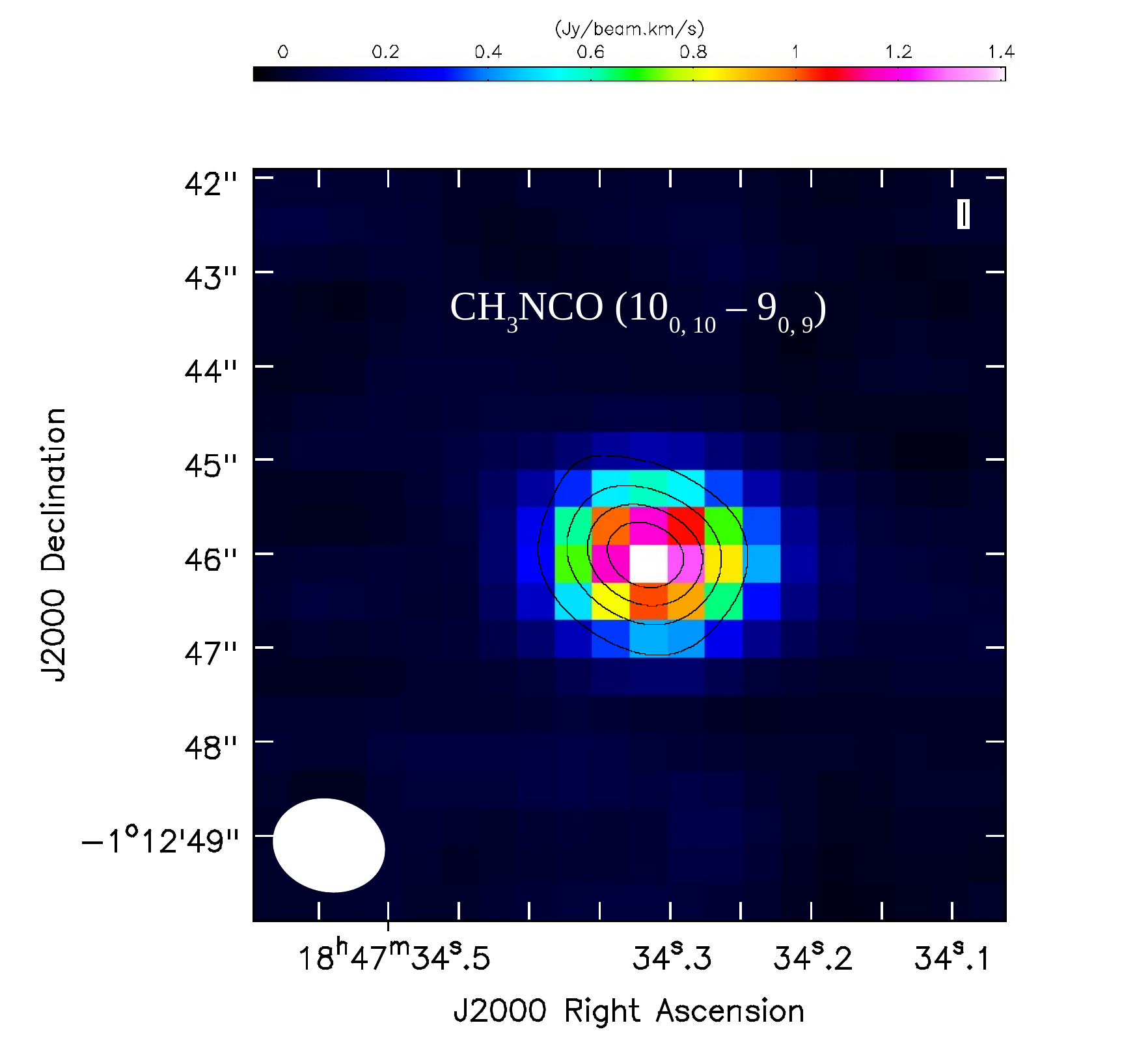}
\end{minipage}
\begin{minipage}{0.35\textwidth}
\includegraphics[width=\textwidth]{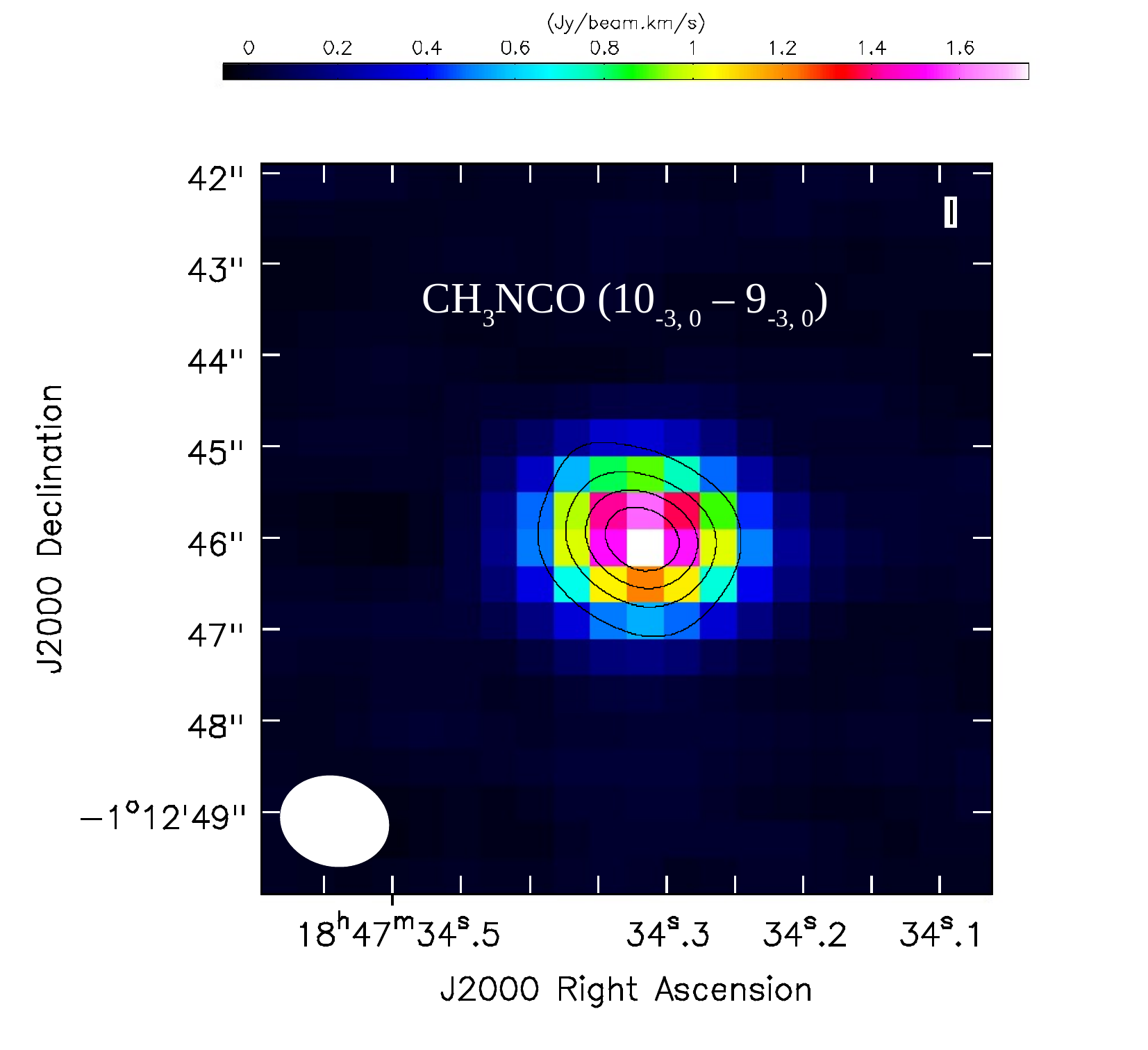}
\end{minipage}
\caption{{Moment 0 of the integrated intensity distribution (color) of CH$_3$NCO transitions is overlaid on the 3.1 mm continuum 
emission (black contours). Contour levels are at 20\%, 40\%, 60 \%, and 80\% of the peak flux of the continuum image. The synthesized beam 
is shown in the lower left-hand corner of each figure.}}
\label{int-int-ch3nco}
\end{figure}

\begin{figure}
\centering
\begin{minipage}{0.35\textwidth}
\includegraphics[width=\textwidth]{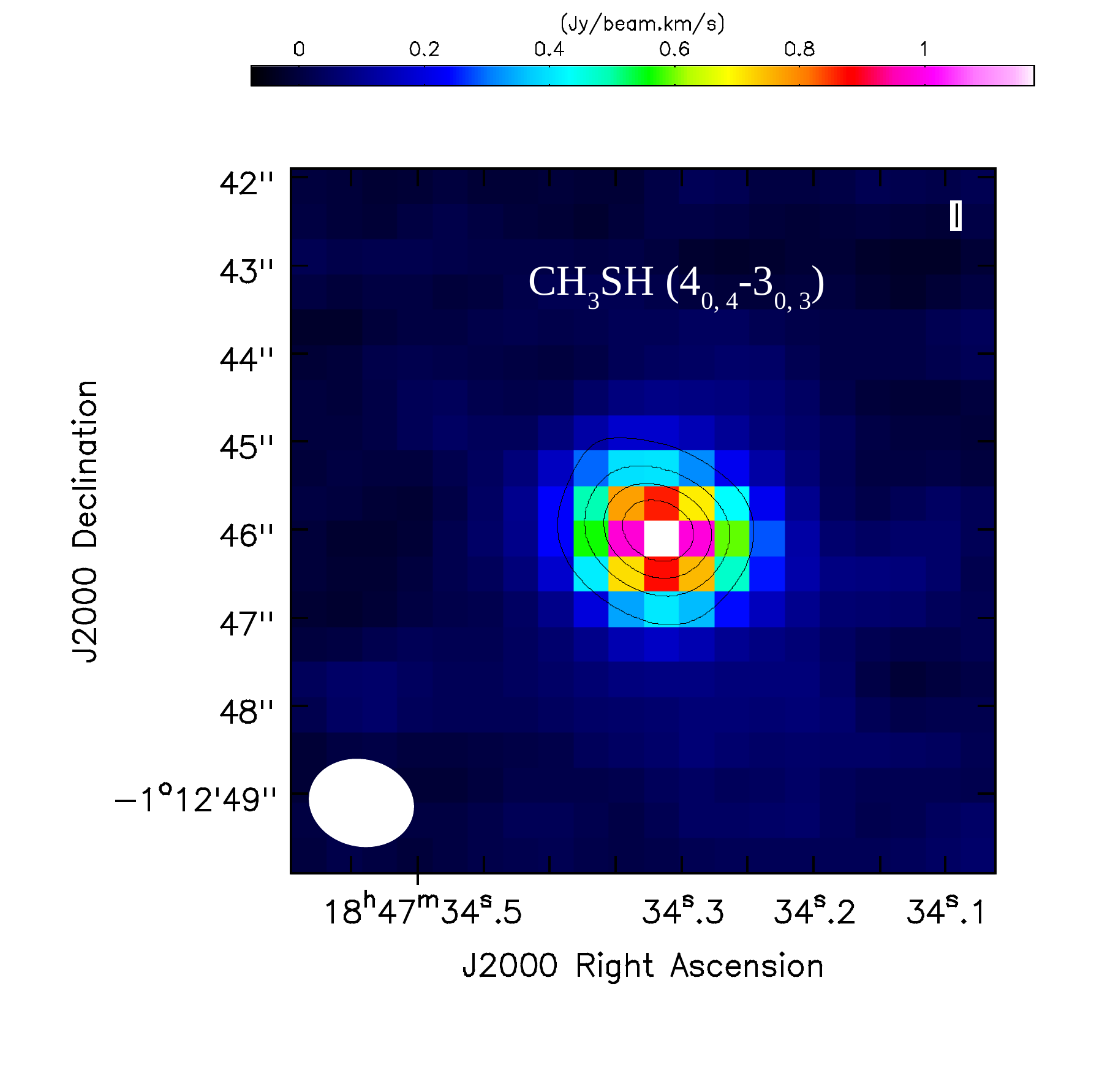}
\end{minipage}
\begin{minipage}{0.35\textwidth}
\includegraphics[width=\textwidth]{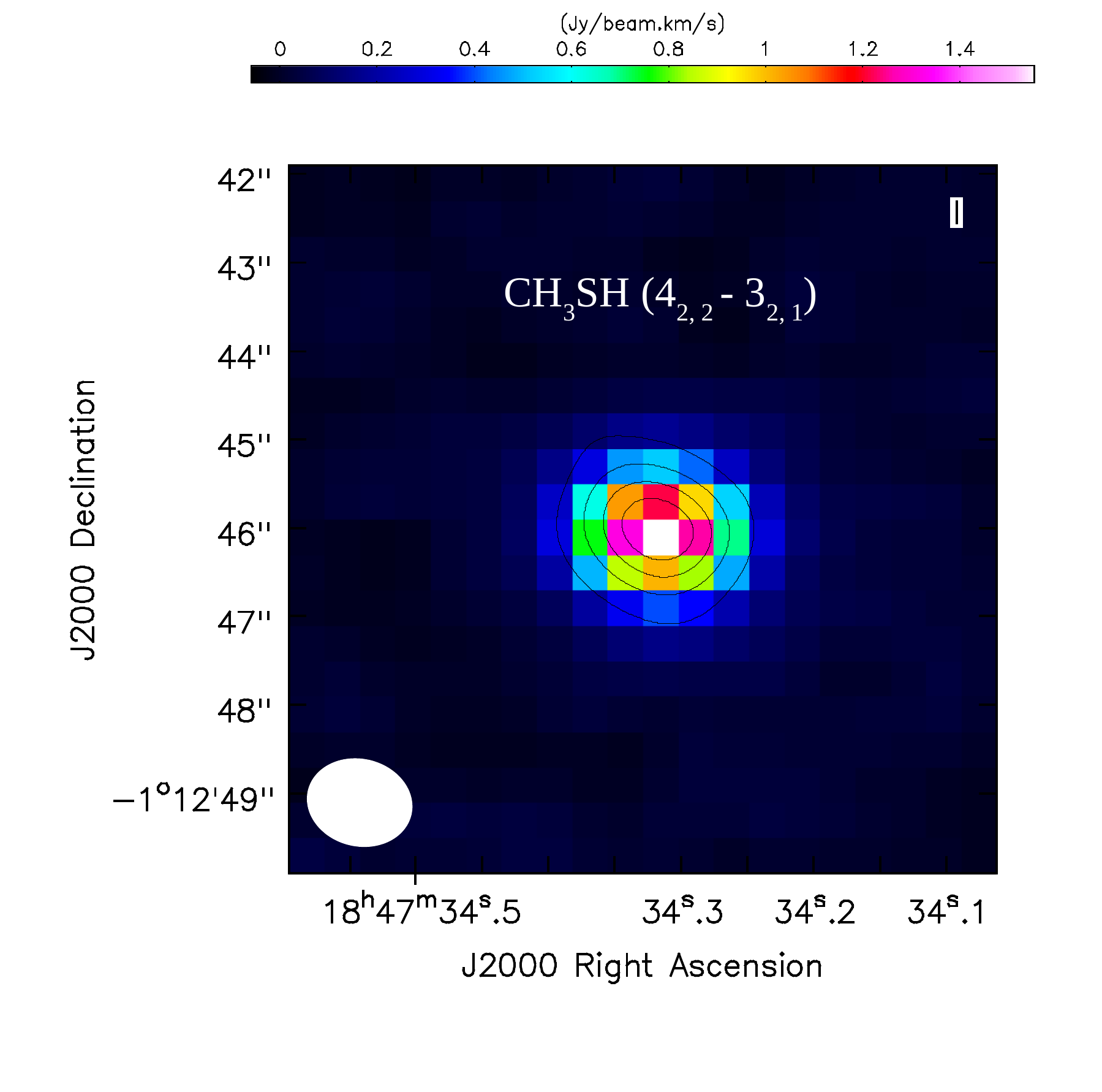}
\end{minipage}
\caption{Moment 0 of the integrated intensity distribution (color) of CH$_3$SH transitions is 
overlaid on the 3.1 mm continuum emission (black contours). Contour levels are at 20\%, 40\%, 60 \%, and 80\% of the peak flux of 
the continuum image. The synthesized beam is shown in the lower left-hand corner of each figure. Multiple transitions (101.13915/101.13965 GHz) 
appear in CH$_3$SH 4$_{0,4}$-3$_{0,3}$ moment maps. Similarly multiple transitions (101.16715/101.16830 GHz) appear in  CH$_3$SH 4$_{2,2}$-3$_{2,1}$ 
moment maps.}
\label{int-int-ch3sh}
\end{figure}

\clearpage
\section{Summary of the emitting regions of the observed species.}
\begin{table}[t]
\centering
\caption{Emitting region of observed transitions of COMs}
\begin{tabular}{ccccc}
\hline
\hline
Species& Transition & Emitting&Upper State\\ 
&Frequency & region & Energy \\
&(GHz)&($''$)&(K)\\
\hline
\hline
CH$_3$SH&101.13915$^{a}$&0.89$\pm$0.06&12.14\\
&101.15999$^{b}$&0.80$\pm$0.02&52.39\\
&101.16830$^{c}$&0.80$\pm$0.05&30.27\\
&&&\\
CH$_3$OH&86.61557&1.35$\pm$0.01&102.70\\
&86.90291&1.35$\pm$0.02&102.72\\
&88.59478&1.29$\pm$0.02&328.28\\
&88.93997&1.27$\pm$0.01&328.28\\
&101.12685&1.14$\pm$0.03&60.73\\
&101.29341&1.21 $\pm$0.02&90.91\\
&101.46980&1.23 $\pm$0.03&109.49\\
&&&\\
CH$_3$NCO&86.68019&0.82$\pm$0.01&22.82\\
&86.68655&0.83$\pm$0.02&34.97\\
&86.78077&0.78$\pm$0.02&46.73\\
&86.80502&0.77$\pm$0.01&46.74\\
&87.01607&0.76$\pm$0.02&58.88\\
 
&&&\\
CH$_3$OCHO&88.6869&1.37$\pm$0.04&44.97\\
&88.72326&1.10$\pm$0.03&44.96\\
&88.84318&1.36$\pm$0.02&17.96\\
&88.85160&1.42$\pm$0.02&17.94\\
&88.86241&1.15$\pm$0.04&207.10\\
&101.37054&1.46$\pm$0.05&59.64\\
&101.41474&1.33$\pm$0.06&59.63\\
&&&\\
SO$_2$&86.63908&0.71$\pm$0.01&55.20\\
H$_2$CO&101.33299&1.31$\pm$0.02&87.57\\
\hline
\hline
\end{tabular}
\\
{NOTES: $^{a}$ multiple transitions: 101.13915/101.13965 GHz, $^{b}$ multiple transitions: 101.15999/101.15933/101.16066/101.16069 GHz, 
$^{c}$ multiple transitions: 101.16715/101.16830 GHz.}
\label{table:emitting-regions}
\end{table}

\end{document}